\documentclass[10pt,journal]{IEEEtran} 

\IEEEoverridecommandlockouts

\usepackage{graphicx}
\usepackage{color}
\usepackage{cite}
\usepackage{amsmath}
\usepackage[caption=false]{subfig}
\usepackage{amssymb}
\usepackage{comment}

\usepackage{mathtools}
\usepackage{tabulary}
\usepackage{acronym}
\usepackage{upgreek}
\usepackage{algorithm}
\usepackage{algpseudocode}
\usepackage{gensymb}
\usepackage{bbm}
\usepackage{hyperref}

\newtheorem{remark}{Remark}

\DeclareMathOperator*{\argmax}{arg\!\max}

\newacro{ANN}[ANN]{Artificial Neural Network}
\acrodefplural{ANN}[ANNs]{Artificial Neural Networks}

\newacro{RIS}[RIS]{Reconfigurable Intelligent Surface}
\acrodefplural{RIS}[RISs]{Reconfigurable Intelligent Surfaces}

\newacro{TTI}[TTI]{Transmission Time Interval}

\newacro{FET}[FET]{Field-Effect Transistor}
\acrodefplural{FET}[FETs]{Field-Effect Transistors}

\newacro{MEMS}[MEMS]{Micro-ElectroMechanical System}

\newacro{PIN}[PIN]{Positive-Intrinsic-Negative}

\newacro{D2D}[D2D]{Device-to-Device}
\newacro{eMBB}[eMBB]{enhanced Mobile Broadband}

\newacro{CE}[CE]{Channel Estimation}

\newacro{RF}[RF]{Radio Frequency}

\newacro{mmWave}[mmWave]{millimeter Wave}
\newacro{THz}[THz]{Tera-Hertz}

\newacro{URLLC}[URLLC]{Ultra-Reliable and Low Latency Communications}

\newacro{NR}[NR]{New Radio}
\newacro{5G}[5G]{Fifth Generation}
\newacro{6G}[6G]{Sixth Generation}
\newacro{SNR}[SNR]{Signal-to-Noise Ratio}
\newacro{SINR}[SINR]{Signal-to-Interference-plus-Noise Ratio}
\newacro{TX}[TX]{Transmitter}

\newacro{RX}[RX]{Receiver}
\acrodefplural{RX}[RXs]{Receivers} 

\newacro{TVC}[TVC]{Time Varying Channel}
\acrodefplural{TVC}[TVCs]{Time Varying Channels}

\newacro{BS}[BS]{Base Station}
\newacro{UE}[UE]{User Equipment}
\acrodefplural{UE}[UEs]{Users Equipments}

\newacro{MSE}[MSE]{Mean Squared Error}

\newacro{MAC}[MAC]{Medium Access Control}

\newacro{NMSE}[NMSE]{Normalized Mean Squared Error}
\newacro{EM}[EM]{Electro-Magnetic}
\newacro{CSI}[CSI]{Channel State Information}
\newacro{MISO}[MISO]{Multiple-Input Single-Output}
\newacro{NOMA}[NOMA]{Non-Orthogonal Multiple Access}
\newacro{OFDM}[OFDM]{Orthogonal Frequency-Division Multiplexing}
\newacro{QoS}[QoS]{Quality of Service}

\newacro{UAV}[UAV]{Unmanned Autonomous Vehicle}
\acrodefplural{UAV}[UAVs]{Unmanned Autonomous Vehicles}

\newacro{AoI}[AoI]{Age of Information}

\newacro{IID}[IID]{Independent and Identically Distributed}

\newacro{IoT}[IoT]{Internet of Things}
\newacro{LOS}[LOS]{Line-Of-Sight}
\newacro{NLOS}[NLOS]{Non-Line-Of-Sight}

\newacro{AoA}[AoA]{Angle of Arrival}
\acrodefplural{AoA}[AoAs]{Angles of Arrival}

\newacro{AoD}[AoD]{Angle of Departure}
\acrodefplural{AoD}[AoDs]{Angles of Departure}

\newacro{AWGN}[AWGN]{Additive White Gaussian Noise}

\newacro{SL}[SL]{Supervised Learning}
\newacro{RL}[RL]{Reinforcement Learning}
\newacro{DRL}[DRL]{Deep Reinforcement Learning}

\newacro{MDP}[MDP]{Markov Decision Process}
\acrodefplural{MDP}[MDPs]{Markov Decision Processes} 

\newacro{POMDP}[POMDP]{Partially Observable Markov Decision Process}
\acrodefplural{POMDP}[POMDPs]{Partially Observable Markov Decision Processes}

\newacro{DQN}[DQN]{Deep Q Network}
\newacro{D$^3$QN}[D$^3$QN]{Duelling Double Deep Q Network}
\newacro{PG}[PG]{Policy Gradient}
\newacro{PPO}[PPO]{Proximal Policy Optimization}
\newacro{DDPG}[DDPG]{Deep Deterministic Policy Gradient}
\newacro{UCB}[UCB]{Upper Confidence Bound}
\newacro{GPU}[GPU]{Graphical Processing Unit}
\newacro{PDS}[PDS]{Post Decision State}
\newacro{TD}[TD]{Temporal Difference}
\newacro{MAML}[MAML]{Multi-Agent Reinforcement Learning}
\newacro{LSTM}[LSTM]{Long Short-Term Memory}

\newacro{EE}[EE]{Energy Efficiency}
\newacro{SE}[SE]{Spectral Efficiency}
\newacro{MIMO}[MIMO]{Multiple-Input Multiple-Output}

\newacro{DFT}[DFT]{Discrete Fourier Transform}
\newacro{MRT}[MRT]{Maximum Ratio Transmission}
\newacro{EE}[EE]{Energy Efficiency}

\begin{document}
\title{Pervasive Machine Learning\\ for Smart Radio Environments Enabled by\\ Reconfigurable Intelligent Surfaces}
\author{George~C.~Alexandropoulos,~\IEEEmembership{Senior~Member,~IEEE,} Kyriakos Stylianopoulos,~\IEEEmembership{Student~Member,~IEEE,} Chongwen~Huang,~\IEEEmembership{Member,~IEEE,} Chau Yuen,~\IEEEmembership{Fellow,~IEEE,} Mehdi Bennis,~\IEEEmembership{Fellow,~IEEE,}\\ and M\'{e}rouane Debbah,~\IEEEmembership{Fellow,~IEEE}
\thanks{G. C. Alexandropoulos and K. Stylianopoulos are with the Department of Informatics and Telecommunications, National and Kapodistrian University of Athens, Panepistimiopolis Ilissia, 15784 Athens, Greece. G. C. Alexandropoulos is also with the Technology Innovation Institute, 9639 Masdar City, Abu Dhabi, United Arab Emirates (e-mails: \{alexandg, kstylianop\}@di.uoa.gr).} 
\thanks{C. Huang is with the College of Information Science and Electronic
Engineering, Zhejiang University, 310027 Hangzhou, China, the International
Joint Innovation Center, Zhejiang University, 314400 Haining,
China, and the Zhejiang Provincial Key Laboratory of IPCAN, 310027 Hangzhou, China (e-mail: chongwenhuang@zju.edu.cn).}
\thanks{Chau Yuen is with the Engineering Product Development (EPD) Pillar,
Singapore University of Technology and Design, 487372 Singapore (e-mail:
yuenchau@sutd.edu.sg).}
\thanks{M. Bennis is with Centre for Wireless Communications, University of Oulu, 90014 Oulu, Finland (e-mail: mehdi.bennis@oulu.fi).}
\thanks{M. Debbah is with both the Technology Innovation Institute and the Mohamed Bin Zayed University of Artificial Intelligence, 9639 Masdar City, Abu Dhabi, United Arab Emirates (email: merouane.debbah@tii.ae).}
}
\maketitle

%
%
%
%

\maketitle

\begin{abstract}
The emerging technology of Reconfigurable Intelligent Surfaces (RISs) is provisioned as an enabler of smart wireless environments, offering a highly scalable, low-cost, hardware-efficient, and almost energy-neutral solution for dynamic control of the propagation of electromagnetic signals over the wireless medium, ultimately providing increased environmental intelligence for diverse operation objectives.
One of the major challenges with the envisioned dense deployment of RISs in such reconfigurable radio environments is the efficient configuration of multiple metasurfaces with limited, or even the absence of, computing hardware.
In this paper, we consider multi-user and multi-RIS-empowered wireless systems, and present a thorough survey of the online machine learning approaches for the orchestration of their various tunable components.
Focusing on the sum-rate maximization as a representative design objective, we present a comprehensive problem formulation based on Deep Reinforcement Learning (DRL). We detail the correspondences among the parameters of the wireless system and the DRL terminology, and devise generic algorithmic steps for the artificial neural network training and deployment, while discussing their implementation details. Further practical considerations for multi-RIS-empowered wireless communications in the sixth Generation (6G) era are presented along with some key open research challenges. Differently from the DRL-based status quo, we leverage the independence between the configuration of the system design parameters and the future states of the wireless environment, and present efficient multi-armed bandits approaches, whose resulting sum-rate performances are numerically shown to outperform random configurations, while being sufficiently close to the conventional Deep Q-Network (DQN) algorithm, but with lower implementation complexity. 

\end{abstract}


\IEEEpeerreviewmaketitle

\section{Introduction} \label{sect:Introduction}
While the 3rd Generation Partnership Project (3GPP) is finalizing the Release $17$ \cite{5GAmericas} with enhancements on the \ac{5G} \ac{NR} and the major telecommunications operators are deploying 5G wireless networks around the world \cite{Shafi_5G_all}, academia and industry in wireless communications are already focusing on the Release $18$ \cite{3GPP_R18} (termed also as 5G-Advanced) and working on the definition and identification of requirements and candidate technologies for the \ac{6G} of wireless communications \cite{Samsung}. As per the latest consensus \cite{Saad_6G_2020, Emilio_6G}, \ac{6G} networks will require, among other metrics, higher data rates reaching up to 1 Tbps peak values with extended coverage, $1000\times$ network capacity as well as $10\times$ energy and cost efficiency compared to \ac{5G}, cm-level positioning accuracy, and $10^8/{\rm km}^3$ density of wireless connections. To achieve the envisioned requirements enabling various immersive (e.g., virtual and augmented reality), massive connectivity \cite{Cisco_Report} (e.g., \ac{IoT}) in industry and smart cities),  \ac{URLLC} (e.g., autonomous vehicles and remote surgery), and zero-touch \cite{Ericsson_AI_5G} (i.e., increased network automation) applications, \ac{6G} networks need to transform to a unified communication, sensing, and computing platform incorporating artificial intelligence and machine learning methodologies \cite{5GPPP_AIML}. This platform will highly likely integrate spectra from higher than the \ac{5G} \ac{mmWave} band as well as the \ac{THz} frequencies \cite{Akyildiz_6G_2020}, requiring cost- and power-efficient \ac{RF} front-ends and multi-antenna transceiver architectures, as well as ultra-wideband waveforms and computationally efficient signal processing schemes \cite{Mokh_1,Mokh_2,TR_Magazine}.

The potential of \acp{RIS} for programmable \ac{EM} wave propagation has recently motivated extensive academic and industrial interests, as an enabler for smart wireless environments in the \ac{6G} era \cite{Samsung,liaskos2018new,Marco2019,huang2019reconfigurable,Basar2019,wu2019towards,huang2020holographic,WavePropTCCN,rise6g,alexandg_2021,RISE6G_COMMAG,risTUTORIAL2020}. The \ac{RIS} technology, which typically refers to artificial planar structures with almost passive electronic circuitry (i.e., without any power amplification), is envisioned to be jointly optimized with conventional wireless transceivers~\cite{huang2019reconfigurable,risTUTORIAL2020} in order to significantly boost wireless communications in terms of coverage, spectral and energy efficiency, reliability, and security, while satisfying regulated \ac{EM} field emissions. The envisioned  dense deployment of \acp{RIS} over entities of the wireless environment, other than the actual network devices, is considered  as a revolutionary means to transform them into network  entities with reconfigurable properties, providing increased environmental intelligence for diverse communication objectives \cite{rise6g}. The typical unit element of an \ac{RIS} is the meta-atom, which is usually fabricated to realize multiple discrete phase states corresponding to distinct \ac{EM} responses. By externally controlling the states of the meta-atoms in such metasurfaces, various reflection and scattering profiles can be emulated \cite{huang2020holographic}. The \ac{RIS} technology up to date mainly includes meta-atoms of ultra-low power consumption for tuning \cite{WavePropTCCN}, in the sense that an RIS does not include any power amplifying circuitry. Such metasurfaces can only act as tunable reflectors, and thus, neither receive nor transmit on their own. While almost passive \acp{RIS} can enable smart radio propagation environments, their purely reflective operation in conjunction with the very limited computing and storage  capabilities induce notable network design challenges \cite{rise6g_SRE}. 

One of the major challenges with \ac{RIS}-empowered smart wireless environments including  multiple \acp{BS} and multiple \acp{UE} is \ac{CE}, which most commonly constitutes a prerequisite for the optimized orchestration of the phase responses of the unit elements (i.e., meta-atoms) of the multiple \acp{RIS}. \ac{CE} involves the estimation of multiple channels simultaneously: the direct channels between each \ac{BS} and each \ac{UE}, the channels between each RIS and BS, and the channels between each RIS and each \ac{UE} \cite{nadeem_ce,9133156,9144510,LZAWFM2020,parafac_SAM2020,PARAFAC2021, Deepak_2021,Miaowen2021,Guo_2021,deAraujo_2021,VanChien_2021, Chen_2021b, Guan_2021, Guo_2021a, Shi_2021a,9324910,8879620,8683663,9053695,9081935,9054415,9104260,8937491,Wan_2021,tensor_channel_tracking_2022}. This task becomes more complex when the deployed RISs are equipped with large numbers of unit elements having non-linear hardware characteristics \cite{9324910}. For example, in \cite{8879620}, a general framework for cascaded CE in RIS-assisted \ac{MIMO} systems was introduced by leveraging combined bilinear sparse matrix factorization and matrix completion. A control protocol enabling linear square estimation for the involved channel vectors in the \ac{MISO} case was presented in \cite{8683663}. The authors in \cite{9053695} designed an optimal CE scheme, where the RIS unit elements follow an optimal series of activation patterns. In \cite{9081935}, the authors presented the joint optimal training sequence and reflection pattern to minimize the \ac{MSE} of CE for RIS-aided wireless systems, by recasting the corresponding problem into a convex semi-definite programming one. In \cite{9054415}, an RIS-based activity detection and CE problem was formulated as a sparse matrix factorization, matrix completion, and multiple measurement vector problem that was solved via an approximate message passing algorithm. In \cite{9104260}, a least squares Khatri-Rao factorization algorithm was presented for RIS-assisted MIMO systems. A transmission protocol for CE and RIS phase profile optimization was proposed in \cite{8937491} for RIS-enhanced \ac{OFDM} systems. A holographic version of an \ac{RIS} was designed in \cite{Wan_2021} and its application to \ac{THz} massive \ac{MIMO} systems was investigated, along with a closed-loop CE scheme. Very recently, in \cite{tensor_channel_tracking_2022}, a channel tracking scheme for the uplink of RIS-enabled multi-user MIMO systems was presented, capitalizing on a tensor representation of the received signal and its parallel factor analysis. To avoid the inevitably large overhead of the estimation of high-dimensional channel matrices in \ac{RIS}-empowered smart wireless environments, more efficient approaches based on fast RIS phase profile management are lately being investigated \cite{Fast_Beam_Rui_2020,Jamali2022,RIS_Hierarchical}.

Despite the inevitable challenges for the efficient orchestration of all dynamically reconfigurable entities in smart wireless environments, several applications of \ac{RIS}-empowered wireless communication have been lately considered. Among them belong the adoption of \acp{RIS} for extending signal coverage \cite{comparative_study,HoVan_relay_selection,ying_relay_2020, bjornson_intelligent_2020, nemati_ris-assisted_2020,yang_coverage_2020,zeng_reconfigurable_2021, yildirim_hybrid_2021}, enabling or boosting accurate localization \cite{ma_indoor_2021,zhang_metaradar_2020,elzanaty_reconfigurable_2020,keykhosravi_siso_2021,buzzi_radar_2021,rahal_ris-enabled_2021,abu-shaban_near-field_2020,yang_wireless_2021,sankar_joint_2021,nguyen_reconfigurable_2020,locrxris_all,zerobs_all}, and enabling physical-layer security \cite{Chen_2019_all,Shen_2019_all, Cui_2019_all, Xu_2019_all, Yu_2019_all, Almohamad_2020,Chu_2020_all,Hong_2020,Dong_2020b,Shu_2020,PLS_Kostas}. In \cite{AL-Mekhlafi_2021}, the coexistence of \ac{eMBB} and \ac{URLLC} services in a cellular network that is assisted by an RIS was studied. A passive beamformer that can achieve the asymptotic optimal performance by controlling the properties of the incident wave at the RIS was designed in \cite{Jung_2021a}, under a limited RIS control link and practical reflection coefficients. An asymptotic closed-form expression for the mutual information of a multi-antenna transmitter-receiver pair in the presence of multiple RISs, in the large-antenna limit, was presented in \cite{Moustakas_RIS}. The fundamental capacity limits of RIS-assisted multi-user wireless communication systems were also investigated in \cite{Mu_2021b}. The authors in \cite{You_2021} considered the application of RISs to assist the uplink transmission from multiple UEs to a multi-antenna BS, and devised an optimization framework for jointly designing the transmit covariance matrices and the RIS phase profile with partial \ac{CSI}. In \cite{Xu_2021b,amplifying_RIS_2022}, active RISs were considered, which can adapt the phase and amplify the magnitude of the reflected incident signal simultaneously with the support of an additional power source. A joint optimization of the RIS phase profile and the BS beamforming vector was presented in \cite{Xu_2021b} with the objective to minimize the BS transmit power. In \cite{Yang_2021c}, the problem of energy efficiency optimization for a wireless communication system with distributed RISs was investigated. RIS-empowered \ac{D2D} communications underlaying a cellular network were considered in \cite{Yang_2021d}, in which an RIS was employed to enhance the desired signals and suppress interference between paired \ac{D2D} and cellular links. 

With the increasing capabilities of computing infrastructure and the wide availability of testbed data sets, machine learning has been established as a prominent tool of pattern recognition. Particularly, the advent of deep learning \cite{Goodfellow-et-al-2016} has enabled \acp{ANN} to detect innate structures over training data and infer underlying models without relying on domain knowledge. Accordingly, such data-driven methods have been implemented as alternatives to model-based techniques in various domains of wireless communications with interesting results  \cite{Ericsson_AI_5G,5GPPP_AIML}, effectively shaping the road ahead for 6G smart wireless environments \cite{Wireless_DL_Surevey}.
For instance, in \cite{Xiang2021SelfCalibratingIL, LocalizationWifi2019}, \acp{ANN} were trained to predict the positions of \acp{UE} based on channel observations under a localization framework. In \cite{Xiang_RobustLocalization}, the same problem was treated both as a classification and a regression task, and a more robust methodology was proposed. In a similar manner, the estimation of large \ac{MIMO} channels was facilitated by networks trained on pilot signals in \cite{Huang_CE_2019, Hu2021_DL_CE}. Arguably, in the most common scenario in the literature, deep learning methods have been tasked with learning optimal digital, analog, or hybrid digital/analog beamforming practices \cite{Zapone_UnsupervisedBeamforming, H_Huang_DL_precoding, Li2019_DL_precoding, Sapavath2019_ML_Beamforming_MIMO}.

Very recently, researchers from both academia and industry have exhibited a keen interest in applying deep learning for the design of RIS-empowered smart wireless environments. In the light of the large numbers of free parameters to configure, usually introduced by the phase-tunable RIS meta-atoms, it is expected that there are considerable performance improvements to be exploited with the help of trained \acp{ANN}, compared to conventional non-convex optimization approaches which are usually iterative, complex, and without convergence guarantees. In particular, deep learning has already been successfully assessed over a number of pertinent scopes for smart radio environments, like
rate maximization through analog beamforming \cite{huang2019spawc, Chongwen_Spectrum_Learning, RIS_DL_CNN, Gao_unsupervisedBamforming_RIS},
power maximization \cite{liaskos2019spawc},
compressive sensing estimation \cite{RIS_compressive_sensing}, channel estimation with possibly hybrid reflecting and sensing RISs \cite{huang2019spawc,RIS_compressive_sensing,hardware2020icassp,chongwendrl,Tasos_DNN_CE_2019, RIS_DL_channel_estimation2,ma_Smart_2019,liaskos_ABSense_2019,ma_Smart_2020,HRIS_Mag,HRIS_Nature,HRIS_SPAWC}, secrecy maximization \cite{RIS_DL_Secrecy_rate},
and even real-time imaging \cite{Li2019Nature}.
Most works focus on applications on the physical layer, whereas others like \cite{Chongwen_Spectrum_Learning, MAC_AI_RIS} are intended for deploying deep-learning-based systems on the \ac{MAC} layer. Very recently, \cite{Jointly_Learned_2021} presented a meta-learning approach to jointly design the multi-antenna receiver and configure the RIS reflection coefficients in the uplink of RIS-aided multi-user MIMO systems. Focusing on dynamic rich-scattering conditions in  \cite{DeepRIS_2022}, a framework for training a deep neural network as a surrogate forward model, to capture the stochastic dependence of wireless channels on the RIS configuration, was devised.

The vast majority of the deep-learning-based orchestration methods for RIS-empowered smart wireless environments adheres to the paradigm of supervised learning. Accordingly, a data collection stage must be introduced prior to the deployment of the orchestration system, in order for the model to be sufficiently trained. Clearly, it is often inconvenient to assume a separate training phase for the deployed machine-learning-based systems for various practical wireless communications applications. Not only this has the effect of greater deployment delays, but the trained orchestration algorithm is probably less equipped to deal with potential changes in the environment, resulting in a decoherence between the training data and the ones observed during the deployment phase. Alternatively, in this paper, we opt to focus on the methodology of \ac{RL}, which poses the design objective at hand as an online optimization problem, by training a model through continuous interactions with the environment. Orchestration methods based on \ac{RL} algorithms are arguably better suited to a wide variety of applications in wireless networks, and consequently, in RIS-empowered smart radio environments. First and foremost, they have the ability to adapt and change their behavior patterns in non-stationary settings. In addition, the data collection process is embedded in the underlying \ac{RL} algorithms, which are designed for more efficient exploration of the parameter space, as the training evolves. Motivated by the strengths of \ac{RL} for wireless communications and the relevant lately increasing research interest, in this survey paper, we thoroughly overview the latest advances in \ac{DRL} approaches for smart radio environments enabled by a plurality of \acp{RIS}. The main contributions of the paper are the following:
\begin{itemize}
    \item We present a comprehensive modeling of smart wireless environments comprising a multi-antenna BS, multiple single-antenna UEs, and multiple commonly accessible RISs that are all controlled by the same dedicated network entity (i.e., the RIS orchestration controller). 
    \item A detailed introduction to the \ac{RL} theory is provided with the purpose of explaining the principles behind the most prominent \ac{DRL} algorithms currently in use by the increasingly crowded community working on RIS-aided wireless communications, supplemented by a thorough taxonomy that emphasizes their different characteristics.
    \item A principled application of \ac{DRL} to \ac{RIS}-empowered smart radio environments is presented, detailing the correspondences among the design parameters of the wireless system and the DRL terminology. We consider a general sum-rate maximization problem with individual Quality of Service (QoS) constraints by the UEs, and a generic \ac{RL} algorithmic approach is described as a guideline for concrete implementations.
    \item An elaborate literature overview of applications of \ac{DRL} methods in \ac{RIS}-based wireless systems is presented, with the purpose of discussing common algorithmic practices and use cases. An accompanying discussion focuses on practical aspects, extensions of the provided DRL-based formulation, and key open research challenges.
    \item Guided by the careful examination of the theory of \ac{RL} and the particularities of the \ac{RIS}-empowered smart wireless environments, an alternative methodology is proposed for the sum-rate maximization objective. The problem is cast as a multi-armed bandits setting, which is a simpler version of the commonly considered Markovian-based formulation. 
    \item An extensive numerical investigation of the proposed and benchmark (D)RL algorithms, including the optimal policy, is conducted both in the absence and in the presence of individual QoS constraints, as well as for either the full or partial CSI availability cases.
\end{itemize}

The reminder of the paper is organized as follows. Section~\ref{sect:RIS-environments} introduces the considered RIS-empowered smart wireless environment, channel models, and design optimization formulation, while Section~\ref{subsec:RL} includes a thorough overview of the theory behind key \ac{RL} approaches for wireless communications.
The incorporation of \ac{DRL} in \ac{RIS}-empowered wireless settings is presented in Section~\ref{sect:AI-RIS}, including a detailed \ac{DRL}-based formulation for the general orchestration problem under investigation, and a comprehensive overview of the pertinent literature.
Practical considerations, design challenges, and important research directions are discussed in Section \ref{sec:considerations}.
Section~\ref{sect:Experiments} presents our numerical results for the proposed (D)RL schemes in comparison with benchmarks approaches.
Finally, Section~\ref{sect:Conclusion} concludes the paper.

\textit{Notations:} Vectors and matrices are denoted by boldface lowercase and boldface capital letters, respectively. The vectorization, transpose, and Hermitian transpose of $\mathbf{A}$ are denoted by ${\rm vec}(\mathbf{A})$, $\mathbf{A}^T$, and $\mathbf{A}^H$, respectively, while $\mathbf{I}_{n}$ ($n\geq2$) is the $n\times n$ identity matrix and $\mathbf{0}_{n}$ is an $n$-element column vector with zeros. $[\mathbf{A}]_{i,j}$ is the $(i,j)$-th element of $\mathbf{A}$, $[\mathbf{a}]_i$ and $\|\mathbf{a}\|$ denote $\mathbf{a}$'s $i$-th element and Euclidean norm, respectively, and ${\rm diag}\{\mathbf{a}\}$ represents a square diagonal matrix with $\mathbf{a}$'s elements in its main diagonal. $|a|$ denotes the amplitude of the complex scalar $a$, $\jmath\triangleq\sqrt{-1}$ is the imaginary unit, $\mathbb{E}\{\cdot\}$ is the expectation operator, and $\mathbb{E}_x\{\cdot\}$ specifies the expectation with respect to the random variable $x$. $\mathbf{x}\sim\mathcal{CN}(\mathbf{a},\mathbf{A})$ indicates a complex Gaussian random vector with mean $\mathbf{a}$ and covariance matrix $\mathbf{A}$. $O(\cdot)$ and $\Theta(\cdot)$ are the Big-O and the Big-$\Theta$ notations, respectively. The rest of the notations used throughout the paper are listed in Tables~\ref{table:notation-wireless} and~\ref{table:notation-ml}, while Table~\ref{table:abbreviatons} includes the paper's abbreviations.



\begin{table}[t]
    \centering
    \caption{Notation Used for Wireless Communications.}
    \label{table:notation-wireless}
    \begin{minipage}[t]{\linewidth}%
    \begin{tabulary}{\textwidth}{|c|L|}
    \hline
    \textbf{Symbol} & \textbf{Description} \\
    \hline
    \hline
    $\mathbb{C}$, $\mathbb{R}$             & Complex and real number sets, respectively \\
    \hline
    $K$             & Number of single-antenna UEs; UE$_k$ denotes the $k$-th UE with $k=1,2,\ldots,K$ \\
    \hline
    $N_{\rm T}$             & Number of BS antenna elements \\
    \hline
    $M$                                    & Number of identical RISs; RIS$_m$ denotes the $m$-th RIS with $m=1,2,\ldots,M$\\
    \hline
    $N\triangleq N_{\rm h}N_{\rm v}$                                    & Number of unit elements per RIS with $i=1,2,\ldots,N$: $N_{\rm h}$ placed in the horizontal and $N_{\rm v}$ in the vertical dimensions\\
    \hline
    $N_{\rm tot}\triangleq MN$                              & Total number of phase-tunable elements of all RISs\\
    \hline
    $N_{{\rm group}}$                     & Number of RIS elements whose phase configurations are set together to the same value\\
    \hline
    $N_{{\rm control}}$                   & Total number of individually controllable elements from all \acp{RIS} \\
    \hline
    $\mathbf{H}_m$                         & $N\times N_{\rm T}$ channel matrix for the RIS$_m$-BS link \\
    \hline
    $\mathbf{g}_{k,m}$                     & $N$-element row vector for the UE$_k$-RIS$_m$ channel \\
    \hline
    $\mathbf{h}_k$                         & $N_{\rm T}$-element row vector for the UE$_k$-BS channel \\
    \hline
    $L(d)$                                 & Pathloss attenuation at a certain distance $d$; $d_k$, $d_m$, and $d_{k,m}$ denote the geometrical distances of the UE$_k$-BS, RIS$_m$-BS, and UE$_k$-RIS$_m$ links, respectively    \\
    \hline
    $\kappa_1$, $\kappa_2$                  & Ricean factors of each RIS$_m$-BS and each UE$_k$-RIS$_m$ link, respectively \\
    \hline
    $\mathbf{f}_u(\varphi, \vartheta)$        & Array response row vector for a multi-antenna/multi-element node ${\rm u}\in\{{\rm BS}, {\rm RIS}_m\}$ with azimuth and elevation angles of arrival/departure $\varphi$ and $\vartheta$, respectively\\
    \hline
    $\lambda$                              & Signal wavelength\\
    \hline
    ${\rm d}_{\rm RIS}$, ${\rm d}_{\rm BS}$                              & Spacing of adjacent meta-atoms at each RIS$_m$ and that of adjacent antenna elements at the BS\\
    \hline
    $b$                                    & Phase resolution in bits of each RIS element \\
    \hline
    $\boldsymbol{\phi}_m, \mathbf{\Phi}_m$                      & $N$-element phase profile vector for RIS$_m$ and $\mathbf{\Phi}_m\triangleq{\rm diag}\{\boldsymbol{\phi}_m\}$\\
    \hline
    $\mathcal{F}$                               & Set of discrete phase states at each RIS, i.e., $[\boldsymbol{\phi}_m]_i\in\mathcal{F}$ $\forall$$m,i$\\ 
    \hline
        $P$                                    & Total power budget at the BS for transmission \\
    \hline
    $\mathbf{x}\triangleq\mathbf{Vq}$                           & $N_{\rm T}$-element transmitted signal vector from BS composed by the $N_{\rm T}\times K$ precoding matrix $\mathbf{V}$ and the $K$-element symbol vector $\mathbf{q}\triangleq[q_1\,q_2\,\cdots\,q_K]^T$ with $\mathbb{E}\{\mathbf{q}\mathbf{q}^H\}=\frac{P}{K}\mathbf{I}_K$ \\
    \hline
     $\mathcal{V}$                                    & Set of available $N_{\rm T}\times1$ BS precoding vectors of unit norm for all UEs, i.e., $[\mathbf{V}]_{:,k}\in\mathcal{V}$ $\forall$$k$ \\
    \hline
    $y_k$                                  & Received signal at UE$_k$ in baseband\\
    \hline
    $\nu$                                  & Attenuation factor for the power of all direct UE$_k$-BS links such that $\nu\in[0,1]$\\
    \hline
    $n_k\sim\mathcal{CN}(0,\sigma^2)$                      & Additive White Gaussian Noise (AWGN) at each received signal with zero mean and variance $\sigma^2$ \\
    \hline
    $f$                                   & Carrier frequency \\
    \hline
    $\tilde{R}_k$                         & Received data rate by user $k$ \\
    \hline
    $R^{\rm req}_k$                       & Requested data rate by user $k$ \\
    \hline
    \end{tabulary}
    \end{minipage}
\end{table}

\begin{table}[t]
    \caption{Notation Used for Machine Learning.}
     \label{table:notation-ml}
    \begin{minipage}[t]{\linewidth}%
    \begin{tabulary}{\textwidth}{|c|L|}
    \hline
    \textbf{Symbol} & \textbf{Description} \\
    \hline
    \hline
    $t$                          & Discrete time step \\
    \hline
    $\mathbf{s}_t \in \mathcal{S}$ & An \acs{MDP} state at time $t$ chosen from the available state set $\mathcal{S}$\\
    \hline
    ${\rm dim}(\mathbf{s}_t)$                           & Dimensionality of an MDP state vector $\mathbf{s}_t$\\
    \hline
    $\mathbf{a}, \mathbf{a}_t\in \mathcal{A}$          & \acs{MDP} actions chosen from the available action space set $\mathcal{A}$ with cardinality ${\rm card}(\mathcal{A})$; $\mathbf{a}_t$ denotes the action chosen at time $t$ \\ 
    \hline
    ${\rm Pr}[\mathbf{s}_{t+1}| \mathbf{s}_{t}, \mathbf{a}_t]$ & \acs{MDP} transition probability: observing ${\rm s}_{t+1}$ while being at ${\rm s}_{t}$ and acting with $\mathbf{a}_t$ \\
    \hline
    $r_t \triangleq \mathcal{R}(\mathbf{s}_{t},\mathbf{a}_t)$ & Reward at time $t$ via the reward function \\
    \hline
    $S_t$, $A_t$, $R_t$          & The random variables for the state, action, and reward at time $t$ \\
    \hline
    $T$                          & Final time step of a finite \acs{MDP} \\
    \hline
    $\varpi(\mathbf{a}_t|\mathbf{s}_t)$   & Agent's policy at time $t$, i.e., the likelihood of selecting action $\mathbf{a}_t$ when being at $\mathbf{s}_t$ \\
    \hline
    $G_t^\varpi$                    & Return at time $t$ after using policy $\varpi(\mathbf{a}_t|\mathbf{s}_t)$ \\
    \hline
    $\gamma$                     & Discount factor such that $\gamma \in (0, 1]$ \\
    \hline
    $V^{\varpi}(\mathbf{s})$        & Value function of policy for any $\mathbf{s} \in \mathcal{S}$ \\
    \hline
    $Q^{\varpi}(\mathbf{s}, \mathbf{a})$     & Action value function of policy for any $\mathbf{s} \in \mathcal{S}$ and any action $\mathbf{a} \in \mathcal{A}$ \\ 
    \hline
    $V^*(\mathbf{s})$, $Q^*(\mathbf{s}, \mathbf{a})$ & Optimal value and action value functions \\
    \hline
    $\mathbf{w}$, $\mathbf{w}^-$          & Vectors with the weight parameters of neural networks \\
    \hline
    $Q_{\mathbf{w}}$, $\bar{Q}_{\mathbf{w}^-}$                & A neural network and a target neural network used by \acs{DQN} \\
    \hline
    $\tau$                                        & ``Temperature'' of the soft-update of the weights of the target network used by \acs{DQN} \\
    \hline
        $\hat{G}_{\mathbf{w}}(\mathbf{s})$        & Reward-prediction neural network, with weights $\mathbf{w}$, used by the Neural $\epsilon$-greedy algorithm, having an observation $\mathbf{s}$ as input \\
    \hline
    $J(\cdot)$                   & Abstract objective function of a general \acs{DRL} algorithm \\
    \hline
    $\mathcal{D}$                & Batch of experience tuples \\
    \hline
    $t'$                         & Update interval during training of a \acs{DRL} algorithm. \\
    \hline
    $\mathcal{L}(\mathbf{w})$                  & Objective function of a \acs{DQN} \\
    \hline
        $\hat{\mathcal{L}}(\mathbf{w})$           & Loss function of the Neural $\epsilon$-greedy algorithm \\
    \hline
    $\alpha$                     & Learning rate of gradient descent or ascent \\
    \hline
    $\epsilon$                   & Probability of selecting a random action \\
    \hline
    $\mathcal{U}(\mathcal{A})$  & Selection operation from the discrete set $\mathcal{A}$ with probability $1/{\rm card}(\mathcal{A})$ \\
    \hline
    $J_{{\rm PG}}(\mathbf{w})$  & Objective function of a policy gradient method \\
    \hline
    $\tau$                       & Trajectory of an \acs{MDP} episode \\
    \hline
    $\mathsf{A}(\mathbf{s}_t, \mathbf{a}_t)$       & Advantage value function at state $\mathbf{s}_t$ using action $\mathbf{a}_t$\\
    \hline
    $\mathsf{G}_t(\mathbf{a})$                     & Running average of achieved reward for action $\mathbf{a}$ at time $t$\\
    \hline
    $\mathsf{N}_t(\mathbf{a})$                     & Number of times action $\mathbf{a}$ has been selected at time $t$ \\
    \hline
    $r_i(\mathbf{a})$                     & Reward when action $a$ is selected for the $i$-th time \\
    \hline
    $c$                          & Width of the confidence interval of the \acs{UCB} strategy \\
    \hline
    $\mathbb{I}_{\mathcal{A}}(\mathbf{a})$     & Enumeration operator that maps a discrete action $\mathbf{a}$ to its index in the discrete action space $\mathcal{A}$\\
    \hline
    \end{tabulary}
    \end{minipage}
    
\end{table}

\begin{table}[t]
    \caption{The Abbreviations of the Paper.}
    \label{table:abbreviatons}
    \centering
    \begin{tabular}{|l|l|}
    \hline
    \textbf{Abbreviation} & \textbf{Description} \\
    \hline
        
        \acs{5G} & \acl{5G}  \\ \hline
        \acs{6G} & \acl{6G}  \\ \hline
        \acs{ANN} & \acl{ANN}  \\ \hline
        \acs{AoA} & \acl{AoA}  \\ \hline
        \acs{AoD} & \acl{AoD}  \\ \hline
        \acs{AoI} & \acl{AoI}  \\ \hline
        \acs{AWGN} & \acl{AWGN}  \\ \hline
        \acs{BS} & \acl{BS}  \\ \hline
        \acs{CE} & \acl{CE}  \\ \hline
        \acs{CSI} & \acl{CSI}  \\ \hline
        \acs{D$^3$QN} & \acl{D$^3$QN}  \\ \hline
        \acs{DDPG} & \acl{DDPG}  \\ \hline
        \acs{DFT} & \acl{DFT} \\ \hline
        \acs{DQN} & \acl{DQN}  \\ \hline
        \acs{DRL} & \acl{DRL}  \\ \hline
        \acs{EE} & \acl{EE} \\ \hline
        \acs{EM} & \acl{EM}  \\ \hline
        \acs{eMBB} & \acl{eMBB}  \\ \hline
        \acs{FET} & \acl{FET}  \\ \hline
        \acs{GPU} & \acl{GPU}  \\ \hline
        \acs{IID} & \acl{IID}  \\ \hline
        \acs{IoT} & \acl{IoT}  \\ \hline
        \acs{LOS} & \acl{LOS}  \\ \hline
        \acs{LSTM} & \acl{LSTM}  \\ \hline
        \acs{MAC} & \acl{MAC}  \\ \hline
        \acs{MAML} & \acl{MAML}  \\ \hline
        \acs{MEMS} & \acl{MEMS}  \\ \hline
        \acs{MDP} & \acl{MDP}  \\ \hline
        \acs{MIMO} & \acl{MIMO}  \\ \hline
        \acs{MISO} & \acl{MISO}  \\ \hline
        \acs{mmWave} & \acl{mmWave}  \\ \hline
        \acs{MRT} & \acl{MRT} \\ \hline
        \acs{MSE} & \acl{MSE}  \\ \hline
        \acs{NLOS} & \acl{NLOS}  \\ \hline
        \acs{NMSE} & \acl{NMSE}  \\ \hline
        \acs{NOMA} & \acl{NOMA}  \\ \hline
        \acs{NR} & \acl{NR}  \\ \hline
        \acs{OFDM} & \acl{OFDM}  \\ \hline
        \acs{PG} & \acl{PG}  \\ \hline
        \acs{POMDP} & \acl{POMDP}  \\ \hline
        \acs{PPO} & \acl{PPO}  \\ \hline
        \acs{QoS} & \acl{QoS}  \\ \hline
        \acs{RF} & \acl{RF}  \\ \hline
        \acs{RIS} & \acl{RIS}  \\ \hline
        \acs{RL} & \acl{RL}  \\ \hline
        \acs{SINR} & \acl{SINR}  \\ \hline
        \acs{SNR} & \acl{SNR}  \\ \hline
        \acs{TD} & \acl{TD}  \\ \hline
        \acs{THz} & \acl{THz}  \\ \hline
        \acs{TTI} & \acl{TTI}  \\ \hline
        \acs{UAV} & \acl{UAV}  \\ \hline
        \acs{UCB} & \acl{UCB}  \\ \hline
        \acs{UE} & \acl{UE}  \\ \hline
        \acs{URLLC} & \acl{URLLC}  \\ \hline

    \end{tabular}
\end{table}

\section{The Considered Smart Radio Environment} \label{sect:RIS-environments}
In this section, we present the considered smart wireless propagation environment enabled by multiple \acp{RIS} and introduce an abstract model for the RIS operation as well as the channel models used throughout the paper. We also describe the optimization problem for the joint orchestration of the reflections from the RISs and the BS precoding, under investigation. 

\subsection{Abstract Modeling of RISs}\label{sec:system-model}
The typical hardware implementation of RISs is a planar array of ultra-thin meta-atoms (also known as unit cells or elements), which have multiple digitalized states corresponding to distinct \ac{EM} responses \cite{alexandg_2021}. The meta-atoms are usually printed on a dielectric substrate and their digital control can be achieved by leveraging semiconductor devices, such as  \ac{PIN} diodes, \acp{FET}, and \ac{MEMS} switches (e.g., varactor diodes) \cite{cui2014coding,Tsinghua_RIS_Tutorial}. 

For quasi-free-space beam manipulation, akin to the main scenario currently considered for RISs in the wireless communications literature~\cite{tang2020wireless_arxiv,dai2020reconfigurable_arxiv}, a fine-grained control over the reflected EM field is essential for accurate beamforming. This fact motivated researchers to rely on meta-atoms of sub-wavelength size \cite{huang2020holographic}, despite inevitable strong mutual coupling between meta-atoms (e.g., when the spacing of adjacent meta-atoms is ${\rm d}_{\rm RIS}=\lambda/10$ with $\lambda$ being the signal wavelength). In contrast, in rich scattering environments, the wave energy is statistically equally spread throughout the wireless medium and the ensuing ray chaos implies that rays impact the RIS from all possible, rather than one well-defined, directions. Instead of creating a directive beam, the goal becomes the manipulation of as many ray paths as possible. This manipulation may either aim at tailoring those rays to create constructive interference at a target location or to efficiently stir the field. These manipulations can be efficiently realized with $\lambda/2$-sized meta-atoms, which enable the control of more rays with a fixed amount of electronic components, as compared to RISs equipped with their sub-wavelength counterparts. Additionally, mutual coupling among half-wavelength meta-atoms (i.e., ${\rm d}_{\rm RIS}=\lambda/2$) is weaker, if not negligible.   

In this paper, we consider $M$ identical RISs each consisting of $N\triangleq N_{\rm h}N_{\rm v}$ meta-atoms placed in groups of $N_{\rm h}$ in the horizontal dimension, one below the other, such that $N_{\rm v}$ meta-atoms exist in the vertical dimension. Following recent experimental RIS implementations \cite{Nature13,PhysRevApplied.11.044024}, we adopt the transmission-line circuit model in \cite{Abeywickrama_2020} for the meta-atoms. According to this model, each $i$-th element of each $m$-th RIS (denoted as RIS$_m$ from now), with $i=1,2,\ldots,N$ and $m=1,2,\ldots,M$, behaves like a parallel resonant circuit and its reflection coefficient $[\boldsymbol{\phi}_m]_i$ determines the fraction of the reflected EM wave, due to the impedance discontinuity between the free-space impedance and this element's impedance. Assuming that $b$ is the phase resolution in bits per meta-atom, we consider the following $2^{b}$-element set for each $[\boldsymbol{\phi}_m]_i$:
\begin{equation}\label{eq:phases}
\mathcal{F}\triangleq\{\exp\left(\jmath2^{1-b}\pi f\right)\}^{2^{b}-1}_{f=0},
\end{equation}
which results in a total of $2^{bN}$ phase profiles (also known as phase configurations) per RIS. It is noted that, in practice, holds: \textit{i}) $|[\boldsymbol{\phi}_m]_i|\leq1$ and that the amplitude of each reflection coefficient depends both on its phase value and the angle of the impinging wave \cite{alexandg_2021, Abeywickrama_2020}; and \textit{ii}) $[\boldsymbol{\phi}_m]_i$ is frequency dependent exhibiting a Lorentzian-like frequency response \cite{pulidopolarizability2017} (similar to antennas, an RIS element can be a combination of resonant circuits in series or in parallel \cite{Antonio_2012}). 

\subsection{System Model}\label{sec:System_Model}
\begin{figure}[t]
    \centering
    \includegraphics[width=0.97\columnwidth]{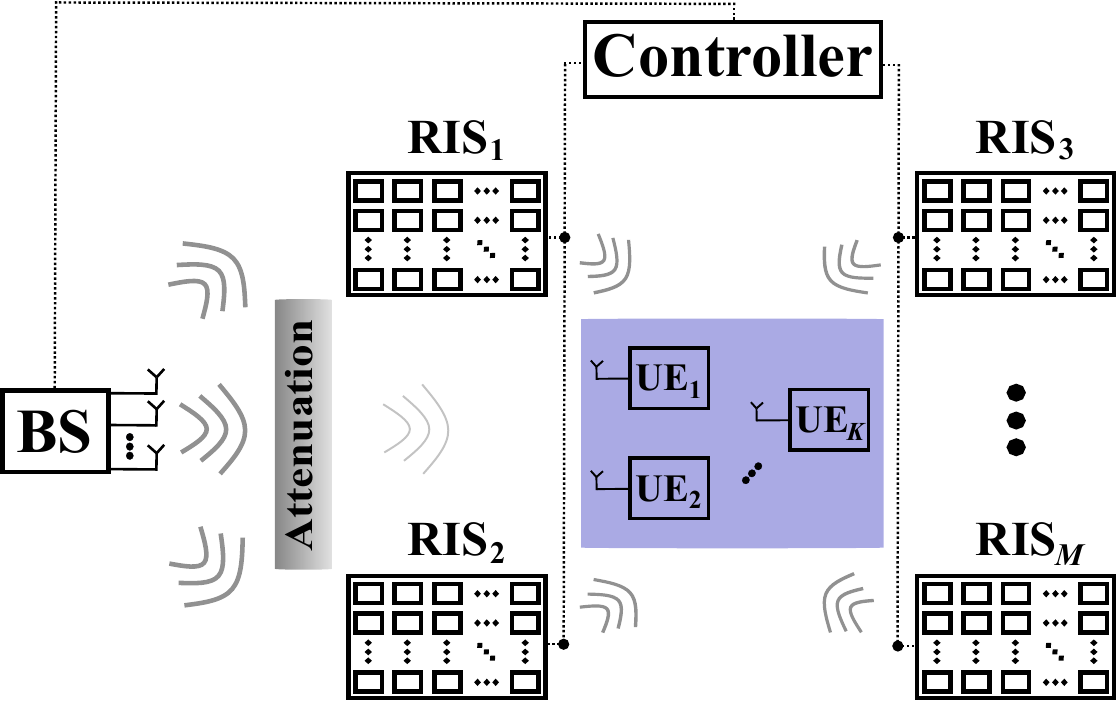}
    \caption{The considered RIS-empowered smart radio environment comprising an $N_{\rm T}$-antenna BS, $M$ identical $N$-element RISs, and $K$ single-antenna UEs. We assume that the RISs smartly control signal propagation in space in order to boost the performance of the targeted UEs which lie in an area where direct communication from the serving BS is highly attenuated.}
    \label{fig:System_Model}\vspace{-0.1cm}
\end{figure}
We consider the downlink wireless communication between a \ac{BS} equipped with $N_{\rm T}$ antenna elements and $K$ single-antenna \acp{UE}, which is empowered by $M$ identical rectangular RISs. All \acp{RIS} are assumed to be connected to the same controller, which is responsible for their joint orchestration in conjunction with the multi-antenna BS. We consider wired connections between each RIS and the controller, while the latter communicates also with the BS via either a wired or an out-of-band wireless link \cite{RISE6G_COMMAG}. 
As depicted in Fig$.$~\ref{fig:System_Model}, we assume that the RISs are placed close to the area where the UEs are located, enabling stronger wireless links (i.e., individual hops) compared to the case where the UEs are linked to the BS via only direct connections. The geometrical distances of the UE$_k$-BS, RIS$_m$-BS, and UE$_k$-RIS$_m$ links $\forall$ $k,m$ are denoted by $d_k$, $d_m$, and $d_{k,m}$, respectively, and we, further, assume that $d_{k,m}\ll d_k$. Without loss of generality, we consider the free-space pathloss model according to which $L(d) \triangleq 20\log_{10}\left(4\pi d\lambda^{-1} \right)$ in dB, which represents the power loss factor at a certain distance $d$.

We assume narrowband downlink transmissions where the multi-antenna BS deploys the complex-valued $N_{\rm T}\times K$ precoding matrix $\mathbf{V}$ to multiplex the information symbols for all $K$ UEs, which are included in the vector $\mathbf{q}\triangleq[q_1\,q_2\,\cdots\,q_K]^T$. These symbols, which are herein assumed to be mutually independent, are usually complex-valued and chosen from a discrete modulation set. In a similar manner to \cite{3GPP,Dahlman_5G_NR}, we consider that $[\mathbf{V}]_{:,k}\in\mathcal{V}$ $\forall$ $k$, where $\mathcal{V}$ represents the set with the available unit-norm $N_{\rm T}\times1$ BS precoding vectors for serving all UEs. By assuming, for simplicity, equal power allocation among the $K$ UE signals\footnote{Multi-user power control (e.g., \cite{alexandg_fd_power_control,Ghazanfari_power_control_2020}) provides an extra degree of freedom for transmission design optimization and is left for future work.}, i.e., $\mathbb{E}\{|q_k|^2\}= P/K$ $\forall$ $k$, the $N_{\rm T}\times 1$ transmitted signal vector $\mathbf{x}\triangleq\mathbf{Vq}$ is usually constrained as $\mathbb{E}\{\|\mathbf{x}\|^2\}\leq P$ (in our case, the equality holds), where $P$ denotes the total transmit power budget. Using the latter definitions and the following complex-valued $N_{\rm T}$-element row vector for the end-to-end channel \cite{huang2019reconfigurable}\footnote{The cascaded end-to-end channel model of \cite{huang2019reconfigurable} for each BS-RIS$_m$-UE$_k$ link is almost uniquely deployed in all RIS phase profile design investigations up to date \cite{risTUTORIAL2020}. Very recently (see \cite{PhysFad} and references therein), physics-inspired channel models that describe EM wave propagation in various complex media (ranging from free space to rich scattering) are being considered, which cannot be in general decomposed in a cascaded matrices form. Nevertheless, the DRL framework presented in this paper can be considered for any RIS-parametrized channel model; its detailed application for the model in \cite{PhysFad} is left for future work.}: 
\begin{equation}\label{eq:b_vector}
    \mathbf{b}_k \triangleq \left(\sqrt{L\left(d_k\right)}\mathbf{h}_k + \sum_{m=1}^M \sqrt{L\left(d_m\right)L\left(d_{m,k}\right)}\mathbf{g}_{m,k} \mathbf{\Phi}_m \mathbf{H}_{m}  \right),
\end{equation}
the baseband received signal at each UE$_k$ can be mathematically expressed as follows:
\begin{equation}\label{eq:receive-signal-model}
\begin{split}
    y_k &\triangleq \mathbf{b}_k \mathbf{x} + n_k,\\
    &= \mathbf{b}_k[\mathbf{V}]_{:,k}q_k+\sum_{i=1,\,i \neq k}^{K} \mathbf{b}_k[\mathbf{V}]_{:,i}q_i+ n_k,
\end{split}    
\end{equation}
where\footnote{We assume that signals experiencing reflections from more than one RISs are highly attenuated due to the multiplicative pathloss. Therefore, we neglect any inter-RIS channel contribution in the received signal model given by \eqref{eq:receive-signal-model}.} $\mathbf{H}_m\in\mathbb{C}^{N\times N_{\rm T}}$, $\mathbf{g}_{k,m}\in\mathbb{C}^{1\times N}$, and $\mathbf{h}_k\in\mathbb{C}^{1\times N_{\rm T}}$ represent the channel gain matrices for each of the links RIS$_m$-BS, UE$_k$-RIS$_m$, and UE$_k$-BS, respectively, while $\mathbf{\Phi}_m$ is defined, using \eqref{eq:phases}, as\footnote{When ${\rm d}_{\rm RIS}\ll\lambda/2$ (i.e., non-negligible mutual coupling), $\mathbf{\Phi}_m$ will be a full or banded matrix. For those cases, $\mathbf{\Phi}_m$ can be expressed as $\mathbf{C}_m{\rm diag}\{\boldsymbol{\phi}_m\}$, where $\mathbf{C}_m$ represents the mutual coupling matrix \cite{alexandg_ESPARs,Gradoni2020} of the RIS$_m$ structure. This matrix can be then absorbed in the channel matrix $\mathbf{g}_{m,k}$ in \eqref{eq:b_vector}, rendering the cascaded channel model in this expression mutual-coupling aware. In a similar manner, correlated fading can be included in each involved channel matrix \cite{Emil_RIS_Correlation,correlated_Weibull,correlated_Nakagami}.} $\mathbf{\Phi}_m\triangleq{\rm diag}\{\boldsymbol{\phi}_m\}$ and $n_k\sim\mathcal{CN}(0,\sigma^2)$ is the \ac{AWGN}. The \ac{SINR} at each UE$_k$, which provides a key performance indicator for the considered multi-user wireless communication system, can be easily obtained from \eqref{eq:receive-signal-model} as follows:
\begin{align}\label{eq:SINR}
    {\rm SINR}_k \triangleq \frac{\left| \mathbf{b}_k[\mathbf{V}]_{:,k}\right|^2}
    {\sum_{i=1,\,i \neq k}^{K}  \left| \mathbf{b}_k [\mathbf{V}]_{:,i}\right|^2+\frac{K\sigma^2}{P}}.
\end{align}
When there's only UE$_k$ present in the RIS-empowered communication system, the ${\rm SINR}_k$ formula boils down to the expression ${\rm SNR}_k\triangleq P\left| \mathbf{b}[\mathbf{V}]_{:,k}\right|^2K^{-1}\sigma^{-2}$, which describes the \ac{SNR} (in bps/Hz) received at this specific UE.


%

\subsection{Channel Models}
We consider frequency-flat fading channel models for all involved wireless links and assume that channels change independently from one discrete \ac{TTI} to another\footnote{To capture the essence of time-varying fading channels, the finite-state Markov channel model can be used \cite{Channel_Time_Correlation} that induces channel correlation in time. Such fading conditions can be intuitively exploited to reduce the overhead of RIS optimization, similar to beam tracking schemes (e.g., \cite{HeTaHaKu14_all,alexandg_correlation,3GPP_NR_ArXiv2018}) in \ac{mmWave} systems with hybrid analog and digital beamforming. This constitutes an interesting research direction for the DRL-based orchestration of RIS-empowered smart radio environments considered in this paper.}. In particular, each channel gain matrix $\mathbf{H}_{m}$ and each channel gain vector $\mathbf{g}_{m,k}$ are modeled as Ricean faded with factors\footnote{Each Ricean factor is defined as the power ratio of the specular component of the specific channel over the random components \cite{J:George_Elsevier_13}.} $\kappa_1$ and $\kappa_2$, respectively. Each former complex-valued $N\times N_{\rm T}$ matrix is given $\forall$$m$ by \cite{Farrokhi_Ricean}:
\begin{equation}\label{Eq:H_m}
\begin{split}
\mathbf{H}_m \triangleq& \sqrt{\frac{\kappa_1}{\kappa_1+1}}\mathbf{f}_{{\rm RIS}_m}^H\left(\varphi_{m}^A, \vartheta_{m}^A\right)\mathbf{f}_{\rm BS}\left(\varphi_m^D, \vartheta_m^D\right)\\&
+\sqrt{\frac{1}{\kappa_1+1}}\mathbf{D}_m,
\end{split}
\end{equation}
where $\varphi_{m}^D$ and $\vartheta_{m}^D$ denote the azimuth and elevation \acp{AoD} of the \ac{LOS} component leaving the BS, which reaches the RIS$_m$ with the azimuth and elevation \acp{AoA} $\varphi_{m}^A$ and $\vartheta_{m}^A$, respectively\footnote{All AoAs/AoDs in the array response vectors are taken with respect to the coordinate system defined by the arriving/departing node \cite{3GPP_NR_ArXiv2018,ULBA2021}.}. The $N$-element row vector $\mathbf{a}_{{\rm RIS}_m}(\varphi,\vartheta)$ for the azimuth and elevation AoAs/AoDs $\varphi$ and $\vartheta$, respectively, at/from the RIS$_m$ can be expressed, for the considered regular rectangular placement of ideal isotropic meta-atoms at each RIS$_m$ structure, as: 
\begin{equation} \label{eq:RIS-array-response}
\mathbf{f}_{{\rm RIS}_m}\left(\varphi,\theta\right) = \mathbf{f}_{\rm el}\left(\theta\right)\otimes\mathbf{f}_{\rm az}\left(\varphi\right),
\end{equation}
where $\mathbf{f}_{\rm el}\left(\theta\right)$ is the complex-valued $N_{\rm v}$-dimension elevation steering vector, defined for $\omega\triangleq2\pi {\rm d}_{\rm RIS}\cos\left(\theta\right)$ as \cite{Alkhateeb_JSTSP_all}:
\begin{equation}\label{eq:a_el_BS}
\mathbf{f}_{\rm el}\left(\theta\right) \triangleq \frac{1}{\sqrt{N_{\rm v}}}\left[e^{\jmath\omega}\,e^{2\jmath\omega}\,\cdots\,e^{N_{\rm v}\jmath\omega}\right],
\end{equation}
with ${\rm d}_{\rm RIS}$ representing the spacing of adjacent meta-atoms in wavelengths in both elevation and azimuth, and $\mathbf{f}_{\rm az}\left(\phi\right)$ is a complex-valued $N_{\rm h}$-dimension vector denoting the azimuth steering vector, which is given using the definition $\psi\triangleq2\pi {\rm d}_{\rm RIS}\sin\left(\phi\right)\cos\left(\phi\right)$ by the expression:
\begin{equation}\label{eq:a_az_BS}
\mathbf{f}_{\rm az}\left(\phi\right) \triangleq \frac{1}{\sqrt{N_{\rm h}}}\left[e^{\jmath\psi}\,e^{2\jmath\psi}\,\cdots\,e^{N_{\rm h}\jmath\psi}\right].
\end{equation}
The $N_{\rm T}$-element row vector $\mathbf{f}_{{\rm BS}}(\varphi,\vartheta)$ for the azimuth and elevation AoDs $\varphi$ and $\vartheta$, respectively, is defined similar to $\mathbf{a}_{{\rm RIS}_m}\left(\varphi,\theta\right)$ by just just replacing ${\rm d}_{\rm RIS}$ with ${\rm d}_{\rm BS}$. Finally, in \eqref{Eq:H_m}, the $N\times N_{\rm T}$ matrix $\mathbf{D}_m$ includes the \ac{NLOS}, i.e., scattering, components of the channel and is modeled as $[\mathbf{D}_m]_{i,j}\sim\mathcal{CN}(0,1)$ $\forall$$i=1,2,\ldots,N$ and $\forall$$j=1,2,\ldots,N_{\rm T}$. Each $N$-element row channel vector $\mathbf{g}_{m,k}$, referring to each UE$_k$-RIS$_m$ link, is modeled similar to $\mathbf{H}_m$, as follows:
\begin{equation}\label{Eq:g_mK}
\mathbf{g}_{m,k} \triangleq \sqrt{\frac{\kappa_2}{\kappa_2+1}}\mathbf{f}_{{\rm RIS}_m}(\varphi_{m,k}, \vartheta_{m,k})+\sqrt{\frac{1}{\kappa_2+1}}\mathbf{d}_{m,k},
\end{equation}
where $\varphi_{m,k}$ and $\vartheta_{m,k}$ denote the azimuth and elevation AoDs of the LOS component between the RIS$_m$ and the position of the UE$_k$. In addition, the $N$-element NLOS row vector $\mathbf{d}_{m,k}$ is modeled as $\mathbf{d}_{m,k}^T\sim\mathcal{CN}(\mathbf{0}_N,\mathbf{I}_N)$.

In accordance with the system model presented in Section~\ref{sec:System_Model}, we assume that all direct 
UE$_k$-BS links are NLOS channels which are power-attenuated by the same attenuation factor $\nu\in[0,1]$. To this end, each $N_{\rm T}$-element row vector $\mathbf{h}_k$ is modeled as Rayleigh faded and is given by:
\begin{equation} \label{eq:h}
    \mathbf{h}_k \sim \mathcal{CN}\left(\mathbf{0}_{N_{\rm T}}, \nu\mathbf{I}_{N_{\rm T}} \right).
\end{equation}
Note that the extreme attenuation cases $\nu=0$ and $\nu=1$ refer to the absence of the direct links and to the non-attenuated direct links, respectively.



\subsection{Orchestration of Smart Radio Environments}
The free parameters of the considered RIS-empowered smart radio propagation environment in Fig$.$~\ref{fig:System_Model}, i.e., the vectors with the reflection coefficients of the multiple RISs and the BS precoding matrix, are amenable to optimization according to certain performance objectives. For example, such smart environments can be programmed to boost the achievable rate or energy efficiency; increase signal coverage; enable localization, sensing, radio mapping, and secrecy; as well as to provide low EM-field exposure to both intended and unintended UEs in an area(s) of interest \cite{rise6g}. 
\par

In this paper, we will be using the maximization of the \textit{achievable sum rate} as our exemplary performance metric to devise the core optimization problem, to derive \ac{RL}-based formulations on that (Section~\ref{sec:DRL_formulation}), and to present pertinent results from extensive numerical simulations (Section~\ref{sect:Experiments}).
Nevertheless, the orchestration of \ac{DRL} in smart radio environments, as presented in this survey, is readily applicable to any of the aforementioned (or other) network utilities, as long as they can be seen as a function of the responses (i.e., the phases in this paper) of the RIS elements. The motivation behind selecting the particular objective is due to its ubiquity in multi-user communications, and subsequently, its extensive use in the comparative literature (Section~\ref{sect:Literature}).
\par
The sum-rate performance objective can be mathematically defined using the ${\rm SINR}_k$ expression in \eqref{eq:SINR} as follows:
\begin{subequations}\label{main_problem}
\begin{align}
    \mathcal{OP}_1: \,\, &\max_{\{\boldsymbol{\phi}_m\}_{m=1}^M,\,\mathbf{V}}  \quad \sum_{k=1}^{K} \tilde{R}_k  \label{eq:sum-rate-problem} \\
    &\quad \quad \text{s.t.}  \quad [\boldsymbol{\phi}_m]_i\in\mathcal{F}~ \forall m,i,\, [\mathbf{V}]_{:,k}\in\mathcal{V}~ \forall k,\\& \quad \quad\quad\quad\hspace{0.1cm} \tilde{R}_k \geq R^{\rm req}_k~ \forall k \label{constraint:rate}, 
\end{align}
\end{subequations}
where $\tilde{R}_k \triangleq \log_2\left(1+{\rm SINR}_k\right)$ represents the instantaneous achievable rate for UE$_k$'s communication link, while $R^{\rm req}_k$ denotes the minimum rate request by this UE. In the sequel, we will treat this design optimization problem both with and without the individual rate constraints in \eqref{constraint:rate}.
It noted that, even under the perfect \ac{CSI} availability case at the controller of the smart radio environment, the latter non-convex optimization problem with the discrete constraints is hard to solve optimally. It is actually a computationally exhaustive combinatorial problem, whose special cases have been recently sub-optimally treated via classical optimization tools in \cite{Huang_GLOBECOM_2019} (the case of a single RIS with a multi-antenna BS) and via Supervised Learning in \cite{Samarakoon_2020} (multiple RISs and a single-antenna BS). In this paper, we propose to tackle this problem via the RL methodology, profiting from its inherit capability to offer online learning and optimization in non-stationary conditions (as usually wireless environments are), while avoiding lengthy training collections. In the sequel, we: \textit{i}) overview key RL approaches and their application for wireless communications; \textit{ii}) present our DRL-based formulation for dealing with $\mathcal{OP}_1$; and \textit{iii}) present the latest advances in DRL schemes for RIS-empowered wireless communication systems.

\section{Overview of Reinforcement Learning}\label{subsec:RL}
\begin{figure*}[t]
    \centering
    \fbox{
    \setlength{\fboxsep}{1pt}%
    \includegraphics[width=0.95\textwidth]{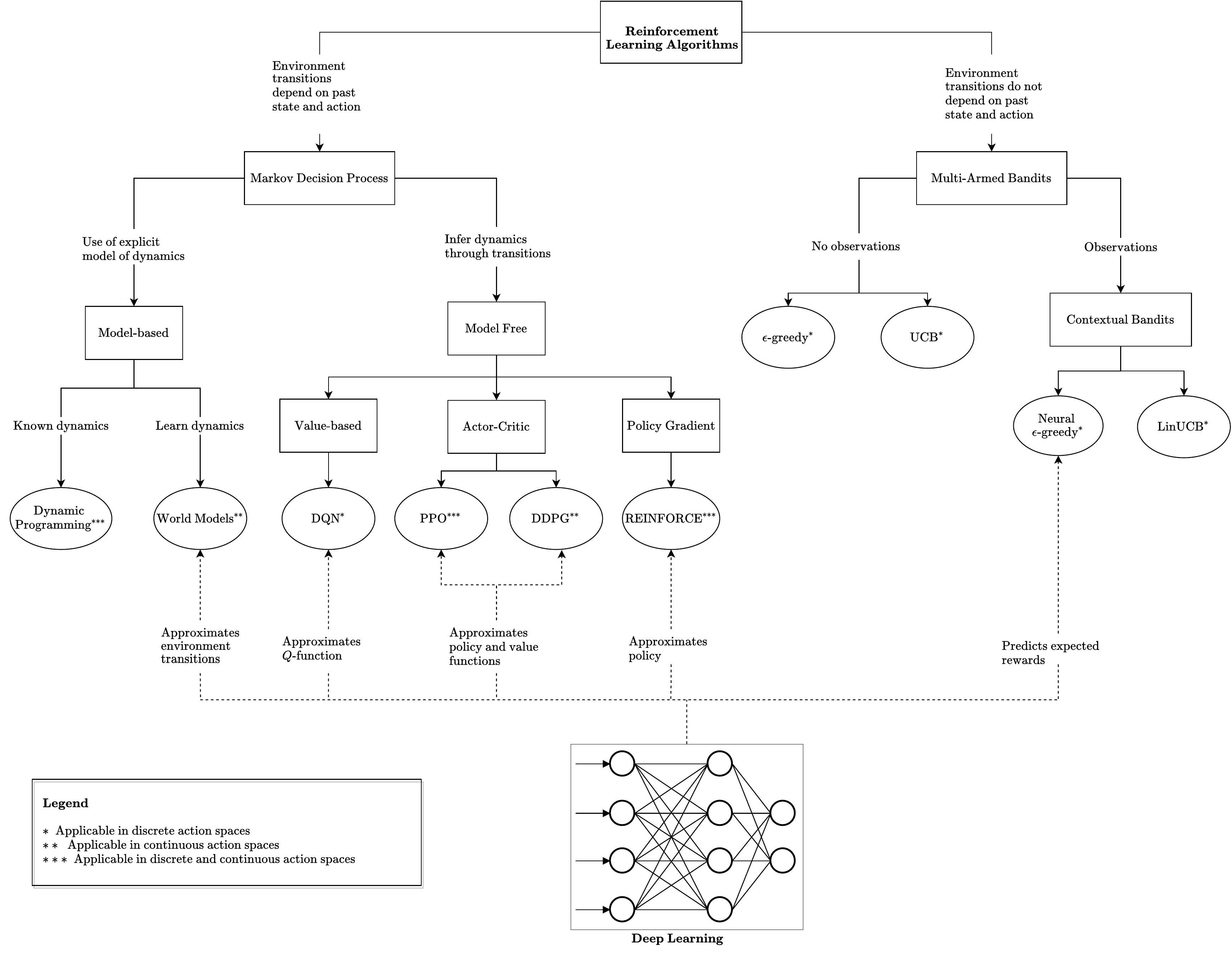}
    }
    \caption{Taxonomy of the overviewed RL algorithms in this survey. Ellipses signify \acs{RL}/\acs{DRL} algorithms, rectangles denote classes of methods, and text between arrows designates a criterion for the classification. The dashed arrows signify \acs{DRL} algorithms and explain how deep learning is utilized in each case. The listed algorithms are additionally marked depending on the type of the action space they admit.}
    \label{fig:rl-taxonomy}
\end{figure*}
Artificial intelligence and machine learning are very broad domains having already enjoyed great success in wireless communications in general, and very recently in representative applications of RIS-empowered communication systems in particular. In this survey paper, the focus is specifically in \ac{RL} approaches. We invite the readers to consult \cite{Wireless20}, \cite{DL_RIS_survey}, and \cite{AIRIS} for tutorials and overviews on deep learning schemes for communication systems based on RISs, which mainly focus on supervised learning methodologies. Herein, we do not attempt to introduce \acp{ANN}, neither the concept of deep learning; we assume though a broad level of familiarity. In fact, we treat \acp{ANN} as layered \textit{black box} function approximators that can be trained via the stochastic gradient descent approach. A detailed taxonomy of the \ac{RL} methods discussed in this section is illustrated in the flowchart of Fig$.$~\ref{fig:rl-taxonomy}.

The mathematical tool of RL \cite{Sutton98reinforcementlearning} is a machine learning domain that deals with the problem of iteratively maximizing a stochastic objective function that decomposes as a discrete sequence of payoffs.
Since such problems are in general non-convex and NP-hard, \ac{RL} approaches fall under the broad umbrella of non-convex and stochastic optimization algorithms. The main difference between RL and classical optimization is that, in the former, the problem is assumed to be learnable in the sense that past iterations provide information that can be exploited to provide more efficient actions/performance in latter stages. The learning part of the RL algorithms is commonly handled by deep, or not, \acp{ANN}. In this section, we provide a brief introduction to the RL theory, starting with the main formalism and problem structure, while including useful definitions in formulating RL-based solutions. We also describe how \acp{ANN} can be used in the context of DRL to obtain approximate solutions, and discuss the main ideas behind the key DRL-based approaches as well as their evaluation methodology. 

\subsection{Markov Decision Processes}
In a typical \ac{RL} problem, a decision making agent interacts with an environment to achieve a predefined goal. Unlike the paradigms of supervised and unsupervised learning, where data sets must be acquired prior to the training phase, \ac{RL} constitutes a \textit{closed-loop} approach. An agent continuously observes the information about the state of the system, usually using its measurement collection equipment, and chooses one of the available actions, that is fed back to the system, influencing its evolution in turn. During this interaction, reward signals are communicated by the environment to the agent, which indicate the system's performance at the time of reward collection. The agent is, thus, tasked to learn actions that maximize its cumulative reward signal.

\begin{figure}[!t]
    \centering
    \includegraphics[scale=0.7]{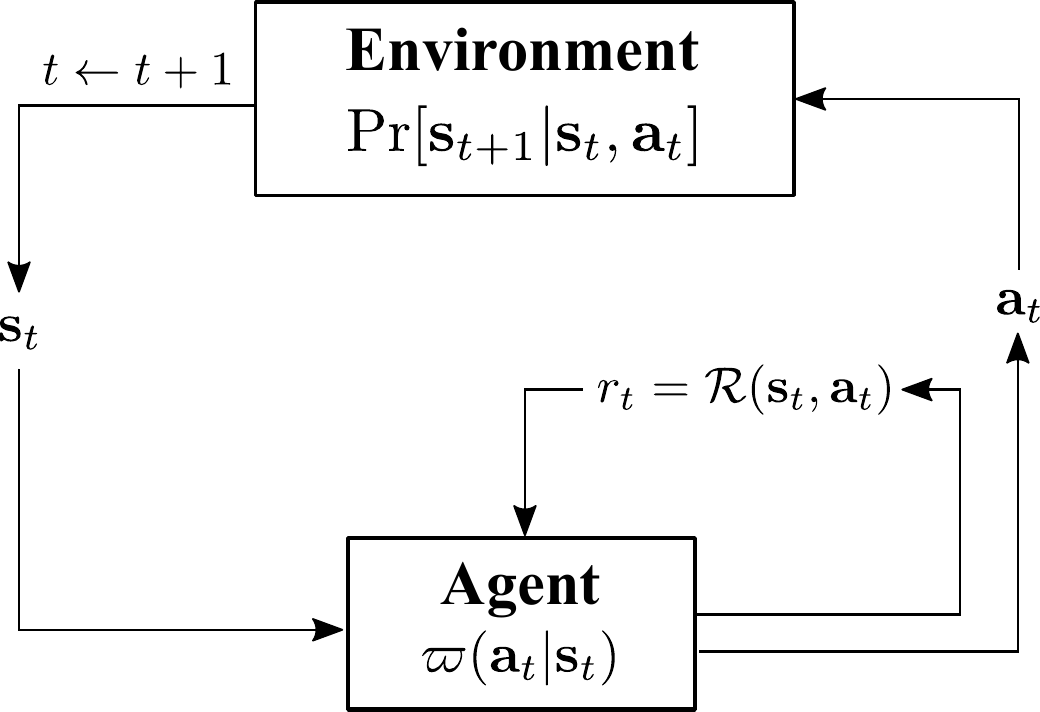}
    \caption{Block diagram of an MDP. The agent observes the state of the environment at each time step $t$ and selects an action according to its policy. The action is fed back to environment which sets the next time step $t+1$. A reward is given as a feedback to the agent. The reward function can be thought either as part of the environment or as external stimulus to the agent.}
    \label{fig:mdp}
\end{figure}
The field of \ac{RL} has its roots in optimal control, with the concept of the \ac{MDP} being the cornerstone of the theory behind sequential decision making \cite{Bertsekas}. Formally, an \ac{MDP} is defined as a $4$-tuple $(\mathcal{S}, \mathcal{A}, \mathcal{P}, \mathcal{R})$ with $\mathcal{S}$ and $\mathcal{A}$ being the \textit{state} and \textit{action} spaces, respectively, while $\mathcal{P}$ constitutes the \textit{transition probability} of the environment and $\mathcal{R}(\cdot)$ is the reward function. Note that the action and state spaces can be either continuous or discrete, though, each case requires different treatment and not all algorithms are applicable to both. At each (discrete) time step $t$, the agent: \textit{i}) observes the state of the environment $\mathbf{s}_t \in \mathcal{S}$; and \textit{ii}) selects an action $\mathbf{a}_t \in \mathcal{A}$, with $\mathcal{A}$ including ${\rm card}(A)$ discrete actions, which determines the instantaneous reward\footnote{It is also conceivable for the reward to depend both on the previous state-action pair as well as the next state. In this case, the reward signal is delayed one time step, i.e., $r_{t+1} = \mathcal{R}(\mathbf{s}_t, \mathbf{a}_t, \mathbf{s}_{t+1})$.} 
$r_t \triangleq \mathcal{R}(\mathbf{s}_t, \mathbf{a}_t)$. At the same time, the environment transitions to the next state $\mathbf{s}_{t+1}$ with probability $\mathcal{P}\triangleq{\rm Pr}[\mathbf{s}_{t+1} | \mathbf{s}_{t}, \mathbf{a}_t]$. We will be using the capital letters $S_t$, $A_t$, and $R_t$ when referring, respectively, to the random vectors for the state and action, and the random variable for the reward at time $t$. Additionally, without loss of generality, we assume an ordering of the actions within $\mathcal{A}$ using the enumerator operator $\mathbb{I}_{\mathcal{A}}(\mathbf{a}) \in \{1,2,\dots,{\rm card}(\mathcal{A})\}$, which returns the index of $\mathbf{a}$ in $\mathcal{A}$ (i.e., it maps an action to its index in the action space). A schematic overview of the \ac{MDP} formulation is depicted in Fig$.$~\ref{fig:mdp}.

The distinctive characteristic of an \ac{MDP} is that the following \textit{Markovian property} holds for the transition probability:
\begin{align}\label{eq:markovian-property}
    &{\rm Pr}\left[ S_{t+1} = \mathbf{s}_{t+1} | \nonumber \right.\\
    & \left.S_{t} = \mathbf{s}_{t}, A_{t} = \mathbf{a}_{t}, S_{t-1} = \mathbf{s}_{t-1}, A_{t-1} = \mathbf{a}_{t-1}, \dots  \right] = \nonumber \\
    & = {\rm Pr}\left[ S_{t+1} = \mathbf{s}_{t+1} | S_{t} = \mathbf{s}_{t}, A_{t} = \mathbf{a}_{t} \right]=\mathcal{P}. 
\end{align}
This property dictates that the next state at time $t+1$ depends exclusively on the state-action pair at time $t$, without being influenced by the history of earlier actions and observations.
In effect, the dependence of $\mathbf{s}_{t+1}$ on $\mathbf{s}_{t}$ describes a time evolution on the environment. The additional dependence on $\mathbf{a}_{t}$ signifies that the agent is able to influence the transitions, and the Markovian property implies that by observing $\mathbf{s}_{t}$, the agent knows all needed to know to predict the next state. In RL implementations, it is convenient to assume the existence of a final time $T$, which may be infinite. In \textit{finite} \acp{MDP}, a subset of the state space constitutes the \textit{terminal} states, for which no further transitions take place. An agent's sequence of interactions from $t=0$ to $T$ is termed an \textit{episode}. Typically, multiple episodes are needed for an agent to learn to select good actions.

Formally, the agent selects actions based on a \textit{policy} function $\varpi:\mathcal{S}\rightarrow{\mathcal{A}}$, which describes the mapping from states to actions. This mapping might be non-stationary (i.e., depend on $t$).
Without the loss of generality, policies can be assumed to be stochastic.
In this case, $\varpi: \mathcal{S}\times\mathcal{A} \rightarrow [0,1]$ describes the probability of selecting an action given a state.
We will use the notation $\varpi(\mathbf{a}|\mathbf{s})$ when referring to likelihood values of actions and $\varpi$ when referring to a policy as a function.
Hence, the objective of typical \ac{RL} problem formulations can be expressed as finding the optimal policy that maximizes the expected \textit{return} from some time step $t$:
\begin{equation}\label{eq:RL-objective}
\mathcal{OP}_2: \,\,\varpi^* \triangleq \operatorname*{argmax}_{\varpi} \mathbb{E}_{\varpi} \left\{  G^{\varpi}_t  \right\}, 
\end{equation}
where the return $G^{\varpi}_t$ is defined as the discounted sum of rewards, given that the agent takes actions following policy $\varpi$:
\begin{equation}
    G^{\varpi}_t \triangleq \sum\limits_{k=0}^{T-t} \gamma^{k} R_{t+k+1} = R_{t+1} + \gamma G^{\varpi}_{t+1} \ .
\end{equation}
In the above expression, $\gamma \in (0, 1]$ denotes the \textit{discount factor} which makes the sum of rewards finite even in the infinite case.
Conceptually, the discount factor imposes a preference on larger immediate rewards versus future ones.
Typical discount values reside in the interval $[0.9, 0.99]$, aiming to promote long-term gains instead of myopic strategies.

The typical \ac{MDP} formalism assumes that the agent is capable to observe the state of the environment instantly. This is evidently an impractical assumption for the majority of real-world applications, and it is especially problematic for wireless communication systems. The theory of \acp{POMDP}, apart from the tuple $(\mathcal{S}, \mathcal{A}, \mathcal{P}, \mathcal{R})$, further supposes an observation space and a probability density function of collecting a particular observation given the true state. The agent sees only a distorted observation of the true state, making the problem of learning the value functions exceedingly hard. Even though there are exact and approximate methods devoted specifically to \acp{POMDP}, in practice, algorithms that have been proposed in the context of \acp{MDP} are also applied to this category, with the assumption that the noisy observations contain enough useful information to sufficiently estimate the target functions. \ac{DRL} approaches, which deploy \acp{ANN} to learn useful representations of the input states and will be described in the sequel, tackle the problem of partial observability. They usually equip the agents with memory units (e.g., by using recurrent networks) to aid them distinguish the true state of the environments they reside in.

\subsection{Value Functions}
It is common for RL algorithms to keep a notion of value functions that estimate the benefit of an agent observing a state or an action-state pair.
Specifically, for a given policy $\varpi$, the \textit{value function} at any state $\mathbf{s}\in \mathcal{S}$ is mathematically defined as
\begin{align}\label{eq:V}
    & V^{\varpi}(\mathbf{s}) \triangleq \mathbb{E}_{\varpi} \left\{ G^{\varpi}_t | S_t = \mathbf{s} \right\}  \\
    & = \sum\limits_{\mathbf{a} \in \mathcal{A}} \varpi(\mathbf{a} | \mathbf{s} ) \sum_{\mathbf{s'} \in \mathcal{S}} {\rm Pr}[\mathbf{s'} | \mathbf{s}, \mathbf{a}] \left( \mathcal{R}(\mathbf{s}, \mathbf{a}, \mathbf{s'}) + \gamma V^{\varpi}(\mathbf{s'}) \right) ,\nonumber
\end{align}
which describes the expected return when the environment at time step $t$ is at state $\mathbf{s}$ and the agent follows policy $\varpi$. 
In the latter expression, discrete action and state spaces were assumed, but the notation can be extended to continuous spaces. Accordingly, the \textit{action value function} at any state $\mathbf{s}\in \mathcal{S}$ and action $\mathbf{a}\in \mathcal{A}$ is given by the expression: 
\begin{align}\label{eq:Q}
    & Q^{\varpi}(\mathbf{s}, \mathbf{a}) \triangleq \mathbb{E}_{\varpi} \left\{ G^{\varpi}_t | S_t = \mathbf{s}, A_t = \mathbf{a} \right\}  \\
    & = \sum_{\mathbf{s'} \in \mathcal{S}} {\rm Pr}[\mathbf{s'} | \mathbf{s}, \mathbf{a}] \left( \mathcal{R}(\mathbf{s}, \mathbf{a}, \mathbf{s'}) + \gamma \sum_{\mathbf{a}' \in \mathcal{A} } \varpi(\mathbf{a}' | \mathbf{s} ) Q^{\varpi}(\mathbf{s'}, \mathbf{a}') \right),\nonumber
\end{align}
which concerns the expected value when following policy $\varpi$ and selecting action $\mathbf{a}$ when observing state $\mathbf{s}$.
The latter two value functions are connected by the following relationship:
\begin{equation}\label{eq:V_Q}
    V^{\varpi}(\mathbf{s}) = \sum_{\mathbf{a} \in \mathcal{A}} \varpi(\mathbf{a}|\mathbf{s}) Q^{\varpi}(\mathbf{s}, \mathbf{a}).
\end{equation}

In order to solve $\mathcal{OP}_2$ appearing in \eqref{eq:RL-objective}, it suffices to obtain the optimal value functions, i.e., find the maximum values for each state (or action-state pair) under any policy. By exploiting the recursive structure of \eqref{eq:V} and \eqref{eq:Q}, the optimal value functions for any state and any state-action pair can be derived through Bellman's \textit{principle of optimality}, which is expressed via the Bellman equations:
\begin{align}\label{eq:bellman-V}
    V^*(\mathbf{s}) =& \max_{\mathbf{a}}\sum_{\mathbf{a}\in \mathcal{A}} \varpi(\mathbf{a} | \mathbf{s} ) \nonumber \\
    &\times\sum_{\mathbf{s'}\in \mathcal{S}} {\rm Pr}[\mathbf{s'} | \mathbf{s}, \mathbf{a}] \left( \mathcal{R}(\mathbf{s}, \mathbf{a}, \mathbf{s'}) + \gamma V^*(\mathbf{s'}) \right)
\end{align}
and
\begin{equation}\label{eq:bellman-Q}
     Q^*(\mathbf{s}, \mathbf{a}) = \sum_{\mathbf{s'}\in \mathcal{S}} {\rm Pr}[\mathbf{s'} | \mathbf{s}, \mathbf{a}] \left( \mathcal{R}(\mathbf{s}, \mathbf{a}, \mathbf{s'}) + \gamma \max_{\mathbf{a}'\in\mathcal{A}}  Q^*(\mathbf{s'}, \mathbf{a}') \right). 
\end{equation}
These equations define a pair of dynamic programming problems that can be solved by backward induction. Upon attaining the optimal value functions, it is straightforward to define the optimal policy as the one that selects the action which maximizes the expected return via the $Q$-function, i.e.:
\begin{equation}\label{eq:optimal-policy}
\mathcal{OP}_3: \,\,    \varpi^*(\mathbf{a}|\mathbf{s}) \triangleq \argmax\limits_{\mathbf{a} \in \mathcal{A}}Q^*(\mathbf{s},\mathbf{a}).
\end{equation}
This optimal solution of the \ac{MDP} suffers, however, from two important drawbacks. First, the transition probabilities of the environment are assumed to be known to the decision maker (i.e., \textit{model-based \ac{RL}}), which is restrictive for real-world applications. Secondly, the dynamic programming solution requires iterating over all possible states or state-action pairs, which is intractable for most demanding practical problems.

\subsection{Deep Reinforcement Learning Algorithms}\label{sec:DRL-algorithms}
Recent advances in the RL field aim to approximate different functions of the \ac{MDP} formalism in order to derive inexact, but efficient, solutions. 
Even though some approaches may target learning directly the transition function of the environment (e.g., ``World Models'' of \cite{WorldModels}), the vast majority of the RL methods employs a \textit{model-free} strategy, performing estimates using the experience collected by the agent.
In general, a \ac{DRL} algorithm is primarily equipped with a policy function that may be directly parameterized (i.e., using an \ac{ANN}) or defined indirectly (i.e., through the maximization of value functions which are parameterized instead).
\par
During training, the agent uses a \textit{collection policy} to interact with the \ac{MDP}, as well as amasses and stores tuples of observations, actions, and rewards.
At certain intervals (potentially at every time step $t$), a variant of gradient descent/ascent algorithm is invoked to learn the parameters of the underlying \acp{ANN}, so that a pertinent objective function over the collected data is optimized.
The collection policy may coincide with the policy intended to be learned (\textit{on-policy} algorithms), or may deviate from it to encourage better exploration (\textit{off-policy} algorithms).
A generalized description of the DRL training process of an agent is provided in Algorithm~\ref{alg:DRL_training}.
The algorithm is purposely given in an abstract form, since the exact scheme varies according to the specifics of each method, but in principle, all algorithms incorporate ANNs designed to observe state vectors and output ``quality assessements'' over actions.
\par
Ordinarily, \ac{DRL} approaches are categorized either as value-based, policy-gradients, or a combination of the two (actor-critic methods). We next present the basic components of certain important methods that are commonly used in the field of wireless communications and \ac{RIS}-empowered communication systems.
A detailed walkthrough of the state-of-the-art in \ac{DRL} approaches can be found in the recent survey \cite{Lazaridis2020SOTA_walkthrough}.
\begin{algorithm}[!t]
\caption{Training Procedure of a DRL Agent}\label{alg:DRL_training}
\begin{algorithmic}[1]
\Require MDP description $(\mathcal{S}, \mathcal{A}, \mathcal{P},\mathcal{R})$, final time step $T$, and for the agent: initial ANN parameters $\mathbf{w}$, policy $\varpi_{\mathbf{w}}$, collection policy $\hat{\varpi}_{\mathbf{w}}$, objective function $J(\cdot)$, and update interval $t'$. 
\State Observe initial state $\mathbf{s}_1$ from the environment.
\For{$t=1,2,\dots,T$}
    \State Use $\hat{\varpi}_{\mathbf{w}}$ to determine action $\mathbf{a}_t$.
    \State Feed $\mathbf{a}_t$ to the environment to observe $\mathbf{s}_{t+1}$ and $r_t$.
    \State Store experience ($\mathbf{s}_t$, $\mathbf{a}_t$, $\mathbf{s}_{t+1}$, $r_t$) to a set $\mathcal{D}$.  
    \If{${\rm mod}(t,t') = 0$}
        \State Use $\mathcal{D}$ to compute $\mathbf{w}^*$ optimizing $J(\mathbf{w})$.
        \State Update as $\varpi_{\mathbf{w}} \gets \varpi_{\mathbf{w}^*}$ and $\hat{\varpi}_{\mathbf{w}} \gets \hat{\varpi}_{\mathbf{w}^*}$.
    \EndIf
\EndFor
\State \Return{Learned policy $\varpi_{\mathbf{w}}$.}
\end{algorithmic}
\end{algorithm}

\subsubsection{Value-Based Approaches}\label{sec:value-based-DRL}
The \ac{DQN} algorithm \cite{DQN} constitutes the most well-established value-based \ac{DRL} method for discrete action spaces. An \ac{ANN} is employed to estimate the optimal $Q$-function of the underlying environment, using data collected from the agent's past interactions. This $Q$-network receives as input a state representation and outputs predictions of the $Q$-values for each action (i.e., the output layer has ${\rm card}(\mathcal{A})$ units). To estimate the action value function, the agent collects experience tuples of the form $(\mathbf{s}_t, \mathbf{a}_t, r_t, \mathbf{s}_{t+1})$, which are stored in a replay buffer. At every iteration, experience tuples are sampled in mini-batches $\mathcal{D}$ from the buffer and the network updates its current estimate of the $Q$-value for the pair $(\mathbf{s}_t, \mathbf{a}_t)$ based on the \ac{TD} error; this error indicates the deviation of the predicted optimal $Q$-value for the next state from the estimated $Q$-value of the current action-state pair. By denoting the \ac{ANN} as $Q_{\mathbf{w}}$, with $\mathbf{w}$ being the concatenated parameter vector with the real-valued network's weights, the loss function to be optimized is given by 
\begin{equation}\label{eq:DQN-loss}
\begin{split}
    \mathcal{L}(\mathbf{w}) \triangleq \sum\limits_{(\mathbf{s}_t, \mathbf{a}_t, \mathbf{s}_{t+1}, r_t)\in\mathcal{D}} & ( r_t + \gamma \max_{\mathbf{a} \in \mathcal{A}} Q_{\mathbf{w}}(\mathbf{s}_{t+1}, \mathbf{a}).  \\& \hspace{0.2cm}-Q_{\mathbf{w}}(\mathbf{s}_t, \mathbf{a}_t) )^2,
\end{split}    
\end{equation}
where the summation is over the experience in the mini-batch. The training of the \ac{ANN} can be handled by any variation of the stochastic gradient descent (e.g., Adam \cite{Adam}, RMSprop \cite{Graves_LSTM_RMSProp}), following the generic gradient update rule for $k=0,1,\ldots$:
\begin{equation}\label{eq:dqn-update}
    \mathbf{w}_{k+1} \leftarrow \mathbf{w}_{k} - \alpha \nabla_{\mathbf{w}}\mathcal{L}(\mathbf{w}_k),
\end{equation}
where $\alpha \in (0, 1]$ is the \textit{learning rate} hyperparameter.
This training procedure is known to converge, under mild assumptions, to the optimal $Q$-function up to a statistical error that reflects  the fundamental difficulty of the problem \cite{DQN_TheoreticalAnalysis}.

DQN is an \textit{off-policy} algorithm according to which the policy the agent uses for collecting experience is not the same with the one to be ultimately learned.
The learned policy is effectively the selection of the action that maximizes the predicted $Q$-values, i.e., as in \eqref{eq:optimal-policy}. During training, the $\epsilon$-\textit{greedy} heuristic is used, in which a random action is occasionally selected instead of the one dictated by the maximum $Q$-value:
\begin{equation}\label{eq:epsilon-greedy}
    \mathbf{a}_t \leftarrow \begin{cases}
    \mathcal{U}(\mathcal{A}) & \text{ with probability } \epsilon \\
    \argmax\limits_{a \in \mathcal{A}}Q(\mathbf{s}_t, a) & \text{ with probability } 1-\epsilon \\
    \end{cases}.
\end{equation}
In this expression, $\mathcal{U}(\mathcal{A})$ represents the selection operation from the discrete set $\mathcal{A}$ with probability $1/{\rm card}(\mathcal{A})$, and the value of $\epsilon$ is used to balance the \textit{exploration-exploitation} dilemma:
By only relying on already explored strategies, the agent is prone to suboptimal performance. Thus, a degree of random exploration is required.
On the contrary, by disregarding previously gained information, the algorithm will take more ineffective actions than those necessary for training. Typically, $\epsilon$ takes values in $[0.01, 0.2]$, while it is also common to start the training procedure with a larger value and then decrease it iteratively.

In \cite{DQN}, the efficacy of DQN was demonstrated by training it to achieve human-like performance in Atari games. An important insight from that work is that DQN training can be fairly unstable, due to frequent updates of the network's parameters. Effectively, the same network is generating the next state target $Q$-values that are used in updating its current $Q$-values, which is known to be prone to oscillations or divergence. To account on that, an identical network to $Q_{\mathbf{w}}$, dubbed the target network $Q_{\mathbf{w}^-}$ with separate parameters $\mathbf{w}^-$, was employed to predict the current estimate of the optimal $Q$-value (i.e., (\ref{eq:DQN-loss})'s term $r_t + \gamma \max_{\mathbf{a}} Q_{\mathbf{w}}(\mathbf{s}_{t+1}, \mathbf{a})$).
While the original $Q$-network is updated during every time step $t$, the target network is meant to be updated at a lower frequency, and its new weights are calculated as a ``soft-copy'' from the current value of $\mathbf{w}$, according to a ``temperature'' hyperparameter $\hat{\tau}$:
\begin{equation}\label{eq:dqn-target-soft-update}
    \mathbf{w}^-_{k+1} \leftarrow (1-\hat{\tau}) \mathbf{w}^-_{k} + \hat{\tau} \mathbf{w}_{k}.
\end{equation}
This trick offers increased stability when the estimates over the states change slowly enough for the policy to be able to adapt to the current estimations.
Often, the accompanying practice of clipping the network's gradient values within a specified range is employed to impose more restraint changes on the network's predictions.
\newline

\subsubsection{Policy-Gradient Approaches}
\ac{PG}-based approaches aim to directly find a policy that maximizes the expected sum of rewards in episodic tasks.
In practice, they adopt a parametric approximation of the policy function, i.e., an \ac{ANN} that maps states to actions. For notation purposes, let us rewrite $\mathcal{OP}_2$'s objective in \eqref{eq:RL-objective} as follows:
\begin{equation}\label{eq:RL-PG-objective}
    J_{{\rm PG}}(\mathbf{\mathbf{w}}) \triangleq \mathbb{E}_{\tau \sim \varpi_{\mathbf{w}}} \left\{ \sum_{t=1}^T R_t \right\} = \mathbb{E}_{\tau \sim \varpi_{\mathbf{w}}} \left\{ G^{\varpi_{\mathbf{w}}}_t\right\},
\end{equation}
where the dependency on the network's parameter vector $\mathbf{w}$ is made explicit. The term $\tau$ denotes a \textit{trajectory}, i.e., a sequence of state-action pairs throughout the episode, generated by following $\varpi_{\mathbf{w}}$.
The optimization of $J(\mathbf{\mathbf{w}})$ is carried out through the gradient ascent iteration ($k=1,2,\ldots$):
\begin{equation}\label{eq:theta-pg-gradient-ascent}
    \mathbf{w}_k \leftarrow \mathbf{w}_k + \alpha \nabla_{\mathbf{w}} J_{{\rm PG}}(\mathbf{w}_k).
\end{equation}
The \textit{Policy Gradient} theorem \cite{SuttonPG} states that the stochastic gradient of the objective function can be computed as
\begin{equation}\label{eq:PG-theorem}
    \nabla_{\mathbf{w}} J_{{\rm PG}}(\mathbf{w}) = \mathbb{E}_{\tau \sim \varpi_{\mathbf{w}}} \left\{\sum\limits_{t=1}^T  \nabla_{\mathbf{w}} G^{\varpi_{\mathbf{w}}}_t\log \left(\varpi_{\mathbf{w}} (\mathbf{a}_t | \mathbf{s}_t)\right) \right\}.
\end{equation}
To approximate this expectation, Monte Carlo sampling can be used resulting in the \textit{on-policy} algorithm REINFORCE \cite{REINFORCE}. Note that the theorem does not consider the discounted case of infinite \acp{MDP} (since the training takes place after the end of the episode), but \ac{PG} algorithms may use discounts to compute the returns of finite \acp{MDP} in practice.

\subsubsection{Actor-Critic Approaches}
A problem with Monte Carlo estimates is that they have high variance with respect to the true gradient of the objective function.
One common technique to reduce the resulting noisy gradient updates is to adopt an \textit{advantage} function in place of $G^{\varpi_{\mathbf{w}}}_t$ in (\ref{eq:PG-theorem}).
The most prominent example is the \textit{advantage value} function, which quantifies the benefit of selecting action $\mathbf{a}_t$ at state $\mathbf{s}_t$ when compared to the (average) value of the state.
Specifically, this function is defined as:
\begin{equation}\label{eq:advantage-V}
\begin{split}
    \mathsf{A}(\mathbf{s}_t, \mathbf{a}_t) &\triangleq Q^{\varpi_{\mathbf{w}}}(\mathbf{s}_t, \mathbf{a}_t) - V^{\varpi_{\mathbf{w}}}(\mathbf{s}_t)\\ &= r_{t+1} + \gamma V^{\varpi_{\mathbf{w}}}(\mathbf{s}_{t+1})-V^{\varpi_{\mathbf{w}}}(\mathbf{s}_t).
\end{split}
\end{equation}
Again, the true value function is unknown, but it can be estimated from experience with an \ac{ANN}. Hence, the actor-critic framework employs two distinct networks: the \textit{actor} which learns the optimal policy using a PG-like method, and the \textit{critic} that is trained to predict the value of the state, allowing for \eqref{eq:advantage-V} to be substituted in \eqref{eq:PG-theorem}.

The \ac{PPO} algorithm \cite{PPO} is a \ac{DRL} method that further deals with the problem of noisy updates, by imposing a proximity constraint. According to this algorithm, the gradient values are clipped within a trust region which discourages the likelihood of selecting any specific action under the new policy. The goal of this strategy is not to deviate substantially from the action's likelihood given by the old policy. Apart from being a relatively straightforward method to implement, \ac{PPO} has also shown great performance in multi-agent training scenarios \cite{DOTA-PPO}.

Designed specifically to handle continuous action spaces, the \ac{DDPG} algorithm \cite{DDPG} is another established actor-critic method. It uses a network (critic) to estimate $Q$-values similar to DQN. To deal with the problem of finding the action that maximizes the value is computationally expensive in the continuous domain. Therefore, a separate action-selection network (actor) is used to determine the value of the selected action. The latter networks are trained concurrently in an off-policy manner to minimize their respective objectives, each one having their parameters fixed when computing the gradients of its counterpart. Target networks are employed in both subproblems to stabilize the training process, as proposed by the DQN algorithm.
\\~\\
\par
\begin{figure*}[t]
    \centering
    \includegraphics[width=\linewidth]{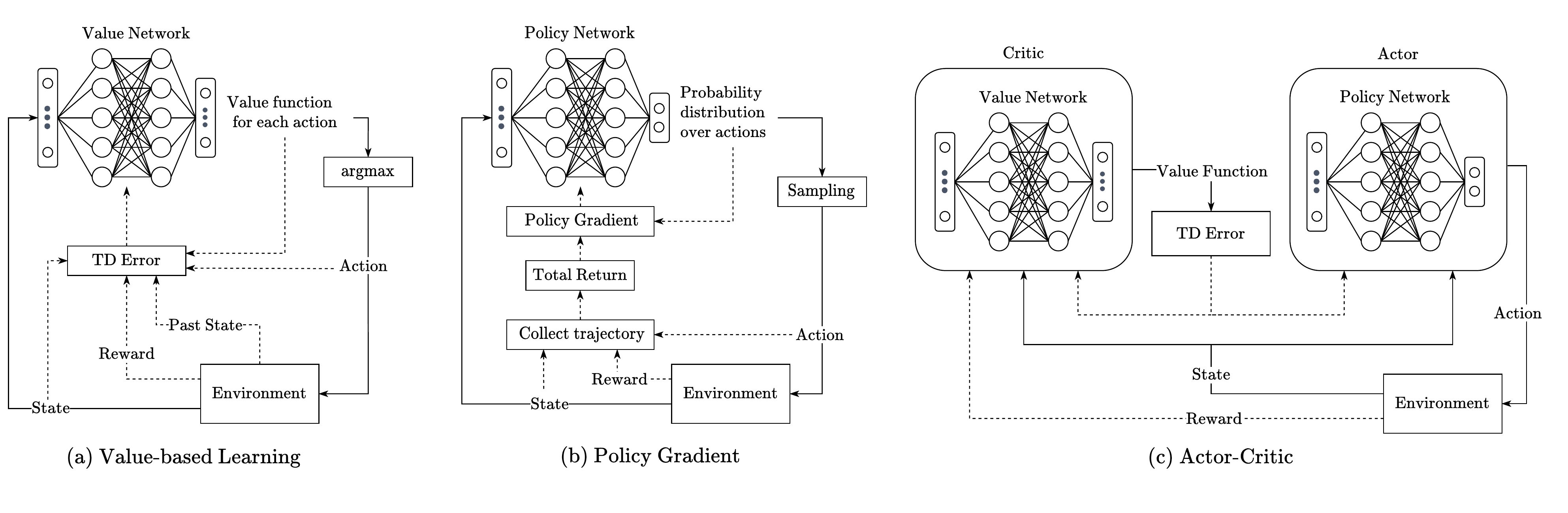}
    \caption{Illustration of the three main approaches of MDP-based DRL methods. Solid lines denote input/output variables of the ANNs and the flow of information during the MDP cycle. Dashed lines denote the flow of information during the training process in each approach.}
    \label{fig:drl_methods_comparison}
\end{figure*}

\remark{The value-based methods are, in general, more straightforward to implement and arguably more interpretable, since they provide evaluations of the different states (and actions) of the MDP. On the contrary, the need for computing one value per state-action pair makes them inefficient in (combinatorially) large and discrete action spaces, even if discretization is applied. On the other hand, PGs can adapt to both of those cases, since a parametric probability distribution is adopted that can allow for efficient action sampling. Despite that, PG updates are usually less sample efficient since trajectories of arbitrary length must be collected through Monte Carlo sampling, and each gradient update affects only the network's belief on the action selected. This is in contrast to value-based methods that may update the estimation of a state's value for all actions simultaneously.
Another difference between these two categories arises from the on-/off-policy distinction of their algorithms. PGs, being on-policy, are trained strictly when exploring the environment. Contrarily, value-based methods are flexible to also be trained completely off-policy (i.e., on previously collected MDP transitions), which might be crucial when online querying of the environment is expensive or unavailable. Finally, actor-critic methods are designed to capitalize on the benefits of the above two categories, since they employ both a value and a policy network. In fact, current state-of-the-art \ac{DRL} methods belong to this class of algorithms, notwithstanding their increased computational needs and added complexity. A schematic overview of the three main DRL approaches discussed in this survey is given in Fig.~\ref{fig:drl_methods_comparison}. The diagrams are kept simple, by excluding less important aspects of the presented methods for the sake of clarity.}

\subsection{Multi-Armed Bandits}\label{sec:Bandits}
As previously mentioned, one of the basic principles of the \ac{MDP} formalism is that the action selected by an agent influences the next state of the environment. Hereinafter, we consider the case where the agent still selects an action to maximize a cumulative reward signal, and potentially receives observations, but its actions have no impact on the evolution of the environment. Such sequential decision making problems are commonly referred to as \textit{multi-armed bandits}. Concretely, at each time step, a decision maker needs to select an action from a discrete action set. The environment internally keeps a separate probability distribution of the payoff (reward) for each action, and provides the agent with a reward sampled from the distribution corresponding to the selected action. The agent is again tasked with maximizing the expected (potentially discounted) reward within a single episode. We note that in the current version of the formalism there are no states present. As a result, the notion of policy is not applicable. Instead, the agent needs to find the most rewarding action(s), as quickly as possible, to maximize its expected payoff.

The multi-armed bandits problem formulation exemplifies the exploration-exploitation dilemma of MDP-based approaches, since the agent needs to try different actions to infer the optimal one(s). Most commonly, the algorithms applicable to this case first keep track of a running average of the achieved reward for each action. This running average serves as an estimate of its true expected value and is expressed as:
\begin{equation}\label{eq:bandit-average}
    \mathsf{G}_{t+1}(\mathbf{a}) \triangleq \frac{1}{\mathsf{N}_t(\mathbf{a})}\sum_{i=1}^{\mathsf{N}_t(\mathbf{a})}r_i^{\mathbf{a}} = \mathsf{G}_t(\mathbf{a}) + \frac{1}{\mathsf{N}_t(\mathbf{a})} \left[ r_{\mathsf{N}_t(\mathbf{a})} - \mathsf{G}_{t}(\mathbf{a}) \right],
\end{equation}
where $\mathsf{N}_t(\mathbf{a})$ is the number of times action $\mathbf{a}$ has been selected up to time $t$. Note that the \ac{MDP} formalism has been slightly altered, by defining the per-action return and by denoting with $r_i^{\mathbf{a}}$ the reward associated with $\mathbf{a}$ when the action is selected for the $i$-th time. This learned information is further combined with an exploration strategy. We briefly describe two of the most common methods:
\begin{enumerate}
    \item[1.] The \textbf{$\epsilon$-greedy} strategy, which was previously described. We explicitly redefine the strategy to incorporate the current multi-armed bandits framework as
    \begin{equation}\label{eq:epsilon-greedy-bandits}
        \mathbf{\mathbf{a}}_t \leftarrow \begin{cases}
    \mathcal{U}(\mathcal{A}) & \text{ with probability } \epsilon \\
    \argmax\limits_{\mathbf{a} \in \mathcal{A}}\mathsf{G}_t(\mathbf{a}) & \text{ with probability } 1-\epsilon \\
    \end{cases}.
    \end{equation}
    \item[2.] The \textbf{\ac{UCB}} heuristic, which takes into account both the maximum estimated values and the uncertainty in those estimates:
    \begin{equation}\label{eq:ucb}
        \mathbf{a}_t \leftarrow \argmax\limits_{\mathbf{a} \in \mathcal{A}} \left\{ \mathsf{G}_t(\mathbf{a}) + c \sqrt{\frac{\ln(t)}{\mathsf{N}_t(\mathbf{a})}} \right\},
    \end{equation}
    where $c>0$ controls the width of the confidence interval, and hence, the amount of exploration. For completeness, when $\mathsf{N}_t(a)=0$, only the confidence bound term is set to $0$.
    It is usually expected from \ac{UCB} to perform better than the naive $\epsilon$-greedy approach, since the former has the advantage of an uncertainty-guided exploration, instead of the random strategy of the latter.
\end{enumerate}
It is noted that the multi-armed bandits problem  concerns both the cases of stationary and non-stationary distributions over the actions. In the latter case, the agent would benefit by giving preference to more recent observations. As a result, the running average of \eqref{eq:bandit-average} can be modified to an exponential recency-weighted average \cite[Chapter 2]{Sutton98reinforcementlearning}.

A more general formulation of the bandits problem admits observations by the agent.
In this \textit{contextual bandits} case, the agent receives an observation vector (termed as the context) from the environment prior to deciding each action, with the intent of being guided to the optimal selection strategy. Therefore, the notion of policies applies to contextual bandits, but since there are no transition probabilities between states and actions, contextual bandits problems are considered relatively easier than the full-MDP ones. It is also worth mentioning that the observation vector is not assumed to fully represent the internal state of the system, but it is rather expected to have a relationship with the reward values.
Conceptually, such bandit settings can be thought of as episodic \ac{MDP} problems with a trajectory of length $1$ (i.e., $T=1$). Due to that, standard \ac{DRL} algorithms can be applied in theory, but bandit-tailored methods are more suited, since the former algorithmic approaches are unnecessarily complex.
\par
Indicative in this contextual bandits category is the LinUCB algorithm \cite{LinUCB} which assumes a linear dependency between observations and actions. Specifically, the expected reward for each action, which depends on the current observation (i.e., context), is modeled as a linear relationship between an unknown coefficient vector and the observation vector.
That coefficient vector is estimated through ridge regression on past observations-rewards data, and an upper bound of the expected payoff is derived.
As such, the \ac{UCB} action-selection strategy is finally employed to determine the chosen action. Unsurprisingly, the assumption of linearity between observations and rewards can be restrictive in many challenging problems; this means that the potential representations of the coefficient vectors are limited. To remedy for that, the authors in \cite{DeepBaysianBandits} proposed the neural linear bandits algorithm, which utilizes an encoding neural network to learn the coefficient vector instead of ridge regression. The network was trained as a regressor to predict the reward given an observation. To attain the encoding vector, the outputs of the final hidden layer were extracted, i.e., the original output layer was dropped. A Thomson sampling action-selection strategy was employed instead of \ac{UCB}, but the approach can be trivially imported to the LinUCB setting.
\par
Another straightforward approach to incorporate deep learning into the contextual bandits formulation would be via a \textit{reward-prediction} \ac{ANN}. Such a network accepts an observation vector as input at time $t$ and outputs as many real numbers as the number of possible actions for the agent. Those numbers are the expected reward values for each action, i.e., $\mathsf{G}_t(\mathbf{a})$ $\forall \mathbf{a} \in \mathcal{A}$. Thus, the $\epsilon$-greedy strategy of \eqref{eq:epsilon-greedy-bandits} can be utilized during the learning process. To this end, the network can be trained using tuples of observation and reward, and the \ac{MSE} between predicted and true rewards can be used as the objective function.
Note that the \ac{MSE} is computed only with respect to the output neuron that corresponds to the selected action, and not to the whole output vector of the network.
We will be referring to this approach as the ``Neural $\epsilon$-greedy'' algorithm in the sequel.
\par
Although \ac{MDP} and \ac{POMDP} formulations are the most popular for wireless communications up to date \cite{gao2021ResourceAllocation, Abdelrahman2020TowardsStandaloneOperation, chongwendrl, Yang2021DRL_for_Secure, Hu2021MetaSensing, Nguyen2021MultiUAVDRL, Feng2020DRL_MISO, Huang2021MultiHop, Lee2020DRL_EE, Liu2021MNOMA_deployment, kim2021multiirsassisted, alhilo2021reconfigurable, Samir2021AgeOfInformation, MehdiDRL}, the multi-armed bandits formulations have a substantial proposition of value, despite their simplicity. 
The multi-armed bandits algorithms are more intuitive and explainable compared to \ac{DRL} methods.
Their converge properties are rather easier to be studied theoretically \cite{Auer2002BanditsAnalysis}, and their effectiveness can be demonstrated empirically \cite{VermorelBanditsEmpirical}, while being able to utilize the immense representational power of deep learning \cite{DeepContextualBandits}.
In the section with the numerical results, different multi-armed bandits algorithms for the orchestration of the considered RIS-empowered smart wireless environments will be evaluated.

\subsection{Algorithmic Evaluation}\label{sebsec:DRL-evaluation}
The theoretical analysis of \ac{DRL} algorithms is an active topic of research. Convergence results have been established for well-known methods under fairly mild conditions (e.g., \cite{DQN_TheoreticalAnalysis} for DQN, \cite{SuttonPG} for PG, and \cite{ActorCriticConvergenceLinearQuadratic} for actor-critic), but they may be asymptotic, consider only specific families of \acp{MDP} (e.g., \cite{ActorCriticConvergenceLinearQuadratic}), or impose constraints on the underlying functions. The fundamental obstacle in attaining general results is that the \ac{MDP} formalism makes almost no assumptions about the underlying problem.
Therefore, the rate and point of convergence are dependent on the characteristics/difficulty of the application.
\par 
A most useful evaluation practice is that of experimental evaluation. \ac{DRL} algorithms are commonly applied to benchmark environments of different levels of difficulty with various characteristics.
Different metrics can be used to assess their performance, namely the cumulative reward achieved and the number of samples (expressed in time steps) required up to a certain performance threshold; the latter provides a notion of \textit{sample efficiency}. In general, off-policy algorithms are expected to be more sample-efficient, since they are able to iterate over previous experiences multiple times.

In terms of execution time per iteration, the DRL algorithms are commonly treated as constant time. In general, the time complexity of a forward pass in an \ac{ANN}(s), which is required for a prediction, is of the order of the number of the training parameters in $\mathbf{w}$ with an added $\Theta({\rm card}(\mathcal{A}))$ cost due to the $\argmax$ operation on discrete spaces, or an equivalent cost of the sampling from a distribution in the continuous case.
The training involves taking a gradient step and back-propagating the gradient through the network, which entails the same complexity as the forward pass, assuming $O(1)$ batch sizes of potential mini-batches.

\section{DRL-Orchestrated Smart Radio Environments}\label{sect:AI-RIS}
In this section, we first present a DRL-based formulation of the design objective introduced in Section~\ref{sect:RIS-environments} for RIS-empowered smart wireless environments, and discuss its applicability for various other performance objectives. Secondly, we provide a detailed overview of the available DRL approaches for wireless communication systems including RISs.

\subsection{DRL Formulation}\label{sec:DRL_formulation}
\begin{figure}[!t]
    \centering
    \includegraphics[scale=0.7]{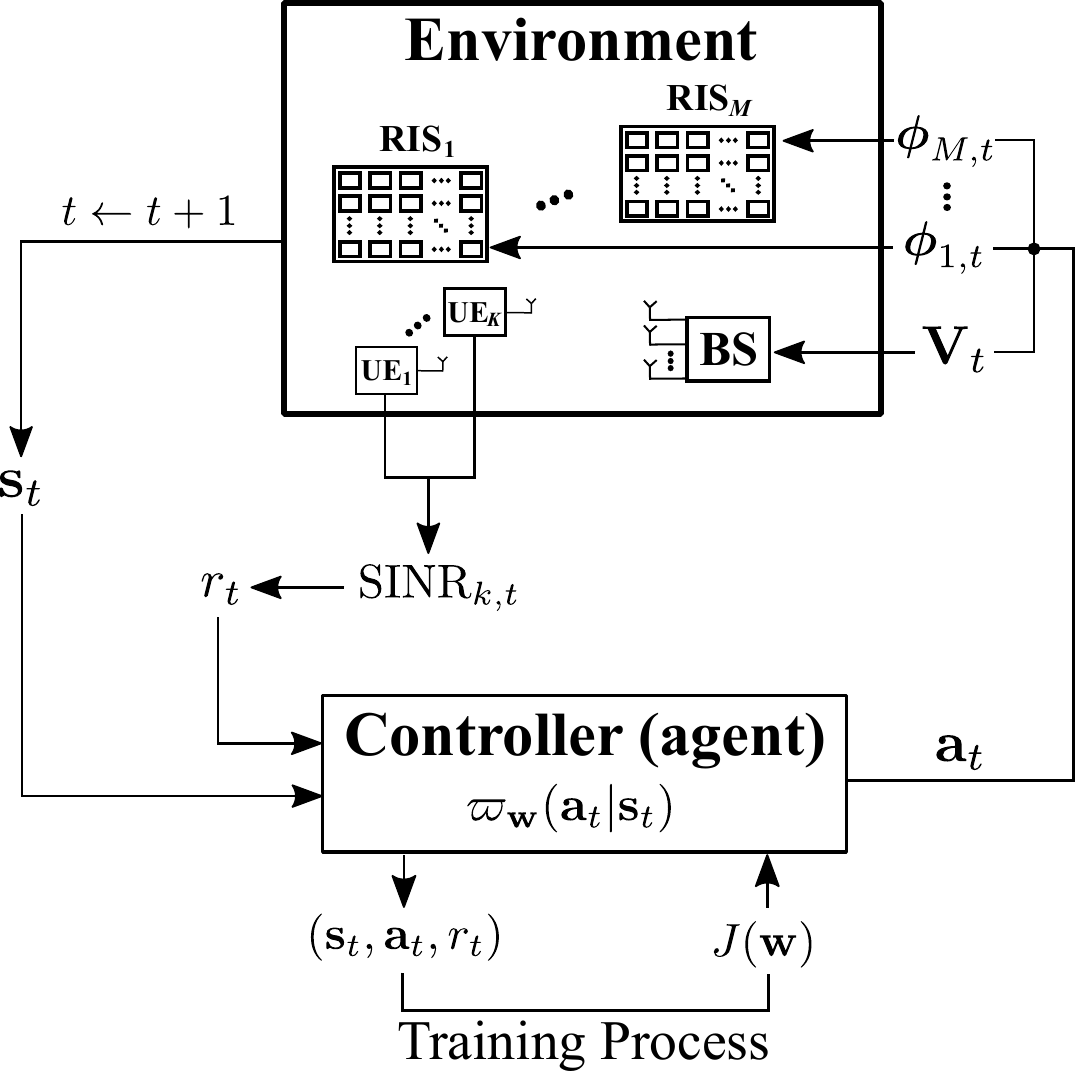}
    \caption{Block diagram of the proposed DRL formulation for the considered RIS-empowered smart radio environment in Section~\ref{sec:System_Model} with the sum-rate maximization design objective.}
    \label{fig:RL_RIS_formulation}
\end{figure}
Following the theoretical foundations of (D)RL, as presented in Section~\ref{subsec:RL}, for the considered system model in Section~\ref{sec:System_Model} and its achievable sum-rate maximization design problem given by \eqref{main_problem}, we map the parameters of the wireless communication system with the DRL ingredients, as illustrated in Fig$.$~\ref{fig:RL_RIS_formulation} and discussed in the sequel:
\begin{itemize}
\item The controller, which is assigned the role to orchestrate the multi-antenna BS, the multiple \acp{RIS}, and the multiple \acp{UE} as per \eqref{main_problem}'s objective, is the \textit{agent} in the DRL formulation, being specifically capable of deciding at each time step $t$ (e.g., per channel coherence time or performance-based requests) the phase profiles for all \acp{RIS} as well as the precoding matrix of the \ac{BS}. We redefine the former as $\{\boldsymbol{\phi}_{m,t}\}_{m=1}^M$ and the latter as $\mathbf{V}_{t}$ to explicitly indicate their dependence on the instant $t$.
\item The possible \textit{actions}, that the controller (i.e., the agent) can take, are determined by the feasible combinations of the free design parameters of the wireless system, which constitute the \textit{action space} $\mathcal{A}$. For the considered system, this space includes $[\boldsymbol{\phi}_m]_i\in\mathcal{F}$ $\forall m,i$ and $[\mathbf{V}]_{:,k}\in\mathcal{V}$ $\forall k$, resulting in ${\rm card}(\mathcal{A})=2^{b N_{\rm tot}}K{\rm card}(\mathcal{V})$. In addition, an action at any time $t$ can be defined by the following complex-valued\footnote{The implementations of the \ac{RL} algorithms are not designed to handle complex-valued numbers. Due to this, it is common to concatenate the real and imaginary parts (or magnitude and phase) of complex-valued vectors into real vectors, having effectively the double dimension.} $KN_{\rm tot}N_{\rm T}$-element column vector:
\begin{equation}\label{eq:action_space_vector}
\mathbf{a}_t \triangleq [\boldsymbol{\phi}_{1,t};\boldsymbol{\phi}_{2,t};\dots;\boldsymbol{\phi}_{M,t}; {\rm vec}(\mathbf{V}_t)]^T.
\end{equation}
\item The wireless propagation environment corresponds to the \textit{environment}\footnote{
The environment in the context of \ac{RL} includes any aspect of a system that is not under the direct control of an agent. For example, in our system model, \ac{UE} properties such as locations, mobility, rate requests, or others constraints may be considered parts of the environment.}, whose state can be naturally mapped to the channel gain coefficients that change per channel coherence time. Consequently, the controller (i.e., the  agent) can observe the environment via performing/collecting estimates of the elements of the channel matrices. The state of the environment at a time instant $t$ can be defined by the following complex-valued column vector with dimensionality ${\rm dim(\mathbf{s}_t)} \triangleq N_{\rm tot}\left(K+N_{\rm T}\right)+KN_{\rm T}$:
\begin{equation}\label{eq:DRL-state}
\mathbf{s}_t \triangleq [{\rm vec}(\tilde{\mathbf{H}}_{t});\mathbf{g}_{m,k,t}^T; \mathbf{h}_{k,t}^T]^T,
\end{equation}
where we have used the definitions:
\begin{align}
\tilde{\mathbf{H}}_{t}&\triangleq[\mathbf{H}_{1,t};\mathbf{H}_{2,t};\,\cdots\,;\mathbf{H}_{M,t}]\in\mathbb{C}^{N_{\rm tot}\times N_{\rm T}},\\
\tilde{\mathbf{g}}_{t}&\triangleq[\mathbf{g}_{1,1,t}^T;\mathbf{g}_{1,2,t}^T;\,\cdots\,;\mathbf{g}_{M,K,t}^T]\in\mathbb{C}^{KN_{\rm tot}\times 1},\\
\tilde{\mathbf{h}}_{t}&\triangleq[\mathbf{h}_{1,t}^T;\mathbf{h}_{2,t}^T;\,\cdots\,;\mathbf{h}_{K,t}^T]\in\mathbb{C}^{KN_{\rm T}\times 1}.
\end{align}
In practice, the controller possesses the estimate $\hat{\mathbf{s}}_t$ of $\mathbf{s}_t$.
\item The controller is capable of inferring the outcome of its actions on the wireless environment at any time instant $t$. This is usually quantified via the \textit{reward} for this instant, which constitutes the feedback of the controller's actions from environment that can be collected by the controller. For the considered sum-rate maximization objective in $\mathcal{OP}_1$, the agent (i.e., the controller) is assumed to be capable of collecting the SINR measurements of each UE$_k$ at any instant $t$, denoted by ${\rm SINR}_{k,t}$. Focusing first on the case where the individual UE rate constraints in \eqref{constraint:rate} are excluded from $\mathcal{OP}_1$, the agent uses the latter measurements to calculate the following reward:
\begin{equation}\label{eq:reward} 
 r_t \triangleq \sum_{k=1}^{K}  \tilde{R}_{k,t}.
\end{equation}
Note that, since the the controller gathers the \ac{CSI} state estimate $\hat{\mathbf{s}}_t$ of the environment at any instant $t$, the SINR values, and hence, the reward for this instant can be estimated without the intervention of the UEs (i.e., without the need to collect feedback from them). Nevertheless, relying on a UE reporting mechanism to generate the reward signal at the agent ensures that the pragmatic reward is captured, which in the previous case would be prone to imperfections due to collection of partial or erroneous CSI (or even insufficient modeling of the channel and the control operation).

\hspace{0.4cm}For the case where $\mathcal{OP}_1$ is considered exactly as it appears in  \eqref{main_problem} (i.e., including the \eqref{constraint:rate} constraints), the reward function needs to revised to penalize actions that result in states where any of the individual rate requirements is not satisfied. There exist various techniques for achieving this goal (e.g., \cite{ijcai2021-614}) , which actually depend on various aspects of the MDP
. In this paper, we choose to reshape the reward function to return $-1$ for each UE$_k$ whose individual ${\rm SINR}_k$ did not satisfy the specified request (i.e., when the specific constraint in \eqref{constraint:rate} was not met). Following this formulation, the agent was considered to compute the reward:
\begin{equation}\label{eq:reward-qos}
 r_t = \begin{cases}
-\sum_{k=1}^{K} \mathbbm{1}_{(\tilde{R}_k < R^{\rm req}_k)}, & \text{if } \exists k: \tilde{R}_k < R^{\rm req}_k\\
\sum_{k=1}^{K}  \tilde{R}_k, &\text{otherwise}
\end{cases},
\end{equation}
where $\mathbbm{1}_{(\cdot)}$ denotes the indicator function. It is noted that, while maximizing the reward in~\eqref{eq:reward-qos} indeed maximizes $\mathcal{OP}_1$, the achieved average rewards can no longer be interpreted as achievable sum rates. This happens because in the definition of \eqref{eq:reward-qos}, negative scores affect the resulted metric. Nevertheless, positive values for rewards indicate that the individual UE rate requests are successfully met; in this case, the resulting reward equals the optimized achievable sum-rate performance.
\item The RIS phase profiles and the BS precoding matrix chosen by the controller at any time instant $t$ (i.e., the agent's action) will have no effect on the channel gain matrices (i.e., the state of the environment) of the next time instant $t+1$. Together with our assumption for \ac{IID} channel realizations (in the time dimension) in Section~\ref{sec:System_Model}, it yields for our DRL formulation that the \textit{Markovian transition probability} in \eqref{eq:markovian-property} is independent of the previous state $\mathbf{s}_t$ and action $\mathbf{a}_t$, i.e., it holds: ${\rm Pr}[\mathbf{s}_{t+1}|\mathbf{s}_t$, $\mathbf{a}_t] = {\rm Pr}[\mathbf{s}_{t+1}]$. This constitutes a special case of the MDP formulation that better adheres to the multi-armed bandits paradigm, as discussed previously in Section~\ref{sec:Bandits}. It is noted, however, that when time-correlated channels are involved (see, e.g., the model in \cite{Channel_Time_Correlation}), the channel state at each time instance is dependent on the preceding channel state(s). Considering that the channel state at each time $t+1$ depends only on the channel state at $t$, as an indicative example of time-correlated channels, the Markovian transition probability is expressed as ${\rm Pr}[\mathbf{s}_{t+1}|\mathbf{s}_t$, $\mathbf{a}_t] = {\rm Pr}[\mathbf{s}_{t+1} | \mathbf{s}_t]$. Even though \ac{MDP}-based algorithms are inherently capable of capturing such time dependencies, the multi-armed bandits outlook can still be adopted. In particular, any algorithm falling into the latter category needs to be applied in ways such that it updates the expected reward estimation by emphasizing the most recent observations. In Section~\ref{sec:results-partial-observability}, the learning capability of the proposed controllers based on multi-armed bandits will be numerically investigated over \ac{IID} and time-correlated channels. 
\end{itemize}

\begin{algorithm}[!t]
\caption{DRL-Based Training Solving $\mathcal{OP}_1$ in \eqref{main_problem}}\label{alg:RIS_DRL_formulation}
\begin{algorithmic}[1]
\Require Number of UEs $K$, number of RISs $M$, initial ANN parameters $\mathbf{w}$, policy $\varpi_{\mathbf{w}}$, collection policy $\hat{\varpi}_{\mathbf{w}}$, objective function $J(\cdot)$, final time step $T$, and update interval $t'$.
\For{$t=1,2,\dots,T$}
    \State Estimate $\mathbf{h}_{k,t}$ $\forall k=1,2,\dots,K$.
    \State Estimate $\mathbf{H}_{m,t}$ and $\mathbf{g}_{m,k,t}$ $\forall m=1,2,\dots,M$ and \Statex\hspace{0.565cm}$\forall k=1,2,\dots,K$. 
    \State The controller formulates the state vector $\hat{\mathbf{s}}_t$ similar to \Statex\hspace{0.565cm}\eqref{eq:DRL-state} using the channel estimates in Steps $2$ and $3$.
    \State The controller decides the action $\mathbf{a}_t$ (i.e., $\boldsymbol{\phi}_{m,t}$ $\forall m$~and \Statex\hspace{0.565cm}$\mathbf{V}_t$) using its policy $\hat{\varpi}_{\mathbf{w}}(\mathbf{a}_t | \hat{\mathbf{s}}_t)$ and shares these \Statex\hspace{0.565cm}settings with the RISs and BS.
    \State Each UE$_k$ measures ${\rm SINR}_{k,t}$ and sends it to the \Statex\hspace{0.565cm}controller.
    \State Using Step $6$, the controller computes the reward $r_t$ \Statex\hspace{0.565cm}and stores the experience tuple $(\hat{\mathbf{s}}_t, \mathbf{a}_t, r_t)$ to a set $\mathcal{D}$. 
    \If{${\rm mod}(t,t') = 0$}
        \State The controller uses $\mathcal{D}$ to compute $\mathbf{w}^*$ that \Statex\hspace{1.105cm}optimizes $J(\mathbf{w})$.  
        \State It then performs the updates $\varpi_{\mathbf{w}} \gets \varpi_{\mathbf{w}^*}$ and \Statex\hspace{1.105cm}$\hat{\varpi}_{\mathbf{w}} \gets \hat{\varpi}_{\mathbf{w}^*}$.
    \EndIf
\EndFor
\State \Return{Learned policy $\varpi_{\mathbf{w}}$.}
\end{algorithmic}
\end{algorithm}
From a system architecture and operation perspective, training a \ac{DRL}-based agent entails a non-negligible infrastructure and computational overhead. At every time instant, CSI measurements need to be performed and collected by the controller. Leveraging the uplink/downlink channel reciprocity in time division duplexing systems, channel acquisition can be implemented by transmitting pilot signals from the UEs in the uplink and applying estimation techniques for the direct and the RIS-assisted channels at the BS and/or the controller sides \cite{Tsinghua_RIS_Tutorial}. With this information at the controller's disposal, the sum-rate maximizing phase profiles for the RISs and the BS precoding matrix are computed and shared with the involved devices. Finally, the \acp{UE} are required to measure their individual SINR values and feedback them to the controller, facilitating its training process in a dynamic manner. The complete DRL training procedure for a time window of length $T$ is described in Algorithm~\ref{alg:RIS_DRL_formulation}.
Note that due to the consideration of IID state transitions, $\hat{\mathbf{s}}_{t+1}$ needs not be stored in any experience tuple (see Step $7$), as in Algorithm~\ref{alg:DRL_training}, since it does not provide any instructive information in learning the environment's dynamics or determining an appropriate action. Moreover, the practical aspect of obtaining a CSI estimate $\hat{\mathbf{s}}_{t}$ in place of $\mathbf{s}_{t}$ poses the current formulation as an \ac{POMDP} problem, rather than the simple \ac{MDP} case. In practice, however, ANN-based algorithms are capable of attaining notable performance, even in the presence of observational noise (i.e., noisy CSI).


After the completion of the training process described in Algorithm~\ref{alg:RIS_DRL_formulation}, the controller configures the elements of the smart wireless environment for downlink communication, according to the learned policy $\varpi_{\mathbf{w}}$. To this end, all steps of this algorithm, except the Steps $6$--$10$ referring to the ANN training, can be used for the orchestration with the difference being that the controller uses $\varpi_{\mathbf{w}}$ instead of $\hat{\varpi}_{\mathbf{w}}$ in Steps $5$ as its policy. Note that the SINR collection is no longer needed at the ANN deployment phase, since no training takes place.  

Although our DRL formulation considers the sum rate as the performance objective, it is general enough to be equivalently applied to any desired performance metric \cite{rise6g} that is a direct function of the current channel coefficients and \ac{RIS} phase profiles (e.g., energy efficiency, low EM-field exposure, secrecy rate, \ac{NMSE}, and localization/sensing accuracy). It is also important to mention that (D)\ac{RL} formulations are by no means limited to observing CSI and controlling standalone \acp{RIS}. A wide variety of more elaborate problems can be conceptualized in which the agent observes different kinds of information (e.g., current \acp{UE} positions) and controls diverse elements of the environment (e.g., power allocation and \acp{UAV} with mounted \acp{RIS}). In the following, we discuss the state-of-the-art in DRL approaches for RIS-empowered wireless communication systems.

\begin{table*}[!t]
    \centering
    \caption{Taxonomy of the state-of-the-art of DRL-Based Approaches for RIS-Empowered Smart Radio Environments.}
    \label{table:DRL_taxonomy}
    \begin{tabulary}{\textwidth}{|c|L|L|L|L|}
        \hline
        \textbf{Reference} & \textbf{Method} & \textbf{Action} & \textbf{State} & \textbf{Reward} (Objective) \\ 
        \hline
        \cite{gao2021ResourceAllocation} & \acs{DQN} & \acs{RIS} phase profile, power allocation & Past action & Difference between current and previous instantaneous rate \\
        \hline 
        \cite{Abdelrahman2020TowardsStandaloneOperation} & DQN & RIS phase profile & Estimated \acs{CSI} & Rate (quantized)  \\
        \hline
        \cite{chongwendrl} & \acs{DDPG} & RIS phase profile, beamforming vector & CSI, past action, transmit power & Sum rate \\
        \hline
        \cite{Yang2021DRL_for_Secure} & DQN (extended) & RIS phase profile, beamforming vector & Past CSI, past rates & Secrecy rate, \acs{QoS} constraints \\
        \hline
        \cite{Hu2021MetaSensing} & \acs{PG}-Based & Single meta-atom phase response & RIS phase profile, index of meta-atom, time counter & Sensing error \\
        \hline
        \cite{Nguyen2021MultiUAVDRL} & DDPG, \acs{PPO} & RIS phase profiles, power allocation & Reflected channels & Energy efficiency \\
        \hline
        \cite{Feng2020DRL_MISO} & DDPG & RIS phase profile & Past SNR value, past action & \acs{SNR} \\
        \hline
        \cite{Huang2021MultiHop} & DDPG & Phase profiles of multiple RISs, beamforming matrix & CSI, past action & Sum rate, power constraints \\
        \hline
        \cite{Lee2020DRL_EE} & DQN & RIS phase profile, ON/OFF meta-atoms, transmit power allocation & Beamforming vectors of UEs, RIS energy level & Energy efficiency \\
        \hline
        \cite{Liu2021MNOMA_deployment} & \acs{D$^3$QN} & RIS phase profile, transmit power allocation, RIS placement & Positions of UEs, past phase profile, past power allocation, past RIS position & Energy efficiency \\
        \hline
        \cite{kim2021multiirsassisted} & DQN & Phase profiles of multiple RISs, transmit power allocation, beamforming vector & Local and neighboring channel powers, neighboring sum rates & Sum rate minus interference \\
        \hline
        \cite{alhilo2021reconfigurable} & PPO & Selection of vehicles to be served & Vehicle positions, velocities, rates & Minimum average rate \\
        \hline
        \cite{Samir2021AgeOfInformation} & PPO & RIS phase profile, UAV height & UAV altitude, rate, and AoI of IoT devices & Negative sum AoI \\
        \hline
        \cite{MehdiDRL} & Quantile Regression & UAV location & Received signal power & Total received data during transmission \\
        \hline
    \end{tabulary}
\end{table*}

\subsection{Literature Overview}\label{sect:Literature}
In this section, we provide a brief overview of the literature in DRL-based orchestration for RIS-empowered smart radio environments. In the largest portion of the overview, the presented approaches consider a system model similar to the one considered in this paper and detailed in Section~\ref{sect:RIS-environments}. More specifically, most of the literature employs a DRL agent that is trained using CSI observations and is assigned the role to learn to configure the phase shifts of the RIS (possibly along with other system parameters), thus making the underlying MDP formulations adhere to the general formalism described in Section~\ref{sec:DRL_formulation}. Each of the discussed studies considers a different variation that is based on the peculiarities of the proposed system model. We begin by describing the approaches that are trained to optimize the UE(s)' achievable rates, and are thus more closely related to this paper's design formulation.
\par
A \ac{DRL} formulation similar to that of Section~\ref{sec:DRL_formulation} was adopted in \cite{chongwendrl}. This work considered continuous RIS phase profiles and precoding matrices, and as a result, the \ac{DDPG} algorithm was deployed to learn the optimal configurations. The transmission power at every time step was embedded in the observation vector, and the underlying policy network was extended with an additional layer that ensured the satisfaction of the problem's power constraint. In \cite{Huang2021MultiHop}, the same design formulation with the extension of power allocation was transferred in a multi-RIS communication system with mobile UEs operating at the \ac{THz} frequency band. It was shown that DDPG is capable to find the RIS phase configurations as well as the BS precoding vectors and power allocation strategies that can compensate for the severe attenuation of \ac{THz} signal propagation, outperforming an alternate beamforming benchmark in terms of total throughput. The authors in \cite{gao2021ResourceAllocation} considered a more elaborate system model that consisted of the consecutive sub-problems of estimating the UEs' mobility patterns, finding their optimal allocation into clusters, computing the RIS phase configuration. Different machine learning algorithms were devised for each case, with DQN set as the RIS orchestrating approach. The overall algorithm received the results of the aforementioned tasks as inputs, apart from the \ac{CSI}-based observation vector defined in Section~\ref{sec:DRL_formulation}, in its aim to maximize the sum-rate performance.
\par
Some works in the area addressed the impracticality of assuming full \ac{CSI} knowledge when deploying DRL-based RIS controllers; this issue will be revisited in the open challenges in Section~\ref{sec:open-challenges}.
The overhead of the channel estimation process in RIS-empowered communication systems was the main focus of the DRL approach in \cite{Abdelrahman2020TowardsStandaloneOperation}. A hybrid RIS was adopted with a small percentage of active elements able to perform channel sensing. As a result, the observation space was comprised of estimates of the true channel state, making the problem a \ac{POMDP}. The authors trained the ubiquitous DQN agent (which is also considered in the numerical evaluation Section~\ref{sect:Experiments} in this paper) with quantized rewards ($+1$ or $-1$ depending on whether the current (single) UE rate exceeded the previous value), demonstrating that DRL methods are applicable even in partial channel knowledge, while requiring minimal training overhead. In a similar manner, the authors of \cite{Feng2020DRL_MISO} investigated the DDPG agent in a scenario with a mobile UE, without relying on CSI knowledge. Instead, the action and reward of the past time step constituted the observation vector. The DRL algorithm was responsible for finding the (continuous) RIS phase shifts that maximize the \ac{SNR}, while the precoder was derived through \ac{MRT} \cite{alexandg_MRT}. A multi-cellular multi-RIS system model was investigated in \cite{kim2021multiirsassisted}, where DRL methods were trained to decide on the BS transmit power and precoder, as well as the RIS phase configuration that maximize the UEs' \ac{SINR} values. The RIS controller (which was a DQN controller) was assumed to have access to only local channel states, and the proposed DRL schemes were evaluated under imperfect \ac{CSI} conditions. In Section~\ref{sect:Experiments}, we follow an approach similar to \cite{Feng2020DRL_MISO} in order to evaluate \ac{DRL} methodologies with and without \ac{CSI}-based observations, by devising a strong LOS scenario with mobile \acp{UE} (i.e., time-evolved channels).
\par
In the context of this survey, we have focused on the sum-rate maximization as the objective of the (D)RL formulations. However, a number of works in the RIS field dealt with the \ac{EE} or the secrecy rate design problems \cite{Lee2020DRL_EE, Nguyen2021MultiUAVDRL, Liu2021MNOMA_deployment, Yang2021DRL_for_Secure}. As discussed in Section~\ref{sec:extensions}, the general formulation in this work can be modified to account for other objectives through the pertinent definition of the reward function. In \cite{Lee2020DRL_EE}, RIS elements that the controller can turn ON or OFF, as part of the action space of the RL formulation, were further considered. In \cite{Nguyen2021MultiUAVDRL}, a multi-\ac{UAV} system model, in which the \acp{UAV} facilitates the downlink transmission in clusters of UEs alongside an RIS, was investigated. The employed PPO algorithm observed channel states and controlled the power allocation at the \acp{UAV}. The DDPG and PPO algorithms were compared against each other, both in a centralized, as well as a decentralized (i.e., \ac{MAML}) system architecture. The \ac{EE} design problem was also studied in \cite{Liu2021MNOMA_deployment} for \ac{MISO} \ac{NOMA} networks with mobile UEs. Apart from the power allocation, the considered design problem accounted for displacements of the RIS, and both were embedded into the action space of the RL formulation. The proposed approach leveraged the \ac{D$^3$QN} variant of the DQN algorithm, in which the $Q$-network was tasked to learn both the value function as well as the advantage function to compose the $Q$-value for each state-action pair, paired with a decaying $\epsilon$-greedy exploration policy. Finally, the authors in \cite{Yang2021DRL_for_Secure} adopted the DQN algorithm under the previously described design formulation, with the purpose of maximizing the secrecy rate of the legitimate UEs under an eavesdropping scenario.
\par
Although the DRL-based design formulation in this work is quite general to account for a large number of applications, it is worth mentioning that more elaborate systems do not adhere to this framework. Like \cite{Nguyen2021MultiUAVDRL}, the works \cite{alhilo2021reconfigurable, MehdiDRL, Samir2021AgeOfInformation} considered mobile RISs and opted for traditional optimization algorithms for their orchestration, which differs from the general formulation presented herein. Specifically, the authors of \cite{alhilo2021reconfigurable} dealt with the problem of scheduling mobile UEs. In \cite{MehdiDRL}, the RIS was positioned onto a \ac{UAV} which was proposed to be controlled by the Quantile Regression DRL algorithm \cite{QuantileRegression}, targeting coverage improvement and data-rate boosting. The altitude of the \ac{UAV} along with the RIS phase configuration to optimize the \ac{AoI} across \ac{IoT} devices was considered in \cite{Samir2021AgeOfInformation}. Diversely, the authors in \cite{Hu2021MetaSensing} proposed a DRL-based sensing system in a RIS-empowered wireless environment, for which the \ac{MDP} formulation involved sequentially selecting the phase shifts of the RIS to identify the intended objects. A \ac{PG}-based approach was designed which worked in conjunction with a supervised learning model that was used for object detection.
\par
It is worth mentioning that the choice between the most popular DRL-based agents that are utilized by the wireless communications community (i.e., DQN and DDPG), depends foremost on the nature of the RIS elements used in the underlying system models, since the aforementioned algorithms are designed specifically for either discrete or continuous action spaces, respectively.
A more detailed taxonomy of the proposed DRL methods is presented in Table~\ref{table:DRL_taxonomy}, in terms of their respective RL formulations.
The majority of the literature is focused on MISO systems in the presence of a single RIS and multiple UEs, whereas many works incorporate additional tasks to the action space of the DRL formulation (e.g., BS spatial processing and power allocation). Finally, we would like to note that many of the considered system models in the aforementioned previous art are characterized by \ac{IID} channel realizations (i.e., states of the environment) and available actions (namely, RIS phase configuration and BS precoding matrix selection), which do not influence future states of the system. Therefore, they adhere to the multi-armed bandits formalism firstly considered this paper, although they have been examined under the MDP-based framework. In Section~\ref{sect:Experiments} that follows including the numerical evaluations, we make use of the newly introduced formulation to assess the performance of a contextual bandits DRL agent, which will be shown to perform similar to the DQN algorithm in a number of scenarios, and with lower implementation complexity.

\section{Practical Considerations and Open Challenges}\label{sec:considerations}
Without intending to be exhaustive within the fast-evolving area of DRL-based orchestration for RIS-empowered smart wireless environments, we present in this section some key practical considerations and open challenges arising from our design formulation previously described in Section~\ref{sec:DRL_formulation}.

\subsection{Extensions with Practical Considerations}\label{sec:extensions}
In various practical cases, time-varying channels are more likely to appear instead of the considered, in Section~\ref{sec:DRL_formulation} as well as in the majority of the available literature, IID channel realizations. Time evolution re-instates the dependency of each state $\mathbf{s}_{t+1}$ on $\mathbf{s}_{t}$ (and potentially on more previous states) in the transition probability, implying that the underlying DRL-based algorithms must be able to store and be trained on transitions (i.e., tuples consisting of actions, observations, and rewards) of a greater length than $1$, as appearing in Algorithm~\ref{alg:RIS_DRL_formulation}.

In another respect, in the considered orchestration formulation, a centralized approach has been adopted, in which the controller is the sole agent, with the intention to facilitate all \acp{UE}. It is, however, conceivable for such a setup to be extended and studied in more realistic distributed manners, e.g.: \textit{i}) with the \acp{UE} being the agents, having potentially conflicting goals; and \textit{ii}) with multiple controllers, thus again multiple agents, each controlling a single RIS or a subset of the RISs or groups of meta-atoms in different RISs. The latter considerations will naturally lead to \ac{MAML} and game-theoretic formulations, which are out of the scope of this paper, although they constitute exciting practical considerations. In the latter case, the activation and assignment of RISs to controllers may be dynamic, yielding self-organizing schemes of spatially distributed RISs for meeting demanding design objectives \cite{rise6g_SRE}, like interference management/mitigation.

\subsection{Open Research Challenges}\label{sec:open-challenges}
Apart from the lack of theoretical guarantees on the performance of the presented (D)RL approaches in Section~\ref{subsec:RL}, there are plentiful of other open problems and challenges related to their application. Works described in the relevant literature overview in Section~\ref{sect:Literature} address some of the following issues, however, we list the ones we consider most important here to provide a synopsis of the potential research directions.

\subsubsection{Environmental Observations and Control Overhead}
The \ac{RL} formulations assume that the agent is capable of querying the environment about observations and reward signals. In wireless communication systems, this entails, on the one hand, the availability of sensing equipment and the adoption of relevant signal processing schemes in order to primarily measure the channel states as well as the desired performance indicators (e.g., communication rates). On the other hand, separate channels to and from the system controller need to be established for the control information exchange. Evidently, there are trade-offs between the costs associated with the quality and frequency of CSI estimation, as well as the achievable performance with the specific \ac{RL} implementation. Another aspect to consider is the channel occupancy due to the potential control and CSI observation exchange processes (in-band or out-of-band \cite{RISE6G_COMMAG,rise6g}), and the associated degradation in achievable performance. Up to date, in the vast majority of the DRL-based orchestration approaches for RIS-empowered smart wireless environments, it is implied that penalties induced by such system control operations are negligible. While this is a reasonable assumption for certain studies, the quantification of the such overheads on the overall performance is, to our view, a critical issue and presents an important research direction, especially in complex system setups, such as the multi-RIS setup considered in this paper.

\subsubsection{Simulations versus Real-World Deployment}
From a practical standpoint, the numerical evaluation of wireless communication schemes is easier to be performed via simulations of the signal propagation environment. It is, therefore, a long-standing question on whether the employed channel models and operation setups capture all important aspects of the underlying physical systems. This is especially true for the emerging field of \acp{RIS} and RIS-empowered smart wireless environments, for which the exact EM models are well less understood at the moment (see, e.g., \cite{PhysFad} and references therein). As a result, the transfer of policies learned in simulated environments to physical systems needs to be treated with care by the RIS community. Crucially, some important progress has been made in the \ac{DRL} area. For example, the domain adaptation problem has been recently dealt with for a robotic arm manipulation in \cite{openai2019Rubik}, where further randomness was imposed on the simulated environments.

\subsubsection{Combinatorially Large Action Spaces}
The RISs to be deployed in future wireless communication systems are expected to contain a large number of phase-tunable unit elements. The same is true about the number of the \ac{BS} antenna elements, and thus the dimension of the precoding vectors. Recall for our DRL formulation in Section~\ref{sec:DRL_formulation} that, when the controller of multiple RISs plays the role of the agent, the cardinality of the action space is exponential to the total number of the \ac{RIS} meta-atoms. Clearly, the problem of attaining optimal performance may become intractable for large enough RISs (or BS antenna elements), or may require considerable training periods.
To remedy for that, the deployed orchestration algorithms may need to resort to domain-specific optimization parts along with \ac{DRL} components, in order to reduce the exponential search space. Another algorithmic approach would be to consider \ac{DRL} algorithms that are specifically tailored to large action spaces (e.g., \cite{Tavakoli2018ActionBranching,DulacArnold2015DeepRL}), or actions consisting of binary vectors \cite{QNetsforBinaryActions} since most practical \ac{RIS} hardware design involve low phase resolution meta-atoms (e.g., $b=1$ and $2$ \cite{alexandg_2021}). In the very recent work \cite{Kyriakos_large_space_2022}, an RIS with $1$-bit phase resolution meta-atoms was modeled as a binary-element vector, resulting in an action space that is linear to the total number of RIS elements. Leveraging this modeling that enables individual tuning of each meta-atom, modified versions of DQN and DDPG were devised.

\subsubsection{RIS Operation Capabilities and Computational Requirements}
An important issue to consider in the envisioned RIS-empowered smart wireless environments is the physical location of the controller for the \acp{RIS}, and consequently, the part(s) of the network connected intelligence equipped with computational capabilities. Taking into consideration the demanding requirements of ANN training (i.e., \ac{GPU}), it is rather assumed that the required computation may be offloaded to a dedicated mobile edge computing server. This is, admittedly, the most general approach, though it requires establishing pertinent communication links, as discussed in Section~\ref{sec:DRL_formulation}. Conversely, combining the DRL-based intelligent controller of the \ac{RIS} with the actual RIS may greatly reduce the overhead of the control information exchange, by also taking advantage of the recently proposed hybrid \ac{RIS} hardware architectures with embedded sensing capabilities \cite{huang2019spawc,RIS_compressive_sensing,hardware2020icassp,chongwendrl,Tasos_DNN_CE_2019, RIS_DL_channel_estimation2,ma_Smart_2019,liaskos_ABSense_2019,ma_Smart_2020,HRIS_Mag,HRIS_Nature,HRIS_SPAWC}. In such configurations of smart wireless environments, far fewer control information exchanges for realizing DRL will be required, albeit with the added cost of the computationally and sensingly autonomous \ac{RIS}. Again, such overheads are not yet extensively studied, especially for multi-\ac{RIS} deployment scenarios.

\section{Numerical Results and Discussion}\label{sect:Experiments}
Having conducted computer simulations for the multi-RIS-enabled smart wireless environment included in Section~\ref{sect:RIS-environments}, we present in this section the performance evaluation of the proposed \ac{DRL} methodology, presented in detail in Section~\ref{sec:DRL_formulation}, which has as an orchestration objective the sum-rate maximization criterion expressed in \eqref{main_problem}. We have elaborated on a number of practical aspects for the design of the proposed (D)\ac{RL} algorithms as well as the benchmark schemes, including hyperparameter selection and performance evaluation strategies.

We also present a \textit{novel} multi-armed bandits methodology to solve the design problem at hand, leveraging the assumption of IID transition probabilities discussed in Section~\ref{sec:DRL_formulation} (due to the considered channel model which assumes independence of the channel realizations in time). Recall that, since the RIS phase profiles and the BS precoding matrix, which need to be designed at every coherent time step by the orchestration controller, affect only the current time frame, the online orchestration problem conceptually decomposes to individual sum-rate maximization problems per time step. This has the effect of making the optimal policy ``myopic,'' i.e., the greedy strategy which selects the RIS phase profiles and the BS precoder that maximize the sum rate in each current channel realization vector $\mathbf{s}_t$. Consequently, the design of the optimization problem can be reduced to predicting the sum rate/reward, given the current \ac{CSI} state $\mathbf{s}_t$, for all available actions. This outlook naturally casts the orchestration problem of the multi-RIS smart wireless environment to a contextual bandits setting, in which the goal is to obtain a model capable to capture the distribution of the reward signals. To solve this problem, we employ the \textit{Neural $\epsilon$-greedy} agent, which was briefly discussed in Section \ref{sec:Bandits} and presented in detail below.

\subsection{Proposed and Benchmark Methods}\label{sec:considered-methods}
\subsubsection{Neural $\epsilon$-Greedy}
The core of the Neural $\epsilon$-greedy algorithm is a reward-prediction ANN, represented by $\hat{G}_{\mathbf{w}}(\cdot)$ and parameterized by the network's weights $\mathbf{w}$, which is tasked to predict the expected sum-rate performance values for all RIS phase profiles and BS precoding matrices, given a \ac{CSI} observation. In particular, at each coherent time instant $t$, the CSI observation vector $\mathbf{s}_t \in \mathcal{S}$ is fed to the ANN, whose output $\hat{\mathbf{r}}_t\triangleq\hat{G}_{\mathbf{w}}(\mathbf{s}_t)$ is a real-valued vector with ${\rm card(\mathcal{A})}$ elements. The $i$-th element, with $i=1,2,\ldots,{\rm card(\mathcal{A})}$, of the latter vector signifies the estimation of the expected reward, if the $i$-th action of the action space $\mathcal{A}$ is selected. The $\epsilon$-greedy exploration strategy selects the action that corresponds to the maximum element of $\hat{\mathbf{r}}_t$ with probability $1-\epsilon$. 
\begin{figure}[t]
    \centering
    \includegraphics[width=\linewidth,trim={0 1.4cm 2.9cm 0},clip]{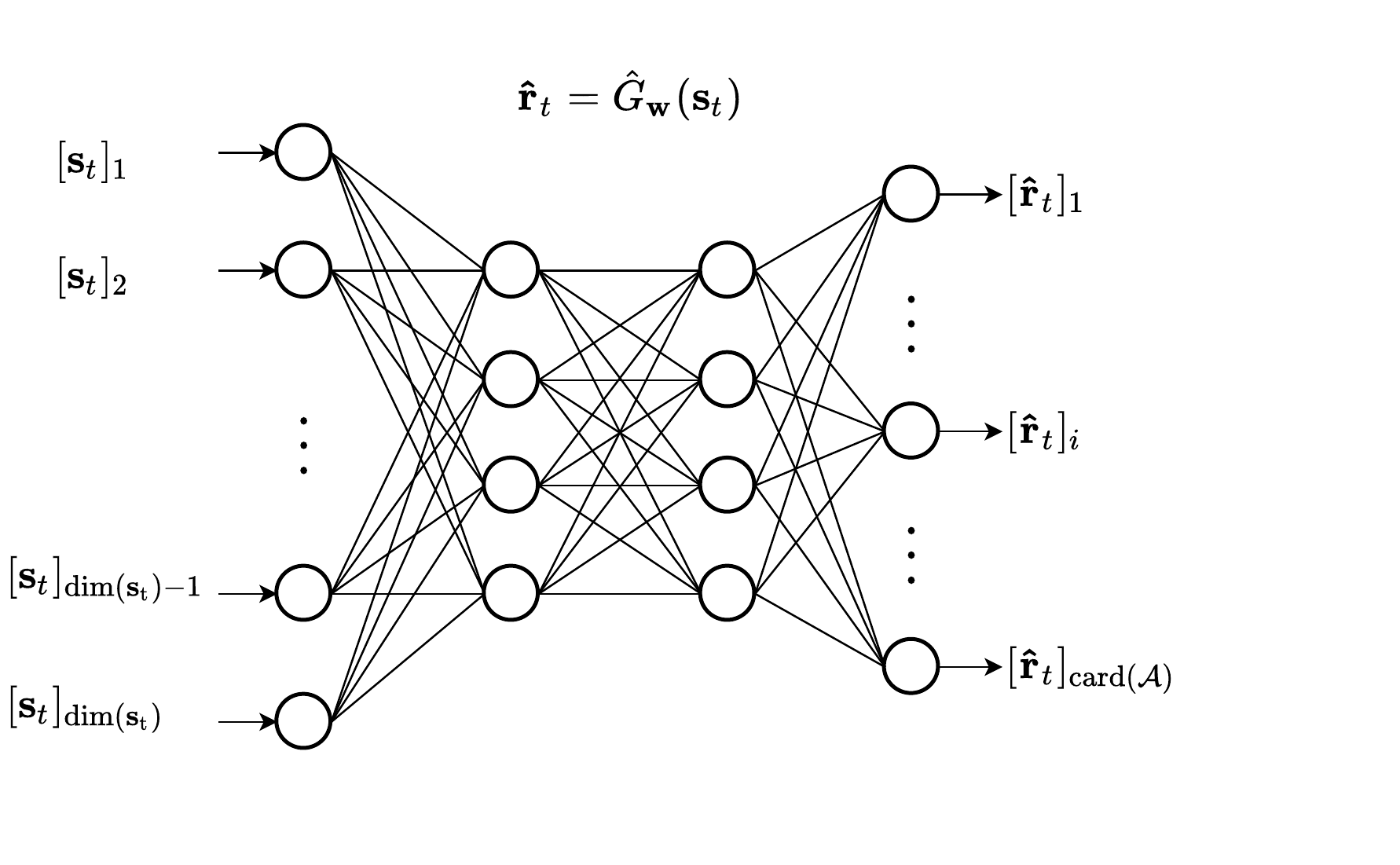}
    \caption{The reward-prediction ANN used by the Neural $\epsilon$-greedy algorithm. The network receives an observation $\mathbf{s}_t$ as input and outputs a vector of reward predictions $\hat{\mathbf{r}}_t$, the $i$-th element of which corresponds to an estimation of the expected reward if the action with index $i$ is selected.}
    \label{fig:reward-prediction-NN}
\end{figure}
\begin{table}[!t]
    \centering
    \caption{Hyperparameter Values for the Neural $\epsilon$-Greedy Algorithm.}
    \begin{tabular}{|l|c|}
        \hline
        $\epsilon$                         & $0.3$        \\
        \hline
        $t'$                               & $32$        \\
        \hline
        $\alpha$                           & $0.001$      \\
        \hline
        Convolutional layer units          & $64$, $64$   \\
        \hline
        Convolution kernel width           & $5$, $5$     \\
        \hline
        Hidden fully connected layer units & $32$, $32$   \\
        \hline
        Max-pooling patch width            & $4$, $4$     \\
        \hline
        Dropout probability                & $0.2$        \\
        \hline
    \end{tabular}
    \label{tab:NeG-hyper-parameters}
\end{table}

The proposed reward-prediction ANN for the orchestration controller (i.e., the agent in our DRL formulation) is illustrated in Fig$.$~\ref{fig:reward-prediction-NN} and its exact architecture is provided in Table~\ref{tab:NeG-hyper-parameters}. As shown, the ANN is a feedforward neural network with two convolutional layers (each preceding a max-pooling operation), followed by two fully connected layers with Rectified Linear Unit (ReLU) activation functions and a final output layer. The dropout technique was used for regularization after each layer apart from the last. The convolutional structure of the ANN is motivated by the fact that the dimensionality ${\rm dim(\mathbf{s}_t)}$ of the state vector is linear to the size and number of RISs, leading to high-dimensional input vectors. Furthermore, since adjacent elements in the channel vectors/matrices correspond to physical properties of adjacent antenna elements, spatial correlation \cite{Emil_RIS_Correlation,correlated_Weibull,correlated_Nakagami,alexandg_MRT} is expected to be present in the ANN's input state vector, which is appropriately treated using convolution and max-pooling operations. The values for the hyperparameters of the proposed Neural $\epsilon$-greedy bandits algorithm are given in Table~\ref{tab:NeG-hyper-parameters}. Their values have been set empirically, upon exploring a small amount of different combinations of them. Notice that the value of the probability $\epsilon$ of selecting a random action is uncharacteristically high. This is in light of the very stochastic nature of the wireless channel realizations, with the purpose of making the controller better explore the action space and to discourage it to converge early to potentially sub-optimal actions.

The training of the controller involves fitting the ANN on the most recent batch of collected experiences. Assuming a batch size of $t'$ coherent time steps, the controller spends them to explore the environment and store experience tuples of the form $(\mathbf{s}, \mathbf{a}, r)$ into a batch set $\mathcal{D}$. At the $t'$-th time step, the ANN is partially trained via single-step gradient descent to minimize the \ac{MSE} between the actual and the predicted rewards on the batch. Note, however, that although the network outputs a vector of predictions, only the true (scalar) reward (i.e., the one corresponding to the action ultimately selected) is made known to the controller. As a result, the residual differences of the \ac{MSE} calculation are taken only with respect to the output neuron that corresponds to the selected action. Concretely, the loss function of the ANN was designed as:
\begin{equation}\label{eq:loss-neural_e_greedy}
    \hat{\mathcal{L}}(\mathbf{w}) \triangleq \sum\limits_{(\mathbf{s}, \mathbf{a}, r) \in \mathcal{D} } \left( \left[\hat{G}_{\mathbf{w}}(\mathbf{s})\right]_{\mathbb{I}_{\mathcal{A}}(\mathbf{a})} - r \right)^2,
\end{equation}
where $\hat{\mathbf{r}} = \hat{G}_{\mathbf{w}}(\mathbf{s})$ is the ANN's prediction vector with the ${\rm card(\mathcal{A})}$ reward values, therefore, $[\hat{\mathbf{r}}]_{\mathbb{I_{\mathcal{A}}(\mathbf{a})}}$ is a scalar prediction of the reward for the action $\mathbf{s}$ actually taken. The proposed algorithmic steps for training the ANN-based reward-prediction controller are included in Algorithm~\ref{alg:neural_e_greedy}.
\newline

\begin{algorithm}[!t]
\caption{Neural $\epsilon$-Greedy Algorithm Solving $\mathcal{OP}_1$ in \eqref{main_problem}}\label{alg:neural_e_greedy}
\begin{algorithmic}[1]
\Require Probability $\epsilon$ of selecting a random action, reward-prediction ANN $\hat{G}_{\mathbf{w}}$ with initial parameters ${\mathbf{w}}$, learning rate $\alpha$, final time step $T$, and update interval $t'$.
\State Initialize $\mathcal{D} \gets \emptyset$.
\State Observe initial state $\mathbf{s}_1$ from the environment.
\For{$t=1,2,\dots,T$}
    \State Get the prediction $\hat{\mathbf{r}}_t = \hat{G}_{\mathbf{w}}(\mathbf{s}_t)$.
   \State $\mathbf{a}_t \gets \begin{cases}
   \mathcal{U}(\mathcal{A}) & \text{ with probability } \epsilon \\
  \argmax\limits_{\mathbf{a} \in \mathcal{A}}\{[\hat{\mathbf{r}}_t]_{\mathbb{I}_{\mathcal{A}}(\mathbf{a})}\} & \text{ with probability } 1-\epsilon \\
  \end{cases}
  $\;
  \State Feed $\mathbf{a}_t$ to the environment to observe $\mathbf{s}_{t+1}$ and $r_t$.
    \State Store $\mathcal{D} \gets \mathcal{D} \cup \{(\mathbf{s}_t, \mathbf{a}_t, r_t)\}$.  
    \If{${\rm mod}(t,t') = 0$}
        \State Compute $\hat{\mathcal{L}}(\mathbf{w})$ from \eqref{eq:loss-neural_e_greedy}.
        \State Perform $\mathbf{w'} \gets \mathbf{w} - \alpha \nabla_\mathbf{w} \hat{\mathcal{L}}(\mathbf{w})$.
        \State Update $\hat{G}_{\mathbf{w}} \gets \hat{G}_{\mathbf{w'}}$.
        \State Reset $\mathcal{D} \gets \emptyset$.
    \EndIf
\EndFor
\State \Return{Trained network $\hat{G}_{\mathbf{w}}$}.
\end{algorithmic}
\end{algorithm}

\subsubsection{DQN}
The performance of the proposed multi-armed bandits algorithm has been compared to the well-known \ac{DQN} algorithm, which was described in Section~\ref{sec:value-based-DRL} and serves as a representative MDP-based DRL approach. Note that, even though the formulation of Section~\ref{sec:DRL_formulation} lends itself to a contextual bandits application scenario, MDP-based approaches are still applicable, since bandits problems can be seen as special cases of \acp{MDP} with time horizon $T=1$. The intention behind such a comparison is to numerically investigate whether the added computational complexity, in terms of ANN requirements and hyperparameter tuning, brought by a state-of-the-art MDP-based \ac{DRL} algorithm offers any benefits over the more simplistic approach based on multi-armed bandits.

For a fair complexity comparison with the proposed DRL approach, we incorporated in \ac{DQN} a $Q$-network identical to the reward-prediction \ac{ANN} of the Neural $\epsilon$-greedy algorithm. It is noted here that DQN uses two independent copies of this network. The considered parameters for the \ac{DQN} algorithm are given in Table~\ref{tab:DQN-hyper-parameters}. The values of the parameters for both methods were determined by a hyper-optimization tuning algorithm over a set of possible values (specifically, Bayesian optimization \cite{NIPS_bayes_opt}), considering a simple scenario for computational purposes.

\begin{table}[!t]
    \centering
    \caption{Hyperparameter Values for the DQN Algorithm.}
    \begin{tabular}{|l|c|}
        \hline
        $\epsilon$                         & $0.3$        \\
        \hline
        $t'$                               & $1$        \\
        \hline
        $\alpha$                           & $0.002$      \\
        \hline
        Batch size                         & $128$           \\
        \hline
        Gradient clipping threshold        & $(-1000, 1000)$ \\
        \hline
        Target network update interval     & $100$ \\
        \hline
        Target network temperature $\hat{\tau}$  & $0.18$ \\ 
        \hline
        Convolutional layer units          & $64$, $64$   \\
        \hline
        Convolution kernel width           & $5$, $5$     \\
        \hline
        Hidden fully connected layer units & $32$, $32$   \\
        \hline
        Max-pooling patch width            & $4$, $4$     \\
        \hline
        Dropout probability                & $0.2$        \\
        \hline
    \end{tabular}
    \label{tab:DQN-hyper-parameters}
\end{table}

\subsubsection{Random and Optimal Schemes}
To ascertain the applicability of the proposed and benchmark DRL methods as intelligent orchestration controllers for the investigated RIS-empowered smart radio environments, we make use of the following baseline and upper-bound approaches: \textit{i}) The ``Random policy'' that chooses the RIS phase profiles and the BS precoders at random at every coherence time interval; and \textit{ii}) the ``Optimal policy'' which computes, at every time step, the best action from the available ones, through exhaustively searching among all available RIS phase profiles and BS precoding matrices for the ones that maximize the achievable sum-rate performance. 
\newline

\subsubsection{UCB}
Acquiring observations in the form of CSI estimations may be a demanding task, in terms of time and computational resources, in certain practical wireless communication systems, especially during the deployment phase of any (D)RL algorithm (in contrast to the training phase). For this reason, it is important to investigate whether the considered (D)RL-based approaches are capable of retaining a remarkable portion of their performance in application scenarios without the demanding availability of CSI observations; for example, using only the feedback of SINR measurements from the \acp{UE} for the reward computation. The proposed multi-armed bandits formulation is actually applicable for this limited-feedback case (i.e., $K$ real values instead of $N_{\rm tot}\left(K+N_{\rm T}\right)+KN_{\rm T}$ complex channel coefficients), as described in Section~\ref{sec:Bandits}. In fact, the SINR values directly yield the reward via \eqref{eq:reward} or \eqref{eq:reward-qos}, 
which implies that, for this simple RL formulation, only the individual rewards (i.e., rates, and consequently, SINRs) determine the decision-making process (i.e., the selected actions).
Note that, for the proposed Neural $\epsilon$-greedy algorithm, we have used the contextual version of multi-armed bandits, which requires both the CSI observations and the feedback of the individual rewards for learning the actions' selection policy. Similarly holds for the DQN algorithm.

For the purpose of comparing the proposed Neural $\epsilon$-greedy algorithm against an observation-free approach, we employ the simple \ac{UCB} heuristic strategy given by \eqref{eq:ucb}, which selects the action that has the highest upper bound estimate of the average gain associated with it. The running averages for each action are then computed via \eqref{eq:bandit-average}, and are initialized randomly to make the algorithm capable to explore the environment for the first few coherent time steps. The \ac{UCB} algorithm requires only a single parameter $c$ that controls the width of the confidence intervals, which was empirically set to $0.6$ for our experimentation process. It is noted that this parameter depends, in general, on various features of the problem (e.g., scale and variances of the reward values).

\subsection{Simulation Setup}\label{sec:simulation-setup}
In all figures that follow, we have considered a \ac{BS} with $N_{\rm T} = 4$ antennas, whose downlink communication with $K=2$ single-antenna \acp{UE} was enabled by $M=2$ identical \acp{RIS}, which were placed close to the \acp{UE}. The positions of the latter were sampled at random and kept fixed throughout the numerical evaluation process. To lay emphasis on the capability of \acp{RIS} to enable wireless communications, we considered the presence of a blocker that obstructed each \ac{BS}-UE$_k$ link, i.e., the respective coefficients of the direct channel vectors $\mathbf{h}_1$ and $\mathbf{h}_2$ were set to zero. The rest of the parameters used in the performance evaluations are included in Table~\ref{tab:setup-parameters}.

As discussed in Section~\ref{sect:RIS-environments}, the design objective of the proposed multi-RIS-empowered smart wireless environment is to maximize the sum-rate performance, i.e., solve $\mathcal{OP}_1$ in \eqref{main_problem}, by allowing the orchestration controller (i.e., the agent) to observe the channel matrices per coherent time interval, and accordingly, configure the profiles of the \acp{RIS} along with the selection of the BS precoding matrix. Cases with and without the constraint in \eqref{constraint:rate} for the individual rate requests have been investigated.    
We have considered a number of setups varying the total number of \ac{RIS} unit elements (i.e., meta-atoms), in order to primarily investigate the performance of our DRL approach as the action space increases. The simulated BS precoding codebook comprised $4$ unit-power beams, i.e., ${\rm card}(\mathcal{V})=4$, which correspond to the columns of the $2 \times 2$ \ac{DFT} matrix. We allocated the first (last) two columns of this matrix as the available precoding vectors for the first (second) UE. Similar to other works in the field (e.g., \cite{Huang_GLOBECOM_2019, Samarakoon_2020}), we have also made the assumption that the meta-atoms of each \ac{RIS} are controlled in groups of $N_{\rm group} = 16$, so that all elements of the group are set with the same configuration. We will be referring to the total number of meta-atoms that can be independently controlled as $N_{\rm control}$ throughout this section (i.e., $N_{\rm control} \triangleq N_{\rm tot} / N_{\rm group}$). It is noted that this implementation decision is expected to have a small adverse effect on the attained performance of all simulated methods, however, it vastly decreases the dimensionality of the combinatorial action space, rendering the ANN training, realized at the controller of the smart wireless environment, computationally tractable.
\begin{table}[t]
    \centering
    \caption{Parameters' Setting Used in the Simulation Results.}
    \begin{tabular}{|l|c|}
        \hline
        \ac{BS} coordinates (m)               & $(10,5,2)$  \\
        \hline
        RIS$_1$ coordinates (m)               & $(7.5,  13, 2)$    \\
        \hline
        RIS$_2$ coordinates (m)               & $(12.5, 13, 2)$    \\
        \hline
        \ac{UE}$_1$ coordinates (m)           & $(8.775, 14.394,  1.634)$    \\
        \hline
        \ac{UE}$_2$ coordinates (m)           & $(9.648, 13.281,  1.632)$    \\
        \hline
        $N_{\rm T}$                           & $4$ \\
        \hline
        $K$                                   & $2$   \\
        \hline
        ${\rm card}(\mathcal{V})$                     & $4$      \\
        \hline
        $\kappa_1$, $\kappa_2$                & $30$ dB \\
        \hline
        $P$                                   & $40$ dBm \\
        \hline
        $\sigma^2_k$ (equal for all \acp{UE}) & $-110$ dBm \\
        \hline
        $f$                                   & $5$ GHz \\
        \hline
        $N_{\rm group}$                       & $16$    \\
        \hline 
        $N_{\rm tot}$                         & $\{32, 48, 64, 80, 160\}$ \\
        \hline
        $N_{\rm control}$                     & $\{2, 4, 6, 8, 10 \}$ \\
        \hline
        ${\rm card}(\mathcal{A})$             & $\{16, 64, 256, 1024, 4096\}$\\
        \hline
        ${\rm dim}(\mathbf{s}_t)$             & $\{ 400, 784, 1168, 1552, 1936 \}$ \\
        \hline
    \end{tabular}
    \label{tab:setup-parameters}
\end{table}

For both our simulated \ac{DRL}-based orchestration controllers, namely the Neural $\epsilon$-greedy and DQN algorithms, we have used a training period (i.e., number of IID channel realizations, each corresponding to a coherent time step) equal to the number of actions ${\rm card(\mathcal{A})}$ multiplied by $50$ in each trial. Their performance was then assessed by computing the average sum rate over $300$ time steps, during which the controller acted with its deterministically greedy learned policy (i.e., without selecting a random action with probability $\epsilon$). To avoid extreme initializations for the ANN parameters of the DRL approaches (see Algorithm~\ref{alg:neural_e_greedy}) with the above evaluation strategy, we have performed averaging over $5$ different trials. Note that depending on the intended application, the training process might continue indefinitely, in which case, the controller needs to be evaluated at certain intervals using its exploration policy. For the baseline random and the optimal schemes, we have computed the average sum rates at the beginning of each trial using $300$ IID channel realizations.

\subsection{Sum-Rate Results versus the Size of the Action Space}
As discussed in Section~\ref{sec:DRL_formulation}, and particularly from the definition of the action space vector in \eqref{eq:action_space_vector}, the action space increases exponentially with the number of the phase-tunable RIS meta-atoms. This implies that the number of observations required for the DRL-based approaches to associate observations to approximately optimal actions will ultimately lead to intractable training times. Consider a toy example, using the parameters in Table~\ref{tab:setup-parameters}, with only one group of meta-atoms per RIS, i.e., $N_{\rm tot}=32$ elements, and $16$ different actions, taking also the selection of the BS precoding matrix under consideration. Adding one phased-controllable group in each RIS, up to a total of $10$ groups, leads to $160$ meta-atoms and an action space including $4096$ choices. Obviously, the action space can be extensively large even with RISs having meta-atoms with the lowest possible phase resolution $b=1$.

\begin{figure}[t]
    \centering
    \includegraphics[width=\linewidth]{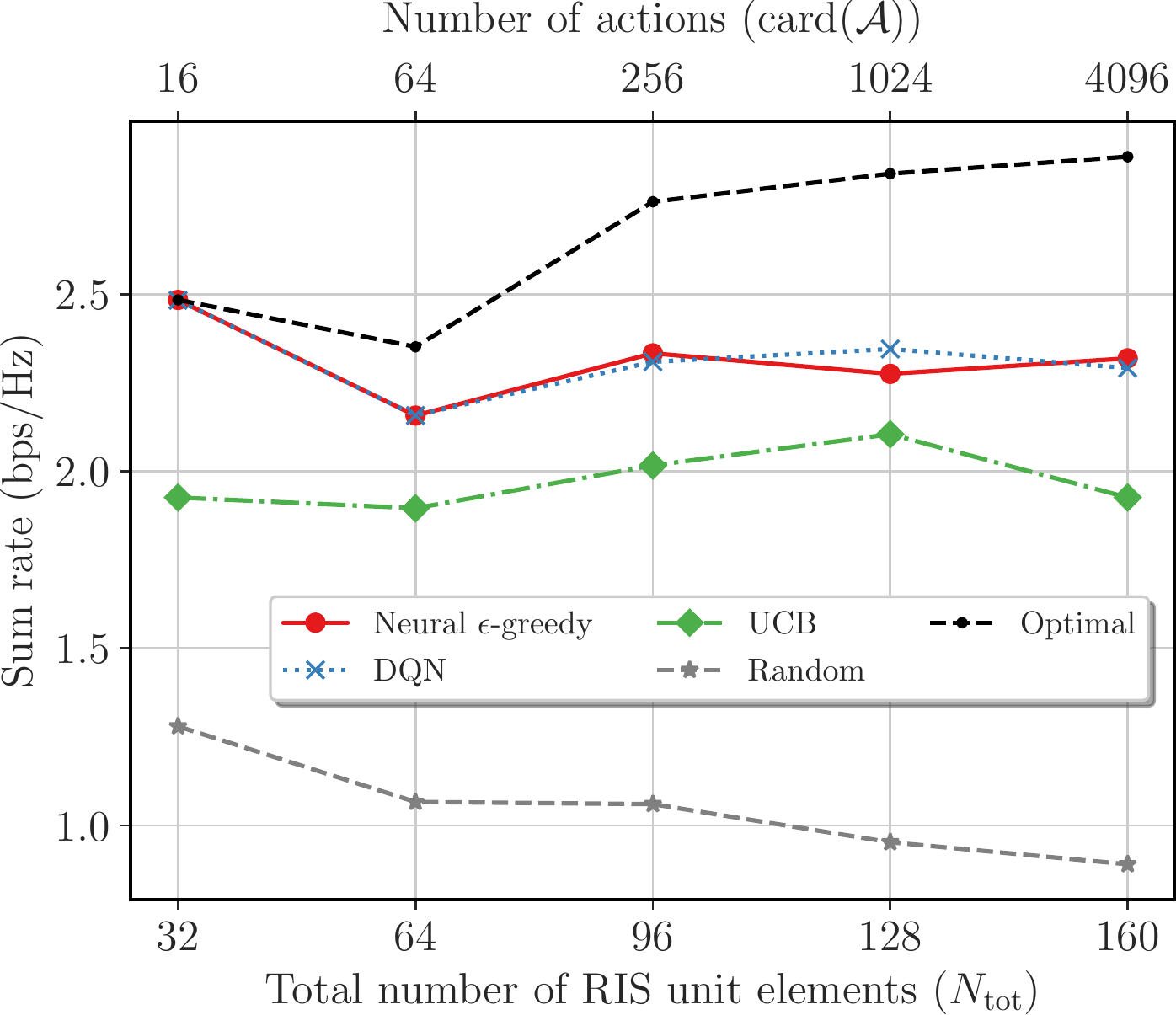}
    \caption{Average sum-rate performance in bps/Hz versus the total number of RIS elements, $N_{\rm tot}$, for each of the $M=2$ identical RISs and the size of the action space, ${\rm card}(\mathcal{A})$, achievable by the Neural $\epsilon$-greedy, DQN, and UCB algorithms. The sum-rate curves of the random selection and optimal schemes are also demonstrated.}
    \label{fig:bandit-performance-lineplot}
\end{figure}
\begin{figure}[t]
    \centering
    \includegraphics[width=\linewidth]{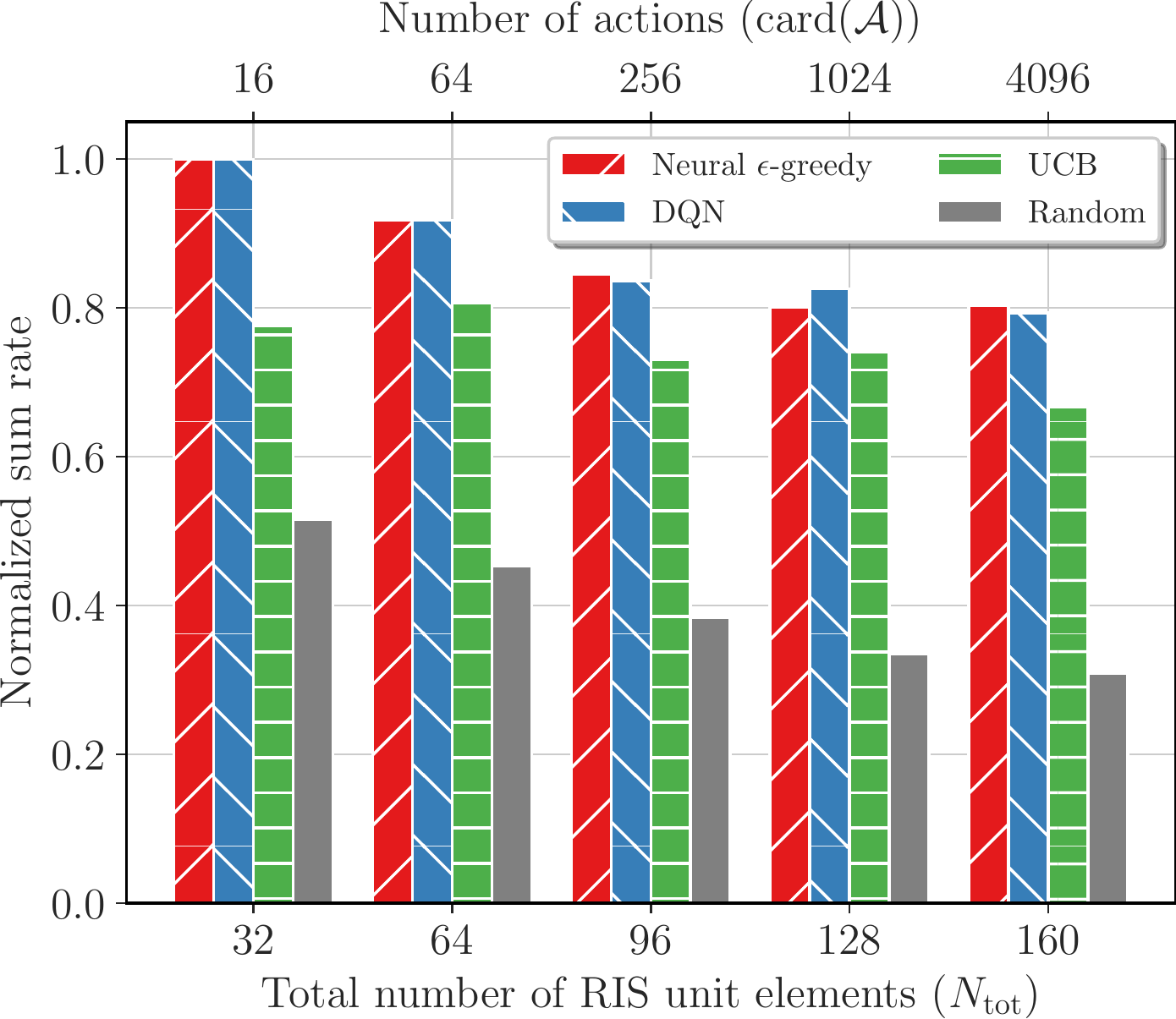}
    \caption{The average sum rates of the Neural $\epsilon$-greedy, DQN, and UCB algorithms, as well as of the random selection scheme illustrated in Fig.~\ref{fig:bandit-performance-lineplot} when normalized over the sum-rate performance of the optimal scheme.}
    \label{fig:bandit-performance-barplot}
\end{figure}

The achievable average sum rates for the simulation setup described in Table~\ref{tab:setup-parameters}, using all considered orchestration methods for solving $\mathcal{OP}_1$ without the constraint in \eqref{constraint:rate} for the individual rate requests, are depicted in Fig$.$~\ref{fig:bandit-performance-lineplot} as functions of the number $N_{\rm tot}$ of the meta-atoms for each RIS and the size ${\rm card}(\mathcal{A})$ of the action space. The same results are also included in Fig$.$~\ref{fig:bandit-performance-barplot} in a normalized manner, by dividing each method's sum rate by the rate of the optimal scheme for all considered $N_{\rm tot}$ and ${\rm card}(\mathcal{A})$ values. As observed in both figures, all \ac{RL} methods far outperform the random selection policy for all considered cases. It is also evident that, both \ac{DRL} methods are capable of maintaining sum rates higher than about $80\%$ of the optimal rates in all cases, while being able to converge to the optimal RIS phase configurations in the simplest case of $32$ meta-atoms per RIS. In fact, the proposed Neural $\epsilon$-greedy method and \ac{DQN} exhibit identical performance, confirming our hypothesis: in smart wireless environments with IID channel realizations, bandit algorithms are akin to state-of-the-art MDP-based \ac{DRL} approaches. Interestingly enough, the performance of our simple UCB method is capable to consistently outperform the baseline random selection policy, while being reasonably close to both \ac{DRL} algorithms, even without relying on CSI observations. Nevertheless, a decreasing trend in the relative performance (w.r.t. the optimal sum rates) can be inferred from Fig$.$~\ref{fig:bandit-performance-barplot} with increasing values for $N_{\rm tot}$ and ${\rm card}(\mathcal{A})$, which we expect to continue in cases where the sizes of the RISs increase.
\begin{figure}[t]
    \centering
    \includegraphics[width=\linewidth]{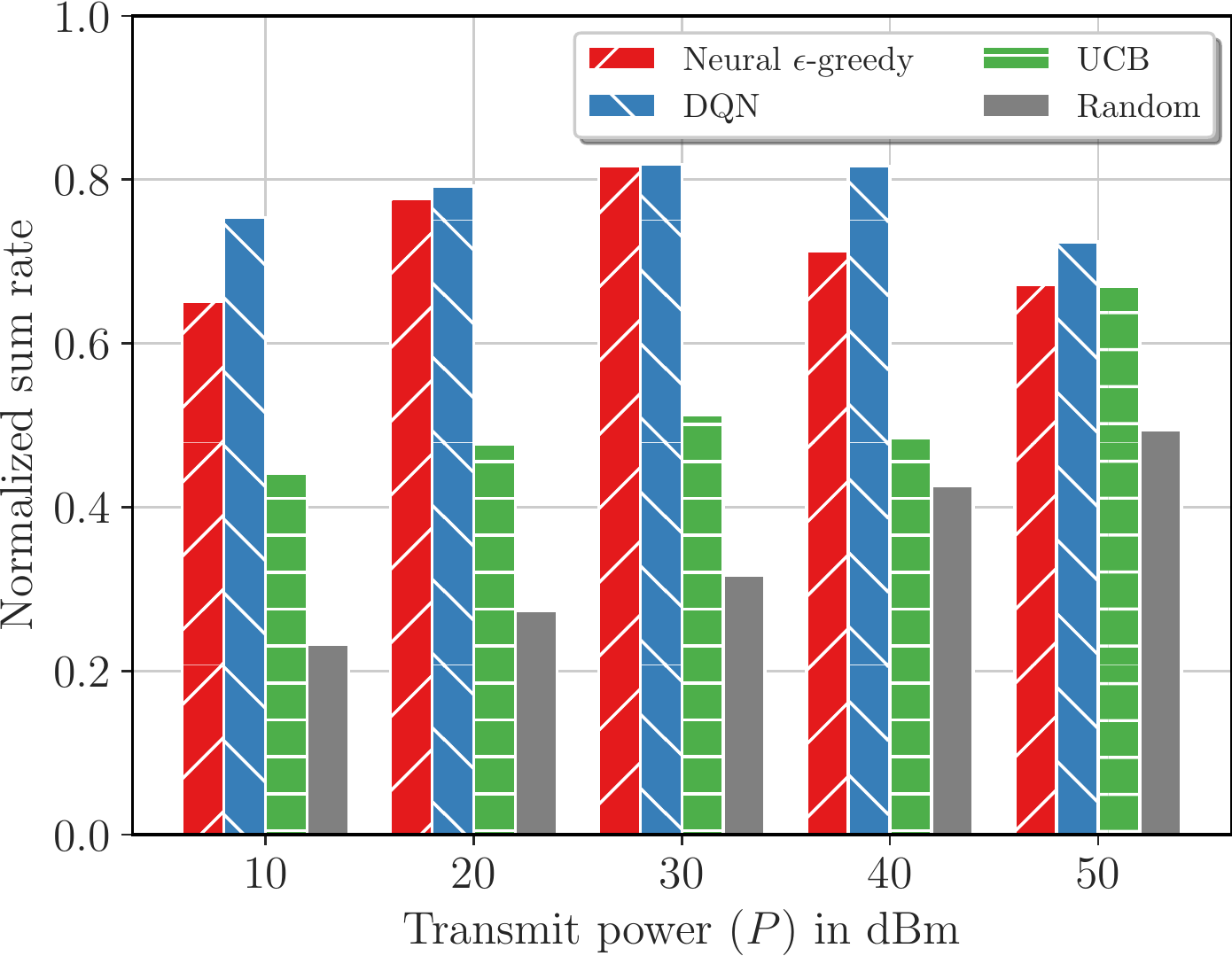}
    \caption{Average sum-rate performance in bps/Hz versus the transmit power $P$ in dBm for the Neural $\epsilon$-greedy, DQN, and UCB algorithms, as well as the random selection scheme, normalized over the sum-rate performance of the optimal scheme. We have considered $M=2$ identical RISs with each having $N_{\rm tot} = 32$ meta-atoms, resulting in an action space of size ${\rm card}(\mathcal{A})=64$.}
    \label{fig:rate-vs-P}
\end{figure}

\subsection{Sum-Rate Results versus the Transmit Power}
The observation and action spaces, defined in our DRL formulation in Section~\ref{sec:DRL_formulation}, are independent of the choice of the BS transmit power budget $P$. However, it is apparent from the SINR expression in \eqref{eq:SINR} that the transmit power affects the reward values (see \eqref{eq:reward} and \eqref{eq:reward-qos}) collected by the controller. Similar to Fig$.$~\ref{fig:bandit-performance-barplot}, in Fig.~\ref{fig:rate-vs-P}, we demonstrate the normalized achievable sum-rate performance versus different $P$ values and $N_{\rm tot}={\rm card}(\mathcal{A})=64$ for the Neural $\epsilon$-greedy, DQN, and UCB approaches, as well as the random selection scheme. It is depicted that the considered \ac{DRL} algorithms (i.e., the proposed Neural $\epsilon$-greedy and DQN) perform similarly, achieving around $75$\% of the optimum performance, nearly irrespective of the tested values for $P$. It is also shown that the performance of our UCB algorithm improves with increasing $P$, ranging from around $45$\% of the optimum sum-rate at $P=10$ dBm to around $70$\% when $P=50$ dBm. Interestingly, for the latter large $P$ value, our simple RL method attains almost equal sum rate with both considered DRL methods. This implies that the average values of favorable actions become more easily discriminated from the expected rates of the on-average unfavorable ones. Finally, it can be observed that all considered (D)RL algorithms outperform the random selection scheme, whose normalized sum rate depends on the transmit power $P$, but is always below the $45$\% of the optimum achievable sum-rate performance.

\subsection{Reward Results versus the Size of the Action Space}\label{sec:QoS-experiments}
\begin{figure}[t]
    \centering
    \includegraphics[width=\linewidth]{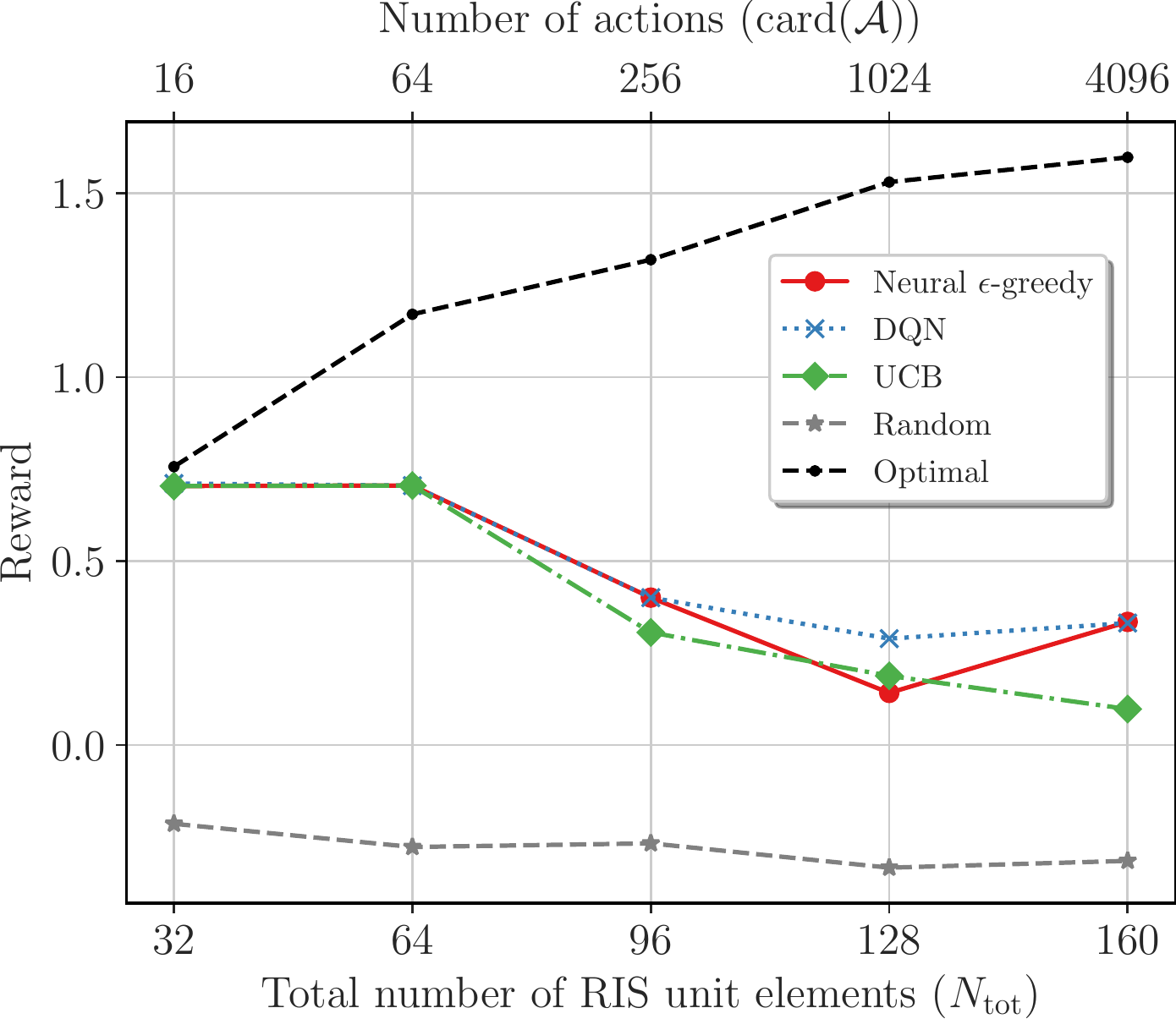}
    \caption{Average rewards using \eqref{eq:reward-qos} for all proposed and benchmark orchestration methods versus the total number of RIS meta-atoms, $N_{\rm tot}$, for each of the $M=2$ identical RISs and the size of the action space, ${\rm card}(\mathcal{A})$.}
    \label{fig:QoS-vs-N}
\end{figure}

By solving $\mathcal{OP}_1$ with the constraint in \eqref{constraint:rate} for the per UE rate requests, Fig$.$~\ref{fig:QoS-vs-N} illustrates the average reward values obtained using~\eqref{eq:reward-qos} as functions of the $N_{\rm tot}$ for each RIS and ${\rm card}(\mathcal{A})$, similar to Fig$.$~\ref{fig:bandit-performance-lineplot}. In this figure, we used the same simulation parameters with Figs$.$~\ref{fig:bandit-performance-lineplot} and~\ref{fig:bandit-performance-barplot} and set both $R^{\rm req}_1$, and $R^{\rm req}_2$ appearing in \eqref{constraint:rate} equal to $0.4$ bps/Hz; this rate value corresponds to about half the optimal achievable rate when $N_{\rm tot}=32$. As clearly shown, the achieved reward with the random selection policy falls below zero for all investigated cases, indicating that the selected combinations of RIS phase profiles and BS precoders do not meet the UEs' rate requests. It is also demonstrated that the optimal policy attains lower average rewards than the achievable sum rates in Fig$.$~\ref{fig:bandit-performance-lineplot}, which witnesses that, at certain channel realizations, the UE rate requests cannot be satisfied. For this problem formulation including the constraints \eqref{constraint:rate}, it is interestingly shown in the figure that the performance of the proposed UCB method is at the same levels with that of the DRL methods, i.e., the proposed Neural $\epsilon$-greedy algorithm and DQN. We attribute this to the fact that the scale of the reward values is now larger, due to the introduction of negative values in \eqref{eq:reward-qos}, thus facilitating the detection of actions of poor quality. More importantly, a decreasing trend is apparent on the performance of the considered methods, signifying that the respective orchestration mechanisms are less efficient than those in Fig$.$~\ref{fig:bandit-performance-lineplot} in learning the environment. It can be finally concluded from Fig$.$~\ref{fig:QoS-vs-N}, due to the positive average reward values, that all methods, expect the baseline random scheme, are successful in satisfying the per UE rate constraints.

\subsection{Orchestration with Partial CSI Observability}\label{sec:results-partial-observability}
The observation of the evolving states of the wireless environment from the orchestration controller, via the collection of the CSI vectors in \eqref{eq:DRL-state}, is one of the key practical challenges with the proposed DRL formulation, as presented in Section~\ref{sec:open-challenges}. The role of the controller is to associate the CSI observations to desirable RIS phase configurations.
In practice, the associated computational cost and latency induced by the exchange of pilot signals for reliable CSI estimation are important considerations in deploying the proposed multi-RIS-empowered smart wireless environment. Therefore, it is essential for the envisioned DRL-based controller to rely on more efficient types of environmental observations.

In this subsection, we compare the performance of the considered orchestration methods between the original full CSI availability case and another case of partial CSI observability, according to which the wireless propagation environment includes strong LOS components and the controller possesses the \acp{AoD} between each RIS and the UEs. The \acp{AoD} between the BS and each RIS are also assumed known due to the known placement of the RISs and the BS position.
In the context of the latter partial CSI case, we do not deal with the problem of accurately estimating the angular channel parameters, instead we assume computationally and sensingly autonomous RISs (e.g., \cite{hardware2020icassp,liaskos_ABSense_2019,HRIS_Nature,locrxris_all}).

Up to this point (see Section~\ref{sec:simulation-setup} with the simulation setup), static \acp{UE} with fixed positions have been considered. To properly evaluate the AoD-based partial CSI observability case, we consider the following mobility scenario: two UEs are moving with a standard walking speed of $1.4$ ${\rm m/s}$ in straight-line trajectories, angled at $45 \degree$ and $-45 \degree$, respectively, with respect to the $x$ axis. When their displacement from their starting positions exceeds $2 {\rm m}$, they turn around and move to the opposite directions. The channel coherence time in this setup is approximately $6$ ${\rm ms}$, which constitutes one time step in our DRL formulation. The rest of the simulation parameters and the algorithmic settings remain as described in the previous subsections, with the following exception regarding the structure of the ANNs employed by the DRL algorithms. Since, the dimension of the observation space is small enough, the two convolutional and max-pooling layers are not needed. Instead, they are all replaced by one fully connected layer of $32$ units. Since the sum rates depended on the positions of the UEs, we evaluated the considered algorithms (including the benchmark schemes) at $15$ evenly-spaced intervals during the training process, each consisted of $300$ time steps (i.e., CSI realizations), and we considered the average achievable sum-rate performance across all intervals.

\begin{figure}[t]
    \centering
    \subfloat[$N_{\rm tot} =64$\label{fig:rates-vs-observation-type-4}]{%
      \includegraphics[width=0.45\textwidth]{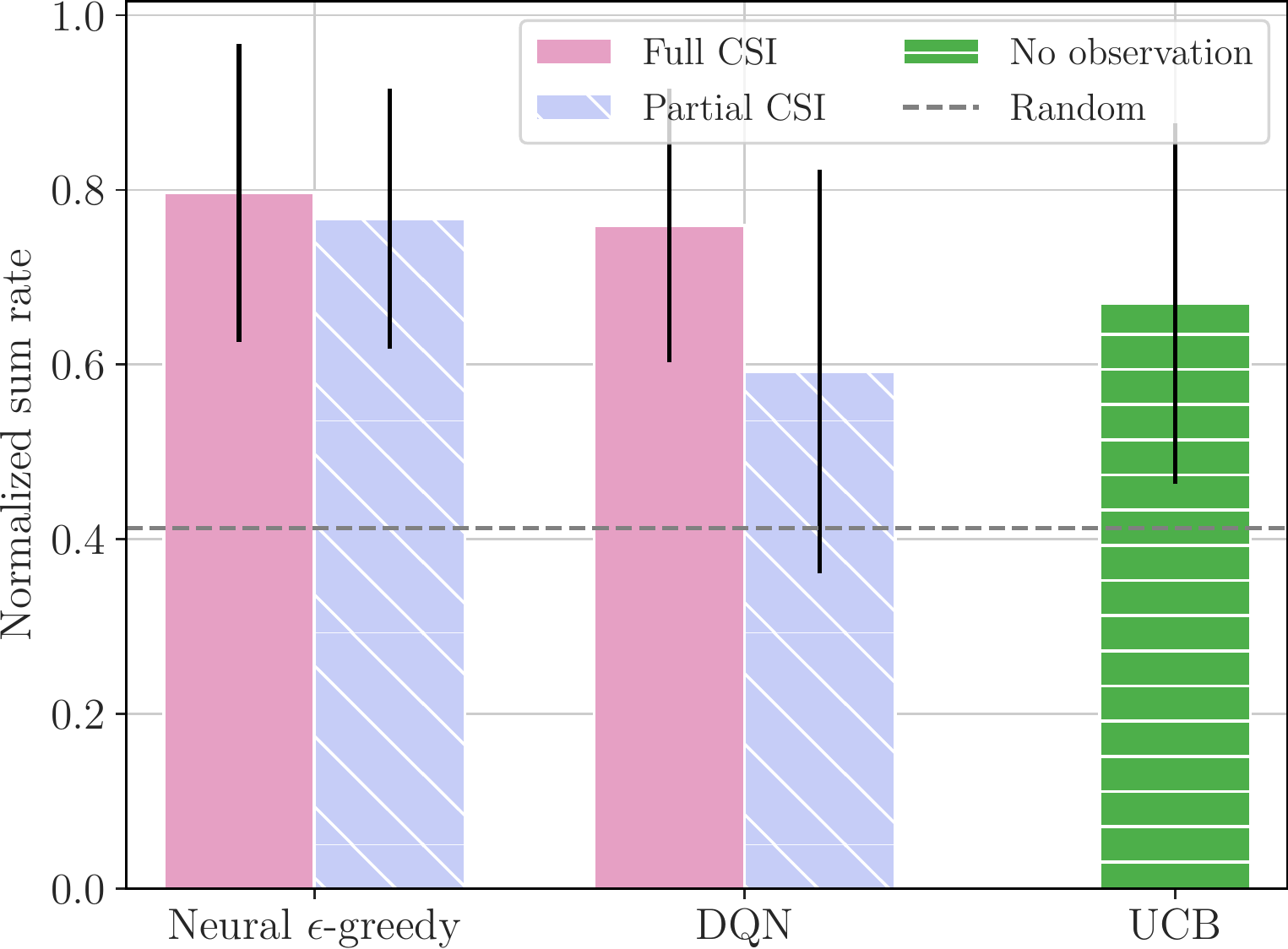}
    }
    \hfill
    \subfloat[$N_{\rm tot} =128$\label{fig:rates-vs-observation-type-8}]{%
      \includegraphics[width=0.45\textwidth]{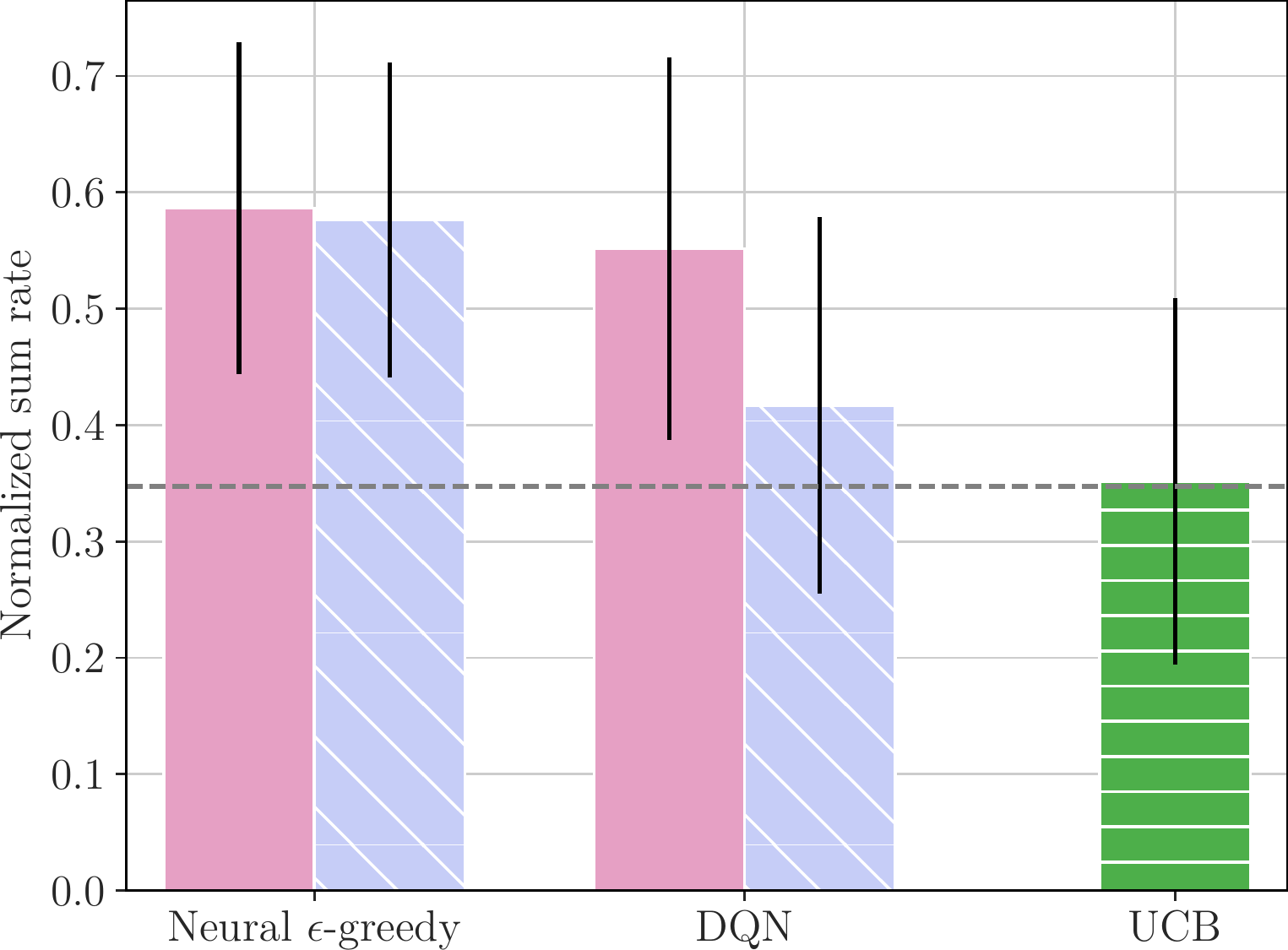}
    }
    \caption{The average sum rates of the Neural $\epsilon$-greedy, DQN, and UCB algorithms, as well as of the random selection scheme, when normalized versus the sum-rate performance of the optimal scheme. The DRL algorithms have been evaluated for both the full and partial CSI observation cases, considering a scenario with two mobile UEs and two values for the total number of RIS meta-atoms, $N_{\rm tot}$. The vertical error bars in the figure indicate the respective standard deviations.}
    \label{fig:rates-vs-observation-type}
\end{figure}

The comparison of the sum rates between the full and partial CSI observation cases, when considering the DRL methods Neural $\epsilon$-greedy and DQN, are illustrated in Fig$.$~\ref{fig:rates-vs-observation-type} for the setups with $N_{\rm tot}=\{64,128\}$ total numbers of RIS unit elements. The performance of the proposed UCB method is also included, which constitutes an RL algorithm that does not rely on any type of environmental observation. More specifically, the individual rewards serve as feedback on the selected actions. As depicted in the figure, the normalized performances of the proposed Neural $\epsilon$-greedy and the DQN algorithms are similar when CSI observations are involved. Interestingly, the former is also shown to perform equally well when the observations are consisted of AoDs. Conversely, DQN exhibits a noticeable decrease in performance when using partial, instead of full CSI, observations. It is finally shown that the normalized performance of the UCB method decreases more rapidly, compared to the DRL methods, as $N_{\rm tot}$ increases.

\subsection{Methods' Execution Time Comparison}
To complement the asymptotic time complexities of the DRL algorithms discussed in Section~\ref{sebsec:DRL-evaluation}, we herein quantify the execution time of each considered method in the simulation setup presented in Section~\ref{sec:simulation-setup}. For this evaluation, we have used a desktop computer with an Intel i5-8400 processor, $16$ GB of RAM, and an NVidia GTX-1060 with a $6$ GB VRAM, and the DRL algorithms were implemented in Tensorflow\footnote{The code for the performance results presented in this paper is available at: \url{https://github.com/NoesysLab/DRL_RIS_Tutorial}.}. The mean execution time in time steps per second for all simulated methods is given in Table~\ref{tab:execution-time}. As it can be concluded from the comparison of the two DRL algorithms, DQN is slower than Neural $\epsilon$-greedy, since the former adopts an additional (second) neural network. However, both algorithms exhibit performance metrics of the same order of magnitude. It is also shown that the UCB method is considerably faster, rivaling the performance of the baseline random selection scheme. Evidently, the optimal scheme (i.e., exhaustive search) runs faster than both DRL methods in the setups with the smaller action spaces, although it becomes prohibitively expensive as the sizes of the RISs increase. A further interesting result that can be inferred from the table is that the execution time of both DRL methods does not increase significantly for large numbers $N_{\rm tot}$ of the RISs' meta-atoms. This behavior is justified by the parallel computing capabilities of the utilized GPU.
\begin{table}[t]
    \centering
        \caption{Execution Time of the Considered Methods in Time Steps per Second.}
    \label{tab:execution-time}
    \begin{tabular}{|c|c|c|c|c|c|}
    \hline
$N_{{\rm tot}}$ &   Neural $\epsilon$-greedy & DQN & UCB & Optimal & Random \\ \hline
32 &   $72$  &   $45$  &  $2603$   &    $493$  &   $3012$  \\       \hline
64 &   $61$  &   $33$  &  $2100$   &   $112$   &   $2489$   \\ \hline
96 &   $58$  &   $31$  &  $1532$   &   $28$    &   $1744$   \\ \hline
128 &   $55$  &   $25$  &  $726$    &    $6$    &   $661$   \\ \hline
160 &  $48$  &   $24$  &  $651$    &    $1$    &   $188$    \\ \hline
    \end{tabular}
\end{table}
\section{Conclusions}\label{sect:Conclusion}

In this paper, focusing on the emerging RIS technology for programmable propagation of information-bearing signals in the era of beyond 5G wireless communications, we studied the dynamic orchestration of the promising, but challenging, multi-RIS-empowered smart wireless environments by means of state-of-the-art \ac{DRL} methods. Our comprehensive treatment included a description of a generic model for such communication systems, which encompasses various crucial aspects of reconfigurable radio wave propagation. A thorough theoretical introduction of the \ac{RL} area was provided to illuminate the basic principles, categorization, and applicability scenarios of the most prominent \ac{DRL} algorithms, with emphasis on those deployed in the wireless communications field. We then described the exemplar sum-rate maximization problem via a quite general \ac{DRL} formulation, that aligns itself with the existing literature. Relevant works were discussed and taxonomized, in order to give a clear overview of the adoption of the \ac{DRL} approaches designed by the community working on \acp{RIS} and RIS-empowered smart wireless environments. The literature overview has been accompanied by a discussion on fundamental challenges, key opportunities, and important future directions of this evolving research area.

Apart from the survey intention of this paper, our technical contribution lied in the alternative treatment of the online sum-rate maximization objective as a multi-armed bandits problem, instead of the MDP-inspired DRL formulations typically found in the field. The proposed framework leveraged the fact that, at any time instant, the design parameters of the multi-RIS-enabled smart wireless environment (i.e., the RIS phase profiles and the BS precoding matrix) have no effect on the state of the environment (i.e., the channel gain matrices) of the next time instant. Our extensive numerical evaluations showcased that the proposed multi-armed bandits orchestration controller is capable of attaining increased performance in a number of setups. Interestingly, this achievable performance is similar to DQN-based orchestration, but with simpler implementation, and consequently, faster execution time. We also presented a simple UCB strategy, selecting the action that has the highest upper bound estimate of the average gain associated with it, which did not require CSI observability, but relied on the feedback of SINR measurements for the reward computation. It was demonstrated that this simple orchestration scheme performs sufficiently close to the considered DRL algorithms in certain scenarios, while always outperforming the baseline random selection scheme.


\section*{Acknowledgements}
This work has been supported by the EU H2020 RISE-6G project under grant number 101017011. The work of Prof. Yuen was supported by the Singapore Ministry of Education Tier 2 MOE-000168-01. The work of Prof. Huang was supported by the China National Key R\&D Program under grant 2021YFA1000500, the National Natural Science Foundation of China under grant 62101492, the Zhejiang Provincial Natural Science Foundation of China under grant LR22F010002, the National Natural Science Fund for Excellent Young Scientists Fund Program (Overseas), the Ng Teng Fong Charitable Foundation in the form of ZJU-SUTD IDEA Grant, the Zhejiang University Education Foundation Qizhen Scholar Foundation, and the Fundamental Research Funds for the Central Universities under grant 2021FZZX001-21.

\bibliographystyle{IEEEtran}
\bibliography{used_references}

\begin{thebibliography}{100}
\providecommand{\url}[1]{#1}
\csname url@samestyle\endcsname
\providecommand{\newblock}{\relax}
\providecommand{\bibinfo}[2]{#2}
\providecommand{\BIBentrySTDinterwordspacing}{\spaceskip=0pt\relax}
\providecommand{\BIBentryALTinterwordstretchfactor}{4}
\providecommand{\BIBentryALTinterwordspacing}{\spaceskip=\fontdimen2\font plus
\BIBentryALTinterwordstretchfactor\fontdimen3\font minus
  \fontdimen4\font\relax}
\providecommand{\BIBforeignlanguage}[2]{{%
\expandafter\ifx\csname l@#1\endcsname\relax
\typeout{** WARNING: IEEEtran.bst: No hyphenation pattern has been}%
\typeout{** loaded for the language `#1'. Using the pattern for}%
\typeout{** the default language instead.}%
\else
\language=\csname l@#1\endcsname
\fi
#2}}
\providecommand{\BIBdecl}{\relax}
\BIBdecl

\bibitem{5GAmericas}
``{3GPP} releases 16 \& 17 \& beyond,'' White Paper, 5G Americas, Jan. 2021.

\bibitem{Shafi_5G_all}
M.~Shafi, A.~F. Molisch, P.~J. Smith, T.~Haustein, P.~Zhu, P.~D. Silva,
  F.~Tufvesson, A.~Benjebbour, and G.~Wunder, ``{5G: A} tutorial overview of
  standards, trials, challenges, deployment, and practice,'' \emph{{IEEE} {J}.
  {S}el. {A}reas {C}ommun.}, vol.~35, no.~6, pp. 1201--1221, Jun. 2017.

\bibitem{3GPP_R18}
\BIBentryALTinterwordspacing
3GPP. (2022) Release 18. [Online]. Available:
  \url{https://www.3gpp.org/release18}
\BIBentrySTDinterwordspacing

\bibitem{Samsung}
``The next hyper- connected experience for all,'' White Paper, Samsung 6G
  Vision, Jun. 2020.

\bibitem{Saad_6G_2020}
W.~Saad, M.~Bennis, and M.~Chen, ``A vision of {6G} wireless systems:
  {A}pplications, trends, technologies, and open research problems,''
  \emph{IEEE Network}, vol.~34, no.~3, pp. 134--142, Jun. 2020.

\bibitem{Emilio_6G}
E.~Calvanese~Strinati and S.~Barbarossa, ``{6G} networks: {B}eyond {S}hannon
  towards semantic and goal-oriented communications,'' \emph{Elsevier Comput.
  Netw.}, vol. 190, pp. 1--17, May 2021.

\bibitem{Cisco_Report}
``Cisco annual internet report (2018–2023),'' White Paper, Cisco, Mar. 2020.

\bibitem{Ericsson_AI_5G}
``Accelerating the adoption of {AI} in programmable {5G} networks,'' White
  Paper, Ericsson, Jul. 2021.

\bibitem{5GPPP_AIML}
``{AI} and {ML} – {E}nablers for beyond {5G} networks,'' White Paper, 5G PPP
  Technology Board, May 2021.

\bibitem{Akyildiz_6G_2020}
I.~F. Akyildiz, A.~Kak, and S.~Nie, ``{6G} and beyond: {T}he future of wireless
  communications systems,'' \emph{IEEE Access}, vol.~8, pp. 133\,995--134\,030,
  Jul. 2020.

\bibitem{Mokh_1}
A.~Mokh, J.~de~Rosny, G.~C. Alexandropoulos, R.~Khayatzadeh, M.~Kamoun,
  A.~Ourir, A.~Tourin, and M.~Fink, ``Time reversal for multiple access and
  mobility: {A}lgorithmic design and experimental results,'' in \emph{Proc.
  {IEEE WCNC}}, Austin, USA, Apr. 2022.

\bibitem{Mokh_2}
A.~Mokh, J.~de~Rosny, G.~C. Alexandropoulos, R.~Khayatzadeh, A.~Ourir,
  M.~Kamoun, A.~Tourin, and M.~Fink, ``Time reversal precoding at {subTHz}
  frequencies: {E}xperimental results on spatiotemporal focusing,'' in
  \emph{Proc. {IEEE CSCN}}, Thessaloniki, Greece, Dec. 2021.

\bibitem{TR_Magazine}
G.~C. Alexandropoulos, A.~Mokh, R.~Khayatzadeh, J.~de~Rosny, M.~Kamoun,
  A.~Ourir, A.~Tourin, M.~Fink, and M.~Debbah, ``Time reversal for {6G}
  wireless communications: {N}ovel experiments, opportunities, and
  challenges,'' \emph{IEEE Veh. Technol. Mag.}, under revision, 2022.

\bibitem{liaskos2018new}
C.~Liaskos, S.~Nie, A.~Tsioliaridou, A.~Pitsillides, S.~Ioannidis, and I.~F.
  Akyildiz, ``A new wireless communication paradigm through software-controlled
  metasurfaces,'' \emph{IEEE Commun. Mag.}, vol.~56, no.~9, pp. 162--169, Sep.
  2018.

\bibitem{Marco2019}
M.~Di~Renzo, M.~Debbah, D.-T. Phan-Huy, A.~Zappone, M.-S. Alouini, C.~Yuen,
  V.~Sciancalepore, G.~C. Alexandropoulos, J.~Hoydis, H.~Gacanin, J.~de~Rosny,
  A.~Bounceu, G.~Lerosey, and M.~Fink, ``Smart radio environments empowered by
  reconfigurable {AI} meta-surfaces: {A}n idea whose time has come,''
  \emph{EURASIP J. Wireless Commun. Netw.}, vol. 2019, no.~1, pp. 1--20, May
  2019.

\bibitem{huang2019reconfigurable}
C.~Huang, A.~Zappone, G.~C. Alexandropoulos, M.~Debbah, and C.~Yuen,
  ``Reconfigurable intelligent surfaces for energy efficiency in wireless
  communication,'' \emph{IEEE Trans. Wireless Commun.}, vol.~18, no.~8, pp.
  4157--4170, Aug. 2019.

\bibitem{Basar2019}
E.~{Basar}, M.~{Di Renzo}, J.~{De Rosny}, M.~{Debbah}, M.~{Alouini}, and
  R.~{Zhang}, ``Wireless communications through reconfigurable intelligent
  surfaces,'' \emph{IEEE Access}, vol.~7, pp. 116\,753--116\,773, 2019.

\bibitem{wu2019towards}
Q.~Wu and R.~Zhang, ``Towards smart and reconfigurable environment:
  {I}ntelligent reflecting surface aided wireless network,'' \emph{IEEE Commun.
  Mag.}, vol.~58, no.~1, pp. 106--112, Jan. 2020.

\bibitem{huang2020holographic}
C.~Huang, S.~Hu, G.~C. Alexandropoulos, A.~Zappone, C.~Yuen, R.~Zhang,
  M.~Di~Renzo, and M.~Debbah, ``Holographic {MIMO} surfaces for 6{G} wireless
  networks: {O}pportunities, challenges, and trends,'' \emph{IEEE Wireless
  Commun.}, vol.~27, no.~5, pp. 118--125, Oct. 2020.

\bibitem{WavePropTCCN}
G.~C. Alexandropoulos, G.~Lerosey, M.~Debbah, and M.~Fink, ``Reconfigurable
  intelligent surfaces and metamaterials: {T}he potential of wave propagation
  control for {6G} wireless communications,'' \emph{IEEE ComSoc TCCN
  Newslett.}, vol.~6, no.~1, pp. 25--37, Jun. 2020.

\bibitem{rise6g}
E.~Calvanese~Strinati, G.~C. Alexandropoulos, V.~Sciancalepore, M.~Di~Renzo,
  H.~Wymeersch, D.-T. Phan-Huy, M.~Crozzoli, R.~D'Errico, E.~De~Carvalho,
  P.~Popovski, P.~Di~Lorenzo, L.~Bastianelli, M.~Belouar, J.~E. Mascolo,
  G.~Gradoni, S.~Phang, G.~Lerosey, and B.~Denis, ``Wireless environment as a
  service enabled by reconfigurable intelligent surfaces: {T}he {RISE-6G}
  perspective,'' in \emph{Proc. Joint EuCNC \& 6G Summit}, Porto, Portugal,
  Jun. 2021.

\bibitem{alexandg_2021}
G.~C. Alexandropoulos, N.~Shlezinger, and P.~del Hougne, ``Reconfigurable
  intelligent surfaces for rich scattering wireless communications: {R}ecent
  experiments, challenges, and opportunities,'' \emph{IEEE Commun. Mag.},
  vol.~59, no.~6, pp. 28--34, Jun. 2021.

\bibitem{RISE6G_COMMAG}
E.~Calvanese~Strinati, G.~C. Alexandropoulos, H.~Wymeersch, B.~Denis,
  V.~Sciancalepore, R.~D'Errico, A.~Clemente, D.-T. Phan-Huy, E.~D. Carvalho,
  and P.~Popovski, ``Reconfigurable, intelligent, and sustainable wireless
  environments for {6G} smart connectivity,'' \emph{IEEE Commun. Mag.},
  vol.~59, no.~10, pp. 99--105, Oct. 2021.

\bibitem{risTUTORIAL2020}
Q.~Wu, S.~Zhang, B.~Zheng, C.~You, and R.~Zhang, ``Intelligent reflecting
  surface aided wireless communications: {A} tutorial,'' \emph{IEEE Trans.
  Commun.}, vol.~69, no.~5, pp. 3313--3351, May 2021.

\bibitem{rise6g_SRE}
G.~C. Alexandropoulos, M.~Crozzoli, D.-T. Phan-Huy, K.~D. Katsanos,
  H.~Wymeersch, P.~Popovski, P.~Ratajczak, Y.~Bénédic, M.-H. Hamon,
  S.~Herraiz~Gonzalez, R.~D'Errico, and E.~Calvanese~Strinati, ``Smart wireless
  environments enabled by {RISs}: {D}eployment scenarios and two key
  challenges,'' in \emph{Proc. Joint EuCNC \& 6G Summit}, Grenoble, France,
  Jun. 2022.

\bibitem{nadeem_ce}
Q.~{Nadeem}, H.~{Alwazani}, A.~{Kammoun}, A.~{Chaaban}, M.~{Debbah}, and
  M.~{Alouini}, ``Intelligent reflecting surface-assisted multi-user {MISO}
  communication: {C}hannel estimation and beamforming design,'' \emph{IEEE Open
  J. Commun. Society}, vol.~1, pp. 661--680, May 2020.

\bibitem{9133156}
H.~{Liu}, X.~{Yuan}, and Y.~J.~A. {Zhang}, ``Matrix-calibration-based cascaded
  channel estimation for reconfigurable intelligent surface assisted multiuser
  {MIMO},'' \emph{IEEE J. Sel. Areas Comm.}, vol.~38, no.~11, pp. 2621--2636,
  Nov. 2020.

\bibitem{9144510}
L.~{Yang}, F.~{Meng}, Q.~{Wu}, D.~B. {da Costa}, and M.~S. {Alouini},
  ``Accurate closed-form approximations to channel distributions of {RIS}-aided
  wireless systems,'' \emph{IEEE Wireless Commun. Lett.}, vol.~9, no.~11, pp.
  1985--1989, Nov. 2020.

\bibitem{LZAWFM2020}
S.~Lin, B.~Zheng, G.~C. Alexandropoulos, M.~Wen, F.-J. Chen, and S.~Mumtaz,
  ``Adaptive transmission for reconfigurable intelligent surface-assisted
  {OFDM} wireless communications,'' \emph{IEEE J. Sel. Areas Commun.}, vol.~38,
  no.~11, pp. 2653--2665, Nov. 2020.

\bibitem{parafac_SAM2020}
L.~Wei, C.~Huang, G.~C. Alexandropoulos, and C.~Yuen, ``Parallel factor
  decomposition channel estimation in {RIS}-assisted multi-user {MISO}
  communication,'' in \emph{Proc. IEEE SAM}, Hangzhou, China, Jun. 2020.

\bibitem{PARAFAC2021}
L.~Wei, C.~Huang, G.~C. Alexandropoulos, C.~Yuen, and Z.~Zhang, ``Channel
  estimation for {RIS}-empowered multi-user {MISO} wireless communications,''
  \emph{IEEE Trans. Commun.}, vol.~69, no.~6, pp. 4144--4157, Jun. 2021.

\bibitem{Deepak_2021}
B.~Deepak, R.~S.~P. Sankar, and S.~P. Chepuri, ``Channel estimation for
  {RIS}-assisted millimeter-wave {MIMO} systems,'' \emph{[online]
  https://arxiv.org/abs/2011.00900}, 2021.

\bibitem{Miaowen2021}
S.~Lin, B.~Zheng, G.~C. Alexandropoulos, M.~Wen, M.~Di~Renzo, and F.~Chen,
  ``Reconfigurable intelligent surfaces with reflection pattern modulation:
  {B}eamforming design, channel estimation, and achievable rate analysis,''
  \emph{IEEE Trans. Wireless Commun.}, vol.~20, no.~2, pp. 741--754, Feb. 2021.

\bibitem{Guo_2021}
M.~Guo and M.~C. Gursoy, ``Channel estimation for intelligent reflecting
  surface assisted wireless communications,'' \emph{[online]
  https://arxiv.org/abs/2104.01221}, Apr. 2021.

\bibitem{deAraujo_2021}
G.~T. {de Ara{\'u}jo}, A.~L.~F. {de Almeida}, and R.~Boyer, ``Channel
  estimation for intelligent reflecting surface assisted {MIMO} systems: {A}
  tensor modeling approach,'' \emph{IEEE J. Sel. Topics Signal Process.},
  vol.~15, no.~3, pp. 789--802, Apr. 2021.

\bibitem{VanChien_2021}
T.~Van~Chien, H.~Q. Ngo, S.~Chatzinotas, M.~Di~Renzo, and B.~Ottersten,
  ``Reconfigurable intelligent surface-assisted cell-free massive {MIMO}
  systems over spatially-correlated channels,'' \emph{[online]
  https://arxiv.org/abs/2104.08648}, Apr. 2021.

\bibitem{Chen_2021b}
X.~Chen, J.~Shi, Z.~Yang, and L.~Wu, ``Low-complexity channel estimation for
  intelligent reflecting surface-enhanced massive {MIMO},'' \emph{IEEE Wireless
  Commun. Lett.}, vol.~10, no.~5, pp. 996--1000, May 2021.

\bibitem{Guan_2021}
X.~Guan, Q.~Wu, and R.~Zhang, ``Anchor-assisted channel estimation for
  intelligent reflecting surface aided multiuser communication,''
  \emph{[online] https://arxiv.org/abs/2102.10886}, Aug. 2021.

\bibitem{Guo_2021a}
H.~Guo and V.~K.~N. Lau, ``Cascaded channel estimation for intelligent
  reflecting surface assisted multiuser {MISO} systems,'' \emph{[online]
  https://arxiv.org/abs/2108.09002}, Aug. 2021.

\bibitem{Shi_2021a}
X.~Shi, J.~Wang, G.~Chen, and J.~Song, ``Triple-structured compressive
  sensing-based channel estimation for {RIS}-aided {MU}-{MIMO} systems,''
  \emph{[online] https://arxiv.org/abs/2108.13765}, Aug. 2021.

\bibitem{9324910}
N.~{Shlezinger}, G.~C. {Alexandropoulos}, M.~F. {Imani}, Y.~C. {Eldar}, and
  D.~R. {Smith}, ``Dynamic metasurface antennas for {6G} extreme massive {MIMO}
  communications,'' \emph{IEEE Wireless Commun.}, vol.~28, no.~2, pp.
  106–--113, Apr. 2021.

\bibitem{8879620}
Z.~{He} and X.~{Yuan}, ``Cascaded channel estimation for large intelligent
  metasurface assisted massive {MIMO},'' \emph{IEEE Wireless Commun. Lett.},
  vol.~9, no.~2, pp. 210--214, Feb. 2020.

\bibitem{8683663}
D.~{Mishra} and H.~{Johansson}, ``Channel estimation and low-complexity
  beamforming design for passive intelligent surface assisted {MISO} wireless
  energy transfer,'' in \emph{Proc. IEEE ICASSP}, Brighton, UK, May 2019.

\bibitem{9053695}
T.~L. {Jensen} and E.~{De Carvalho}, ``An optimal channel estimation scheme for
  intelligent reflecting surfaces based on a minimum variance unbiased
  estimator,'' in \emph{Proc. IEEE ICASSP}, Barcelona, Spain, May 2020.

\bibitem{9081935}
J.~{Kang}, ``Intelligent reflecting surface: {J}oint optimal training sequence
  and refection pattern,'' \emph{IEEE Commun. Lett.}, vol.~24, no.~8, pp.
  1784--1788, Aug. 2020.

\bibitem{9054415}
S.~{Xia} and Y.~{Shi}, ``Intelligent reflecting surface for massive device
  connectivity: {J}oint activity detection and channel estimation,'' in
  \emph{Proc. IEEE ICASSP}, Barcelona, Spain, May 2020.

\bibitem{9104260}
G.~T. {de Araújo} and A.~L.~F. {de Almeida}, ``{PARAFAC}-based channel
  estimation for intelligent reflective surface assisted {MIMO} system,'' in
  \emph{Proc. IEEE SAM}, Hangzhou, China, Jun. 2020.

\bibitem{8937491}
B.~{Zheng} and R.~{Zhang}, ``Intelligent reflecting surface-enhanced {OFDM}:
  {C}hannel estimation and reflection optimization,'' \emph{IEEE Wireless
  Commun. Lett.}, vol.~9, no.~4, pp. 518--522, Apr. 2020.

\bibitem{Wan_2021}
Z.~Wan, Z.~Gao, F.~Gao, M.~Di~Renzo, and M.-S. Alouini, ``Terahertz massive
  {MIMO} with holographic reconfigurable intelligent surfaces,'' \emph{IEEE
  Trans. Commun.}, vol.~69, no.~7, pp. 4732--4750, Jul. 2021.

\bibitem{tensor_channel_tracking_2022}
J.~Yuan, G.~C. Alexandropoulos, E.~Kofidis, T.~L. Jensen, and E.~De~Carvalho,
  ``Tensor-based channel tracking for {RIS}-empowered multi-user {MIMO}
  wireless systems,'' Feb. 2022, [Online] https://arxiv.org/abs/2202.08315.

\bibitem{Fast_Beam_Rui_2020}
C.~You, B.~Zheng, and R.~Zhang, ``Fast beam training for {IRS}-assisted
  multiuser communications,'' \emph{IEEE Wireless Commun. Lett.}, vol.~9,
  no.~11, pp. 1845--1849, Nov. 2020.

\bibitem{Jamali2022}
V.~Jamali, G.~C. Alexandropoulos, R.~Schober, and H.~V. Poor,
  ``Low-to-zero-overhead {IRS} reconfiguration: {D}ecoupling illumination and
  channel estimation,'' \emph{IEEE Commun. Lett.}, vol.~16, no.~4, pp.
  932--936, Apr. 2022.

\bibitem{RIS_Hierarchical}
G.~C. Alexandropoulos, V.~Jamali, R.~Schober, and H.~V. Poor, ``Near-field
  hierarchical beam management for {RIS}-enabled millimeter wave multi-antenna
  systems,'' 2022, [Online] https://arxiv.org/abs/2203.15557.pdf.

\bibitem{comparative_study}
G.~C. Alexandropoulos, A.~Papadogiannis, and P.~C. Sofotasios, ``A comparative
  study of relaying schemes with decode-and-forward over {N}akagami-$m$ fading
  channels,'' \emph{J. Comp. Netw. Commun.}, Article ID 560528, 14 pages, 2011.

\bibitem{HoVan_relay_selection}
K.~Ho-Van, P.~C. Sofotasios, G.~C. Alexandropoulos, and S.~Freear, ``Bit error
  rate of underlay decode-and-forward cognitive networks with best relay
  selection,'' \emph{J. Commun. Netw.}, vol.~17, no.~2, pp. 162--171, Apr.
  2015.

\bibitem{ying_relay_2020}
X.~Ying, U.~Demirhan, and A.~Alkhateeb, ``Relay aided intelligent
  reconfigurable surfaces: {A}chieving the potential without so many
  antennas,'' Jun. 2020, [Online] https://arxiv.org/pdf/2006.06644.pdf.

\bibitem{bjornson_intelligent_2020}
E.~Bj{\"o}rnson, {\"O}.~{\"O}zdogan, and E.~G. Larsson, ``Intelligent
  reflecting surface vs. decode-and-forward: {H}ow large surfaces are needed to
  beat relaying?'' \emph{IEEE Wireless Commun. Lett.}, vol.~9, no.~2, pp.
  244--248, Feb. 2020.

\bibitem{nemati_ris-assisted_2020}
M.~Nemati, J.~Park, and J.~Choi, ``{RIS}-assisted coverage enhancement in
  millimeter-wave cellular networks,'' \emph{IEEE Access}, vol.~8, pp.
  188\,171--188\,185, Oct. 2020.

\bibitem{yang_coverage_2020}
L.~Yang, Y.~Yang, M.~O. Hasna, and M.-S. Alouini, ``Coverage, probability of
  {SNR} gain, and {DOR} analysis of {RIS}-aided communication systems,''
  \emph{IEEE Wireless Commun. Lett.}, vol.~9, no.~8, pp. 1268--1272, Aug. 2020.

\bibitem{zeng_reconfigurable_2021}
S.~Zeng, H.~Zhang, B.~Di, Z.~Han, and L.~Song, ``Reconfigurable intelligent
  surface ({RIS}) assisted wireless coverage extension: {RIS} orientation and
  location optimization,'' \emph{IEEE Commun. Lett.}, vol.~25, no.~1, pp.
  269--273, Jan. 2021.

\bibitem{yildirim_hybrid_2021}
I.~Yildirim, F.~Kilinc, E.~Basar, and G.~C. Alexandropoulos, ``Hybrid
  {RIS}-empowered reflection and decode-and-forward relaying for coverage
  extension,'' \emph{IEEE Commun. Lett.}, vol.~25, no.~5, pp. 1692--1696, May
  2021.

\bibitem{ma_indoor_2021}
T.~Ma, Y.~Xiao, X.~Lei, W.~Xiong, and Y.~Ding, ``Indoor localization with
  reconfigurable intelligent surface,'' \emph{IEEE Commun. Lett.}, vol.~25,
  no.~1, pp. 161--165, Jan. 2021.

\bibitem{zhang_metaradar_2020}
H.~Zhang, J.~Hu, H.~Zhang, B.~Di, K.~Bian, Z.~Han, and L.~Song, ``{MetaRadar}:
  {I}ndoor localization by reconfigurable metamaterials,'' Aug. 2020, [Online]
  https://arxiv.org/pdf/2008.02459.pdf.

\bibitem{elzanaty_reconfigurable_2020}
A.~Elzanaty, A.~Guerra, F.~Guidi, and M.-S. Alouini, ``Reconfigurable
  intelligent surfaces for localization: {P}osition and orientation error
  bounds,'' Sep. 2020, [Online] https://arxiv.org/pdf/2009.02818.pdf.

\bibitem{keykhosravi_siso_2021}
K.~Keykhosravi, M.~F. Keskin, G.~Seco-Granados, and H.~Wymeersch, ``{SISO}
  {RIS}-enabled joint {3D} downlink localization and synchronization,'' Feb.
  2021, [Online] https://arxiv.org/pdf/2011.02391.pdf.

\bibitem{buzzi_radar_2021}
S.~Buzzi, E.~Grossi, M.~Lops, and L.~Venturino, ``Radar target detection aided
  by reconfigurable intelligent surfaces,'' Jun. 2021 [Online]
  https://arxiv.org/pdf/2104.00768.pdf.

\bibitem{rahal_ris-enabled_2021}
M.~Rahal, B.~Denis, K.~Keykhosravi, B.~Uguen, and H.~Wymeersch, ``{RIS}-enabled
  localization continuity under near-field conditions,'' Sep. 2021, [Online]
  https://arxiv.org/pdf/2109.11965.pdf.

\bibitem{abu-shaban_near-field_2020}
Z.~Abu-Shaban, K.~Keykhosravi, M.~F. Keskin, G.~C. Alexandropoulos,
  G.~Seco-Granados, and H.~Wymeersch, ``Near-field localization with a
  reconfigurable intelligent surface acting as lens,'' in \emph{Proc. IEEE
  ICC}, Montreal, Canada, Jun. 2021.

\bibitem{yang_wireless_2021}
Z.~Yang, H.~Zhang, B.~Di, H.~Zhang, K.~Bian, and L.~Song, ``Wireless indoor
  simultaneous localization and mapping using reconfigurable intelligent
  surface,'' Jul. 2021, [Online] https://arxiv.org/pdf/2107.01582.pdf.

\bibitem{sankar_joint_2021}
R.~S.~P. Sankar, B.~Deepak, and S.~P. Chepuri, ``Joint communication and radar
  sensing with reconfigurable intelligent surfaces,'' May 2021, [Online]
  https://arxiv.org/pdf/2105.01966.pdf.

\bibitem{nguyen_reconfigurable_2020}
C.~L. Nguyen, O.~Georgiou, and G.~Gradoni, ``Reconfigurable intelligent
  surfaces and machine learning for wireless fingerprinting localization,''
  Oct. 2020, [Online] https://arxiv.org/pdf/2010.03251.pdf.

\bibitem{locrxris_all}
G.~C. Alexandropoulos, I.~Vinieratou, and H.~Wymeersch, ``Localization via
  multiple reconfigurable intelligent surfaces equipped with single receive
  {RF} chains,'' \emph{IEEE Wireless Commun. Lett.}, to appear, 2022.

\bibitem{zerobs_all}
K.~Keykhosravi, G.~Seco-Granados, G.~C. Alexandropoulos, and H.~Wymeersch,
  ``{RIS}-enabled self-localization: {L}everaging controllable reflections with
  zero access points,'' in \emph{Proc. IEEE ICC}, Seoul, South Korea, Jun.
  2022.

\bibitem{Chen_2019_all}
J.~Chen, Y.-C. Liang, Y.~Pei, and H.~Guo, ``Intelligent reflecting surface: {A}
  programmable wireless environment for physical layer security,'' \emph{IEEE
  Access}, vol.~7, pp. 82\,599--82\,612, Jul. 2019.

\bibitem{Shen_2019_all}
H.~Shen, W.~Xu, S.~Gong, Z.~He, and C.~Zhao, ``\BIBforeignlanguage{en}{Secrecy
  rate maximization for intelligent reflecting surface assisted multi-antenna
  communications},'' \emph{\BIBforeignlanguage{en}{IEEE Commun. Lett.}},
  vol.~23, no.~9, pp. 1488--1492, Sep. 2019.

\bibitem{Cui_2019_all}
M.~Cui, G.~Zhang, and R.~Zhang, ``\BIBforeignlanguage{en}{Secure wireless
  communication via intelligent reflecting surface},''
  \emph{\BIBforeignlanguage{en}{IEEE Wireless Commun. Lett.}}, vol.~8, no.~5,
  pp. 1410--1414, Oct. 2019.

\bibitem{Xu_2019_all}
D.~Xu, X.~Yu, Y.~Sun, D.~W.~K. Ng, and R.~Schober, ``Resource allocation for
  secure {IRS}-assisted multiuser {MISO} systems,'' in \emph{Proc. IEEE
  GLOBECOM}, Waikoloa, USA, Dec. 2019.

\bibitem{Yu_2019_all}
X.~Yu, D.~Xu, and R.~Schober, ``Enabling secure wireless communications via
  intelligent reflecting surfaces,'' in \emph{Proc. IEEE GLOBECOM}, Waikoloa,
  USA, Dec. 2019.

\bibitem{Almohamad_2020}
A.~{Almohamad}, A.~M. {Tahir}, A.~{Al-Kababji}, H.~M. {Furqan}, T.~{Khattab},
  M.~O. {Hasna}, and H.~{Arslan}, ``Smart and secure wireless communications
  via reflecting intelligent surfaces: {A} short survey,'' \emph{IEEE Open J.
  Commun. Society}, vol.~1, pp. 1442--1456, Sep. 2020.

\bibitem{Chu_2020_all}
Z.~Chu, W.~Hao, P.~Xiao, and J.~Shi, ``\BIBforeignlanguage{en}{Intelligent
  reflecting surface aided multi-antenna secure transmission},''
  \emph{\BIBforeignlanguage{en}{IEEE Wireless Commun. Lett.}}, vol.~9, no.~1,
  pp. 108--112, Jan. 2020.

\bibitem{Hong_2020}
S.~Hong, C.~Pan, H.~Ren, K.~Wang, and A.~Nallanathan, ``Artificial-noise-aided
  secure {MIMO} wireless communications via intelligent reflecting surface,''
  \emph{IEEE Trans. Commun.}, vol.~68, no.~12, pp. 7851--7866, Dec. 2020.

\bibitem{Dong_2020b}
L.~{Dong} and H.~M. {Wang}, ``Enhancing secure {MIMO} transmission via
  intelligent reflecting surface,'' \emph{IEEE Trans. Wireless Commun},
  vol.~19, no.~11, pp. 7543--7556, Nov. 2020.

\bibitem{Shu_2020}
F.~Shu, J.~Li, M.~Huang, W.~Shi, Y.~Teng, J.~Li, Y.~Wu, and J.~Wang, ``Enhanced
  secrecy rate maximization for directional modulation networks via {IRS},''
  \emph{[Online] https://arxiv.org/abs/2008.05067}, Aug. 2020.

\bibitem{PLS_Kostas}
G.~C. Alexandropoulos, K.~Katsanos, M.~Wen, and D.~B. da~Costa, ``Safeguarding
  {MIMO} communications with reconfigurable metasurfaces and artificial
  noise,'' in \emph{Proc. ICC}, Montreal, Canada, Jun. 2021.

\bibitem{AL-Mekhlafi_2021}
M.~{AL-Mekhlafi}, M.~A. Arfaoui, M.~Elhattab, C.~Assi, and A.~Ghrayeb, ``Joint
  resource allocation and phase shift optimization for {RIS}-aided
  {eMBB}/{URLLC} traffic multiplexing,'' \emph{[online]
  https://arxiv.org/abs/2108.02346}, Aug. 2021.

\bibitem{Jung_2021a}
M.~Jung, W.~Saad, M.~Debbah, and C.~S. Hong, ``On the optimality of
  reconfigurable intelligent surfaces ({RISs}): {P}assive beamforming,
  modulation, and resource allocation,'' \emph{IEEE Trans. Wireless Commun.},
  vol.~20, no.~7, pp. 4347--4363, Jul. 2021.

\bibitem{Moustakas_RIS}
A.~L. Moustakas, G.~C. Alexandropoulos, and M.~Debbah, ``Capacity optimization
  using reconfigurable intelligent surfaces: {A} large system approach,'' in
  \emph{Proc. IEEE GLOBECOM}, Lucca, Italy, Dec. 2021.

\bibitem{Mu_2021b}
X.~Mu, Y.~Liu, L.~Guo, J.~Lin, and N.~{Al-Dhahir}, ``Capacity and optimal
  resource allocation for {IRS}-assisted multi-user communication systems,''
  \emph{IEEE Trans. Commun.}, vol.~69, no.~6, pp. 3771--3786, Jun. 2021.

\bibitem{You_2021}
L.~You, J.~Xiong, Y.~Huang, D.~W.~K. Ng, C.~Pan, W.~Wang, and X.~Gao,
  ``Reconfigurable intelligent surfaces-assisted multiuser {MIMO} uplink
  transmission with partial {CSI},'' \emph{IEEE Trans. Wireless Commun.},
  vol.~20, no.~9, pp. 5613--5627, Sep. 2021.

\bibitem{Xu_2021b}
D.~Xu, X.~Yu, D.~W.~K. Ng, and R.~Schober, ``Resource allocation for active
  {IRS}-assisted multiuser communication systems,'' \emph{[online]
  https://arxiv.org/abs/2108.13033}, Aug. 2021.

\bibitem{amplifying_RIS_2022}
R.~A. Tasci, F.~Kilinc, E.~Basar, and G.~C. Alexandropoulos, ``A new {RIS}
  architecture with a single power amplifier: {E}nergy efficiency and error
  performance analysis,'' \emph{IEEE Access}, to appear, 2022.

\bibitem{Yang_2021c}
Z.~Yang, M.~Chen, W.~Saad, W.~Xu, M.~{Shikh-Bahaei}, H.~V. Poor, and S.~Cui,
  ``Energy-efficient wireless communications with distributed reconfigurable
  intelligent surfaces,'' \emph{IEEE Trans. Wireless Commun.}, to appear, 2021.

\bibitem{Yang_2021d}
G.~Yang, Y.~Liao, Y.-C. Liang, O.~Tirkkonen, G.~Wang, and X.~Zhu,
  ``Reconfigurable intelligent surface empowered device-to-device communication
  underlaying cellular networks,'' \emph{IEEE Trans. Commun.}, to appear, 2021.

\bibitem{Goodfellow-et-al-2016}
I.~Goodfellow, Y.~Bengio, and A.~Courville, \emph{Deep Learning}.\hskip 1em
  plus 0.5em minus 0.4em\relax MIT Press, 2016.

\bibitem{Wireless_DL_Surevey}
A.~Zappone, M.~Di~Renzo, and M.~Debbah, ``Wireless networks design in the era
  of deep learning: {M}odel-based, {AI}-based, or both?'' \emph{IEEE Trans.
  Commun.}, vol.~67, no.~10, pp. 7331--7376, 2019.

\bibitem{Xiang2021SelfCalibratingIL}
C.~Xiang, S.~Zhang, S.~Xu, and G.~C. Alexandropoulos, ``Self-calibrating indoor
  localization with crowdsourcing fingerprints and transfer learning,'' in
  \emph{Proc. IEEE ICC}, Montreal, Canada, Jun. 2021.

\bibitem{LocalizationWifi2019}
S.~BelMannoubi and H.~Touati, ``Deep neural networks for indoor localization
  using wifi fingerprints,'' in \emph{Proc. MSPN}, Mohammedia, Morocco, Apr.
  2019.

\bibitem{Xiang_RobustLocalization}
C.~Xiang, S.~Zhang, S.~Xu, X.~Chen, S.~Cao, G.~C. Alexandropoulos, and V.~K.~N.
  Lau, ``Robust sub-meter level indoor localization with a single {WiFi} access
  point - regression versus classification,'' \emph{IEEE Access}, vol.~7, pp.
  146\,309--146\,321, 2019.

\bibitem{Huang_CE_2019}
C.~Huang, G.~C. Alexandropoulos, A.~Zappone, C.~Yuen, and M.~Debbah, ``Deep
  learning for {UL}/{DL} channel calibration in generic massive {MIMO}
  systems,'' in \emph{Proc. IEEE ICC}, Shanghai, China, May 2019.

\bibitem{Hu2021_DL_CE}
Q.~Hu, F.~Gao, H.~Zhang, S.~Jin, and G.~Y. Li, ``Deep learning for channel
  estimation: {I}nterpretation, performance, and comparison,'' \emph{IEEE
  Trans. Wireless Commun.}, vol.~20, no.~4, pp. 2398--2412, Apr. 2021.

\bibitem{Zapone_UnsupervisedBeamforming}
H.~Huang, W.~Xia, J.~Xiong, J.~Yang, G.~Zheng, and X.~Zhu, ``Unsupervised
  learning-based fast beamforming design for downlink {MIMO},'' \emph{IEEE
  Access}, vol.~7, pp. 7599--7605, 2019.

\bibitem{H_Huang_DL_precoding}
H.~Huang, Y.~Song, J.~Yang, G.~Gui, and F.~Adachi, ``Deep-learning-based
  millimeter-wave massive {MIMO} for hybrid precoding,'' \emph{IEEE Trans. Veh.
  Technol.}, vol.~68, pp. 3027--3032, Mar. 2019.

\bibitem{Li2019_DL_precoding}
X.~Li and A.~Alkhateeb, ``Deep learning for direct hybrid precoding in
  millimeter wave massive mimo systems,'' \emph{Proc. Asilomar Conf. Signals,
  Sys., Comp.}, Nov. 2019.

\bibitem{Sapavath2019_ML_Beamforming_MIMO}
N.~Sapavath, S.~Safavat, and D.~B. Rawat, ``On the machine learning–based
  smart beamforming for wireless virtualization with large‐scale {MIMO}
  system,'' \emph{Trans. Emerg. Telecommun. Technol.}, vol.~30, Aug. 2019.

\bibitem{huang2019spawc}
C.~{Huang}, G.~C. {Alexandropoulos}, C.~{Yuen}, and M.~{Debbah}, ``Indoor
  signal focusing with deep learning designed reconfigurable intelligent
  surfaces,'' in \emph{Proc. IEEE SPAWC}, Cannes, France, Jul. 2019.

\bibitem{Chongwen_Spectrum_Learning}
B.~Yang, X.~Cao, C.~Huang, C.~Yuen, L.~Qian, and M.~Di~Renzo, ``Intelligent
  spectrum learning for wireless networks with reconfigurable intelligent
  surfaces,'' \emph{IEEE Trans. Veh. Technol.}, vol.~70, no.~4, pp. 3920--3925,
  2021.

\bibitem{RIS_DL_CNN}
B.~Sheen, J.~Yang, X.~Feng, and M.~M.~U. Chowdhury, ``A deep learning based
  modeling of reconfigurable intelligent surface assisted wireless
  communications for phase shift configuration,'' \emph{IEEE Open J. Commun.
  Society}, vol.~2, pp. 262--272, Jan. 2021.

\bibitem{Gao_unsupervisedBamforming_RIS}
J.~Gao, C.~Zhong, X.~Chen, H.~Lin, and Z.~Zhang, ``Unsupervised learning for
  passive beamforming,'' \emph{IEEE Commun. Lett.}, vol.~24, no.~5, pp.
  1052--1056, May 2020.

\bibitem{liaskos2019spawc}
C.~Liaskos, A.~Tsioliaridou, S.~Nie, A.~Pitsillides, S.~Ioannidis, and
  I.~Akyildiz, ``An interpretable neural network for configuring programmable
  wireless environments,'' in \emph{Proc. IEEE SPAWC}, Cannes, France, Jul.
  2019.

\bibitem{RIS_compressive_sensing}
A.~Taha, M.~Alrabeiah, and A.~Alkhateeb, ``Enabling large intelligent surfaces
  with compressive sensing and deep learning,'' \emph{IEEE Access}, vol.~9, pp.
  44\,304--44\,321, 2021.

\bibitem{hardware2020icassp}
G.~C. Alexandropoulos and E.~Vlachos, ``A hardware architecture for
  reconfigurable intelligent surfaces with minimal active elements for explicit
  channel estimation,'' in \emph{Proc. IEEE ICASSP}, Barcelona, Spain, May
  2020.

\bibitem{chongwendrl}
C.~{Huang}, R.~{Mo}, and C.~{Yuen}, ``Reconfigurable intelligent surface
  assisted multiuser {MISO} systems exploiting deep reinforcement learning,''
  \emph{IEEE J. Sel. Area. Comm.}, vol.~38, no.~8, pp. 1839--1850, Aug. 2020.

\bibitem{Tasos_DNN_CE_2019}
A.~M. Elbir, A.~Papazafeiropoulos, P.~Kourtessis, and S.~Chatzinotas, ``Deep
  channel learning for large intelligent surfaces aided mm-wave massive {MIMO}
  systems,'' \emph{IEEE Wireless Commun. Lett.}, vol.~9, no.~9, pp. 1447--1451,
  Sep. 2020.

\bibitem{RIS_DL_channel_estimation2}
S.~Khan and S.~Y.~K. Shin, ``Deep-learning-aided detection for reconfigurable
  intelligent surfaces,'' 2019, [Online] https://arxiv.org/pdf/1910.09136.pdf.

\bibitem{ma_Smart_2019}
Q.~Ma, G.~D. Bai, H.~B. Jing, C.~Yang, L.~Li, and T.~J. Cui, ``Smart
  metasurface with self-adaptively reprogrammable functions,'' \emph{Light Sci
  Appl}, vol.~8, no.~1, p.~98, Oct. 2019.

\bibitem{liaskos_ABSense_2019}
C.~Liaskos, G.~Pirialakos, A.~Pitilakis, S.~Abadal, A.~Tsioliaridou,
  A.~Tasolamprou \emph{et~al.}, ``{ABSense}: {S}ensing electromagnetic waves on
  metasurfaces via ambient compilation of full absorption,'' Jul. 2019,
  [Online] https://arxiv.org/pdf/1907.04811.pdf.

\bibitem{ma_Smart_2020}
Q.~Ma, Q.~R. Hong, X.~X. Gao, H.~B. Jing, C.~Liu, G.~D. Bai, Q.~Cheng, and
  T.~J. Cui, ``Smart sensing metasurface with self-defined functions in dual
  polarizations,'' \emph{Nanophotonics}, vol.~9, no.~10, pp. 3271--3278, Sep.
  2020.

\bibitem{HRIS_Mag}
G.~C. Alexandropoulos, N.~Shlezinger, I.~Alamzadeh, M.~F. Imani, H.~Zhang, and
  Y.~C. Eldar, ``Hybrid reconfigurable intelligent metasurfaces: {E}nabling
  simultaneous tunable reflections and sensing for {6G} wireless
  communications,'' 2021, [Online] https://arxiv.org/abs/2104.04690.

\bibitem{HRIS_Nature}
I.~Alamzadeh, G.~C. Alexandropoulos, N.~Shlezinger, and M.~F. Imani, ``A
  reconfigurable intelligent surface with integrated sensing capability,''
  \emph{Scientific Reports}, vol.~11, no. 20737, pp. 1--10, Oct. 2021.

\bibitem{HRIS_SPAWC}
H.~Zhang, N.~Shlezinger, I.~Alamzadeh, G.~C. Alexandropoulos, M.~F. Imani, and
  Y.~C. Eldar, ``Channel estimation with simultaneous reflecting and sensing
  reconfigurable intelligent metasurfaces,'' in \emph{Proc. IEEE SPAWC}, Lucca,
  Italy, Sep. 2021.

\bibitem{RIS_DL_Secrecy_rate}
H.~Shen, W.~Xu, S.~Gong, Z.~He, and C.~Zhao, ``Secrecy rate maximization for
  intelligent reflecting surface assisted multi-antenna communications,''
  \emph{IEEE Commun. Lett.}, vol.~23, no.~9, pp. 1488--1492, Sep. 2019.

\bibitem{Li2019Nature}
L.~Li, H.~Ruan, C.~Liu, Y.~Li, Y.~Shuang, A.~Al{\`u}, C.-W. Qiu, and T.~J. Cui,
  ``Machine-learning reprogrammable metasurface imager,'' \emph{Nature
  Commun.}, vol.~10, no.~1, p. 1082, Mar. 2019.

\bibitem{MAC_AI_RIS}
X.~Cao, B.~Yang, C.~Huang, C.~Yuen, M.~Di~Renzo, Z.~Han, D.~Niyato, H.~V. Poor,
  and L.~Hanzo, ``{AI}-assisted {MAC} for reconfigurable
  intelligent-surface-aided wireless networks: {C}hallenges and
  opportunities,'' \emph{IEEE Commun. Mag.}, vol.~59, no.~6, pp. 21--27, Jun.
  2021.

\bibitem{Jointly_Learned_2021}
L.~Wang, N.~Shlezinger, G.~C. Alexandropoulos, H.~Zhang, B.~Wang, and Y.~C.
  Eldar, ``Jointly learned symbol detection and signal reflection in
  {RIS}-aided multi-user {MIMO} systems,'' in \emph{Proc. Asilomar Conf.
  Signals, Sys., Comp.}, Pacific Grove, USA, Nov. 2021.

\bibitem{DeepRIS_2022}
K.~Stylianopoulos, N.~Shlezinger, P.~del Hougne, and G.~C. Alexandropoulos,
  ``Deep-learning-assisted configuration of reconfigurable intelligent surfaces
  in dynamic rich-scattering environments,'' in \emph{Proc. IEEE ICASSP},
  Singapore, May 2021.

\bibitem{cui2014coding}
T.~J. Cui, M.~Q. Qi, X.~Wan, J.~Zhao, and Q.~Cheng, ``Coding metamaterials,
  digital metamaterials and programmable metamaterials,'' \emph{Light: Science
  \& Applications}, vol.~3, no.~10, pp. e218--e218, 2014.

\bibitem{Tsinghua_RIS_Tutorial}
M.~Jian, G.~C. Alexandropoulos, E.~Basar, C.~Huang, R.~Liu, Y.~Liu, and
  C.~Yuen, ``Reconfigurable intelligent surfaces for wireless communications:
  {O}verview of hardware designs, channel models, and estimation techniques,''
  \emph{ITU Intell. Converged Netw.}, to appear, 2022.

\bibitem{tang2020wireless_arxiv}
W.~Tang, M.~Z. Chen, X.~Chen, J.~Y. Dai, Y.~Han, M.~Di~Renzo, Y.~Zeng, S.~Jin,
  Q.~Cheng, and T.~J. Cui, ``Wireless communications with reconfigurable
  intelligent surface: {P}ath loss modeling and experimental measurement,''
  vol.~20, no.~1, pp. 421--439, Jan. 2021.

\bibitem{dai2020reconfigurable_arxiv}
L.~Dai, B.~Wang, M.~Wang, X.~Yang, J.~Tan, S.~Bi, S.~Xu, F.~Yang, Z.~Chen,
  M.~Di~Renzo, and L.~Hanzo, ``Reconfigurable intelligent surface-based
  wireless communications: {A}ntenna design, prototyping, and experimental
  results,'' \emph{IEEE Access}, vol.~8, pp. 45\,913--45\,923, Mar. 2020.

\bibitem{Nature13}
O.~{Zhu}, J.~{Zhao}, and Y.~{Feng}, ``Active impedance metasurface with full
  360 reflection phase tuning,'' \emph{Scientific reports}, vol.~3, no.~10, pp.
  3059--3064, 2013.

\bibitem{PhysRevApplied.11.044024}
F.~Liu, O.~Tsilipakos, A.~Pitilakis, A.~C. Tasolamprou, M.~S. Mirmoosa, N.~V.
  Kantartzis, D.-H. Kwon, J.~Georgiou, K.~Kossifos, M.~A. Antoniades,
  M.~Kafesaki, C.~M. Soukoulis, and S.~A. Tretyakov, ``Intelligent metasurfaces
  with continuously tunable local surface impedance for multiple reconfigurable
  functions,'' \emph{Phys. Rev. Applied}, vol.~11, p. 044024, Apr 2019.

\bibitem{Abeywickrama_2020}
S.~Abeywickrama, R.~Zhang, Q.~Wu, and C.~Yue, ``Intelligent reflecting surface:
  {P}ractical phase shift model and beamforming optimization,'' \emph{IEEE
  Trans. Commun.}, vol.~68, no.~9, pp. 5849--5863, Sep. 2020.

\bibitem{pulidopolarizability2017}
L.~Pulido-Mancera, P.~T. Bowen, M.~F. Imani, N.~Kundtz, and D.~Smith,
  ``\BIBforeignlanguage{en}{Polarizability extraction of complementary
  metamaterial elements in waveguides for aperture modeling},''
  \emph{\BIBforeignlanguage{en}{Phys. Rev. B}}, vol.~96, no.~23, p. 235402,
  Dec. 2017.

\bibitem{Antonio_2012}
A.~Clemente, L.~Dussopt, R.~Sauleau, P.~Potier, and P.~Pouliguen, ``$1$-bit
  reconfigurable unit cell based on {PIN} diodes for transmit-array
  applications in {X}-band,'' \emph{IEEE Antennas Prop.}, vol.~60, no.~5, pp.
  2260--2269, May 2012.

\bibitem{3GPP}
3GPP, ``{5G; NR; P}hysical layer procedures for data,'' Tech. Rep. 38.214,
  version 16.2.0, Jul. 2020.

\bibitem{Dahlman_5G_NR}
E.~Dahlman, S.~Parkvall, and J.~Sköld, \emph{“5G NR The Next Generation
  Wireless Access Technology}.\hskip 1em plus 0.5em minus 0.4em\relax Athena
  Academic Press, 2018.

\bibitem{alexandg_fd_power_control}
G.~C. Alexandropoulos, M.~Kountouris, and I.~Atzeni, ``User scheduling and
  optimal power allocation for full-duplex cellular networks,'' in \emph{Proc.
  IEEE SPAWC}, Edinburgh, UK, Jul. 2016.

\bibitem{Ghazanfari_power_control_2020}
A.~Ghazanfari, H.~V. Cheng, E.~Björnson, and E.~G. Larsson, ``Enhanced
  fairness and scalability of power control schemes in multi-cell massive
  {MIMO},'' \emph{IEEE Trans. Commun.}, vol.~8, no.~5, pp. 2878--2890, May
  2020.

\bibitem{PhysFad}
R.~Faqiri, C.~Saigre-Tardif, G.~C. Alexandropoulos, N.~Shlezinger, M.~F. Imani,
  and P.~del Hougne, ``Phys{F}ad: {P}hysics-based end-to-end channel modeling
  of {RIS}-parametrized environments with adjustable fading,'' 2022, [Online]
  https://arxiv.org/pdf/2202.02673.pdf.

\bibitem{alexandg_ESPARs}
G.~C. {Alexandropoulos}, V.~I. {Barousis}, and C.~B. {Papadias}, ``Precoding
  for multiuser {MIMO} systems with single-fed parasitic antenna arrays,'' in
  \emph{Proc. IEEE GLOBECOM}, Austin, USA, Dec. 2014.

\bibitem{Gradoni2020}
G.~Gradoni and M.~Di~Renzo, ``End-to-end mutual-coupling-aware communication
  model for reconfigurable intelligent surfaces: {A}n electromagnetic-compliant
  approach based on mutual impedances,'' \emph{IEEE Wireless Commun. Lett.},
  vol.~10, no.~5, pp. 938--942, May 2021.

\bibitem{Emil_RIS_Correlation}
E.~Bj{\"o}rnson and L.~Sanguinetti, ``Rayleigh fading modeling and channel
  hardening for reconfigurable intelligent surfaces,'' \emph{IEEE Wireless
  Commun. Lett.}, vol.~10, no.~4, pp. 830--834, Apr. 2021.

\bibitem{correlated_Weibull}
G.~C. Alexandropoulos, N.~C. Sagias, and K.~Berberidis, ``On the multivariate
  {W}eibull fading model with arbitrary correlation matrix,'' \emph{IEEE
  Antennas Wireless Prop. Lett.}, vol.~6, pp. 93--95, 2007.

\bibitem{correlated_Nakagami}
G.~C. Alexandropoulos, P.~T. Mathiopoulos, and N.~C. Sagias,
  ``Switch-and-examine diversity over arbitrary correlated {N}akagami-$m$
  fading channels,'' \emph{IEEE Trans. Veh. Technol.}, vol.~59, no.~4, pp.
  2080--2087, May 2010.

\bibitem{Channel_Time_Correlation}
P.~Sadeghi, R.~A. Kennedy, P.~B. Rapajic, and R.~Shams, ``Finite-state {M}arkov
  modeling of fading channels: {A} survey of principles and applications,''
  \emph{IEEE Signal Process. Mag.}, vol.~25, no.~5, pp. 57--80, Sep. 2008.

\bibitem{HeTaHaKu14_all}
J.~He, T.~Kim, H.~Ghauch, K.~Liu, and G.~Wang, ``Millimeter wave {MIMO} channel
  tracking systems,'' in \emph{Proc. IEEE GLOBECOM}, Austin, USA, 8-12 Dec.
  2014.

\bibitem{alexandg_correlation}
G.~C. Alexandropoulos and S.~Chouvardas, ``Low complexity channel estimation
  for millimeter wave systems with hybrid {A/D} antenna processing,'' in
  \emph{Proc. IEEE GLOBECOM}, Washington D.C., USA, Dec. 2016.

\bibitem{3GPP_NR_ArXiv2018}
M.~{Giordani}, M.~Polese, A.~Roy, D.~Castor, and M.~Zorzi, ``{A tutorial on
  beam management for 3GPP NR at mmWave frequencies},'' \emph{IEEE Commun.
  Surveys Tuts.}, vol.~21, no.~1, pp. 173--196, 2019.

\bibitem{J:George_Elsevier_13}
G.~C. Alexandropoulos and C.~B. Papadias, ``A reconfigurable iterative
  algorithm for the {$K$}-user {MIMO} interference channel,'' \emph{{S}ignal
  {P}rocess. ({E}lsevier)}, vol.~93, no.~12, pp. 3353--3362, Dec. 2013.

\bibitem{Farrokhi_Ricean}
F.~R. Farrokhi, A.~Lozano, G.~J. Foschini, and R.~Valenzuela, ``Spectral
  efficiency of {FDMA/TDMA} wireless systems with transmit and receive antenna
  array,'' \emph{IEEE Trans. Wireless Commun.}, vol.~1, no.~4, pp. 591--599,
  Oct. 2002.

\bibitem{ULBA2021}
G.~C. Alexandropoulos, I.~Vinieratou, M.~Rebato, L.~Rose, and M.~Zorzi,
  ``Uplink beam management for millimeter wave cellular {MIMO} systems with
  hybrid beamforming,'' in \emph{Proc. {IEEE WCNC}}, Nanjing, China, Apr. 2021.

\bibitem{Alkhateeb_JSTSP_all}
A.~Alkhateeb, O.~E. Ayach, G.~Leus, and R.~W. Heath, Jr., ``Channel estimation
  and hybrid precoding for millimeter wave cellular systems,'' \emph{{IEEE}
  {J.} {S}el. {T}opics {S}ignal {P}rocess.}, vol.~8, no.~5, pp. 831--846, Oct.
  2014.

\bibitem{Huang_GLOBECOM_2019}
C.~{Huang}, G.~C. {Alexandropoulos}, A.~{Zappone}, M.~{Debbah}, and C.~{Yuen},
  ``Energy efficient multi-user {MISO} communication using low resolution large
  intelligent surfaces,'' in \emph{Proc. IEEE GLOBECOM}, Abu Dhabi, UAE, Dec.
  2018.

\bibitem{Samarakoon_2020}
G.~C. Alexandropoulos, S.~Samarakoon, M.~Bennis, and M.~Debbah, ``Phase
  configuration learning in wireless networks with multiple reconfigurable
  intelligent surfaces,'' in \emph{Proc. IEEE GLOBECOM}, Taipei, Taiwan, Dec.
  2020.

\bibitem{Wireless20}
H.~Gacanin and M.~Di~Renzo, ``Wireless 2.0: {T}oward an intelligent radio
  environment empowered by reconfigurable meta-surfaces and artificial
  intelligence,'' \emph{IEEE Veh. Technol. Mag.}, vol.~15, no.~4, pp. 74--82,
  2020.

\bibitem{DL_RIS_survey}
A.~M. Elbir and K.~V. Mishra, ``A survey of deep learning architectures for
  intelligent reflecting surfaces,'' 2020, [Online]
  https://arxiv.org/pdf/2009.02540.pdf.

\bibitem{AIRIS}
S.~{Zhang}, M.~{Li}, M.~{Jian}, Y.~{Zhao}, and F.~{Gao}, ``{AIRIS}:
  {A}rtificial intelligence enhanced signal processing in reconfigurable
  intelligent surface communications,'' May 2021, [Online]
  https://arxiv.org/pdf/2106.00171.pdf.

\bibitem{Sutton98reinforcementlearning}
R.~S. Sutton and A.~G. Barto, \emph{Reinforcement Learning {I}:
  {I}ntroduction}.\hskip 1em plus 0.5em minus 0.4em\relax Cambridge,
  Massachusetts: MIT Press, 1998.

\bibitem{Bertsekas}
D.~P. Bertsekas, \emph{Dynamic Programming and Optimal Control}.\hskip 1em plus
  0.5em minus 0.4em\relax Athena Scientific, 2000.

\bibitem{WorldModels}
D.~Ha and J.~Schmidhuber, ``Recurrent world models facilitate policy
  evolution,'' in \emph{Proc. NeurIPS}, 2018, pp. 2455--2467.

\bibitem{Lazaridis2020SOTA_walkthrough}
A.~Lazaridis, A.~Fachantidis, and I.~Vlahavas, ``Deep reinforcement learning:
  {A} state-of-the-art walkthrough,'' \emph{J. Artif. Intell. Res.}, vol.~69,
  pp. 1421--1471, 2020.

\bibitem{DQN}
V.~Mnih, K.~Kavukcuoglu, D.~Silver \emph{et~al.}, ``Human-level control through
  deep reinforcement learning,'' \emph{Nature}, vol. 518, no. 7540, pp.
  529--533, 2015.

\bibitem{Adam}
D.~P. Kingma and J.~Ba, ``Adam: {A} method for stochastic optimization,'' 2017,
  [Online] https://arxiv.org/pdf/1412.6980.pdf.

\bibitem{Graves_LSTM_RMSProp}
A.~Graves, ``Generating sequences with recurrent neural networks,'' 2014,
  [Online] https://arxiv.org/pdf/1308.0850.pdf.

\bibitem{DQN_TheoreticalAnalysis}
J.~Fan, Z.~Wang, Y.~Xie, and Z.~Yang, ``A theoretical analysis of deep
  {Q}-learning,'' in \emph{Proc. PMLR}, vol. 120, Jun. 2020.

\bibitem{SuttonPG}
R.~S. Sutton, D.~McAllester, S.~Singh, and Y.~Mansour, ``Policy gradient
  methods for reinforcement learning with function approximation,'' in
  \emph{Proc. NIPS}, Denver, USA, Dec. 1999.

\bibitem{REINFORCE}
R.~J. Williams, ``Simple statistical gradient-following algorithms for
  connectionist reinforcement learning,'' \emph{Mach. Learn.}, vol.~8, no.~3,
  pp. 229--256, May 1992.

\bibitem{PPO}
J.~Schulman, F.~Wolski, P.~Dhariwal, A.~Radford, and O.~Klimov, ``Proximal
  policy optimization algorithms,'' 2017, [Online]
  https://arxiv.org/pdf/1707.06347.pdf.

\bibitem{DOTA-PPO}
C.~Berner, G.~Brockman, B.~Chan, V.~Cheung, P.~Debiak \emph{et~al.}, ``Dota 2
  with large scale deep reinforcement learning,'' 2019, [Online]
  https://arxiv.org/pdf/1912.06680.pdf.

\bibitem{DDPG}
T.~Lillicrap, J.~J. Hunt, A.~Pritzel, N.~Heess, T.~Erez, Y.~Tassa, D.~Silver,
  and D.~Wierstra, ``Continuous control with deep reinforcement learning,''
  2016, [Online] https://arxiv.org/pdf/1509.02971.pdf.

\bibitem{LinUCB}
L.~Li, W.~Chu, J.~Langford, and R.~E. Schapire, ``A contextual-bandit approach
  to personalized news article recommendation,'' in \emph{Proc. Int. Conf.
  WWW}, Raleigh, USA, Apr. 2010.

\bibitem{DeepBaysianBandits}
C.~Riquelme, G.~Tucker, and J.~Snoek, ``Deep {B}ayesian bandits showdown: {A}n
  empirical comparison of {B}ayesian deep networks for {T}hompson sampling,''
  in \emph{Proc. Int. Conf. Learn. Represent.}, Vancouver, Canada, May 2018.

\bibitem{gao2021ResourceAllocation}
X.~Gao, Y.~Liu, X.~Liu, and L.~Song, ``Machine learning empowered resource
  allocation in {IRS} aided {MISO-NOMA} networks,'' 2021, [Online]
  https://arxiv.org/pdf/2103.11791.pdf.

\bibitem{Abdelrahman2020TowardsStandaloneOperation}
A.~Taha, Y.~Zhang, F.~B. Mismar, and A.~Alkhateeb, ``Deep reinforcement
  learning for intelligent reflecting surfaces: {T}owards standalone
  operation,'' in \emph{Proc. IEEE SPAWC}, Atlanta, USA, May 2020.

\bibitem{Yang2021DRL_for_Secure}
H.~Yang, Z.~Xiong, J.~Zhao, D.~Niyato, L.~Xiao, and Q.~Wu, ``Deep reinforcement
  learning-based intelligent reflecting surface for secure wireless
  communications,'' \emph{IEEE Trans. Wireless Commun.}, vol.~20, no.~1, pp.
  375--388, Jan. 2021.

\bibitem{Hu2021MetaSensing}
J.~Hu, H.~Zhang, K.~Bian, M.~Di~Renzo, Z.~Han, and L.~Song, ``{MetaSensing}:
  {I}ntelligent metasurface assisted {RF} {3D} sensing by deep reinforcement
  learning,'' \emph{IEEE J. Sel. Areas Commun.}, vol.~39, no.~7, pp.
  2182--2197, Jul. 2021.

\bibitem{Nguyen2021MultiUAVDRL}
K.~K. Nguyen, S.~Khosravirad, L.~D. Nguyen, T.~T. Nguyen, and T.~Q. Duong,
  ``Reconfigurable intelligent surface-assisted multi-{UAV} networks:
  {E}fficient resource allocation with deep reinforcement learning,'' 2021,
  [Online] https://arxiv.org/pdf/2105.14142.pdf.

\bibitem{Feng2020DRL_MISO}
K.~Feng, Q.~Wang, X.~Li, and C.-K. Wen, ``Deep reinforcement learning based
  intelligent reflecting surface optimization for {MISO} communication
  systems,'' \emph{IEEE Wireless Commun. Lett.}, vol.~9, no.~5, pp. 745--749,
  May 2020.

\bibitem{Huang2021MultiHop}
C.~Huang, Z.~Yang, G.~C. Alexandropoulos, K.~Xiong, L.~Wei, C.~Yuen, Z.~Zhang,
  and M.~Debbah, ``Multi-hop {RIS}-empowered terahertz communications: {A
  DRL}-based hybrid beamforming design,'' \emph{IEEE J. Sel. Areas Commun.},
  vol.~39, no.~6, pp. 1663--1677, Jun. 2021.

\bibitem{Lee2020DRL_EE}
G.~Lee, M.~Jung, A.~T.~Z. Kasgari, W.~Saad, and M.~Bennis, ``Deep reinforcement
  learning for energy-efficient networking with reconfigurable intelligent
  surfaces,'' in \emph{Proc. IEEE ICC}, Dublin, Ireland, Jun. 2020.

\bibitem{Liu2021MNOMA_deployment}
X.~Liu, Y.~Liu, Y.~Chen, and H.~V. Poor, ``{RIS} enhanced massive
  non-orthogonal multiple access networks: {D}eployment and passive beamforming
  design,'' \emph{IEEE J. Sel. Areas Commun.}, vol.~39, no.~4, pp. 1057--1071,
  Apr. 2021.

\bibitem{kim2021multiirsassisted}
J.~Kim, S.~Hosseinalipour, T.~Kim, D.~J. Love, and C.~G. Brinton,
  ``Multi-{IRS}-assisted multi-cell uplink {MIMO} communications under
  imperfect {CSI}: {A} deep reinforcement learning approach,'' 2021, [Online]
  https://arxiv.org/pdf/2011.01141.pdf.

\bibitem{alhilo2021reconfigurable}
A.~Al-Hilo, M.~Samir, M.~Elhattab, C.~Assi, and S.~Sharafeddine,
  ``Reconfigurable intelligent surface enabled vehicular communication: {J}oint
  user scheduling and passive beamforming,'' 2021, [Online]
  https://arxiv.org/pdf/2101.12247.pdf.

\bibitem{Samir2021AgeOfInformation}
M.~Samir, M.~Elhattab, C.~Assi, S.~Sharafeddine, and A.~Ghrayeb, ``Optimizing
  age of information through aerial reconfigurable intelligent surfaces: {A}
  deep reinforcement learning approach,'' \emph{IEEE Trans. Veh. Technol.},
  vol.~70, no.~4, pp. 3978--3983, Apr. 2021.

\bibitem{MehdiDRL}
Q.~Zhang, W.~Saad, and M.~Bennis, ``Distributional reinforcement learning for
  mm{W}ave communications with intelligent reflectors on a {UAV},'' 2020,
  [Online] https://arxiv.org/pdf/2011.01840.pdf.

\bibitem{Auer2002BanditsAnalysis}
P.~Auer, N.~Cesa-Bianchi, and P.~Fischer, ``Finite-time analysis of the
  multiarmed bandit problem,'' \emph{Machine Learning}, vol.~47, no.~2, pp.
  235--256, May 2002.

\bibitem{VermorelBanditsEmpirical}
J.~Vermorel and M.~Mohri, ``Multi-armed bandit algorithms and empirical
  evaluation,'' in \emph{ECML}, Berlin, Heidelberg, 2005, pp. 437--448.

\bibitem{DeepContextualBandits}
M.~Collier and H.~Llorens, ``Deep contextual multi-armed bandits,'' 2018,
  [Online] https://arxiv.org/pdf/1807.09809.pdf.

\bibitem{ActorCriticConvergenceLinearQuadratic}
Z.~Yang, Y.~Chen, M.~Hong, and Z.~Wang, ``Provably global convergence of
  actor-critic: {A} case for linear quadratic regulator with ergodic cost,'' in
  \emph{Proc. NeurIPS}, vol.~32, 2019.

\bibitem{ijcai2021-614}
Y.~Liu, A.~Halev, and X.~Liu, ``Policy learning with constraints in model-free
  reinforcement learning: {A} survey,'' in \emph{Proc. Int. Jt. Conf. Artif.
  Intell.}, Montreal, Canada, Aug. 2021, pp. 4508--4515.

\bibitem{alexandg_MRT}
G.~C. Alexandropoulos and M.~Kountouris, ``Maximal ratio transmission in
  wireless {P}oisson networks under spatially correlated fading channels,'' in
  \emph{Proc. IEEE GLOBECOM}, San Diego, USA, Dec. 2015.

\bibitem{QuantileRegression}
W.~Dabney, M.~Rowland, M.~G. Bellemare, and R.~Munos, ``Distributional
  reinforcement learning with quantile regression,'' in \emph{Proc. AAAI Conf.
  Artificial Intell.}, vol.~32, no.~1, New Orleans, USA, 2018.

\bibitem{openai2019Rubik}
OpenAI \emph{et~al.}, ``Solving {R}ubik's cube with a robot hand,'' 2019,
  [Online] https://arxiv.org/pdf/1910.07113.pdf.

\bibitem{Tavakoli2018ActionBranching}
A.~Tavakoli, F.~Pardo, and P.~Kormushev, ``Action branching architectures for
  deep reinforcement learning,'' in \emph{Proc. AAAI Conf. Artificial Intell.},
  vol.~32, no.~1, New Orleans, USA, Feb. 2018.

\bibitem{DulacArnold2015DeepRL}
G.~Dulac-Arnold, R.~Evans, H.~V. Hasselt, P.~Sunehag, T.~Lillicrap, J.~J. Hunt,
  T.~A. Mann, T.~Weber, T.~Degris, and B.~Coppin, ``Deep reinforcement learning
  in large discrete action spaces,'' 2015, [Online]
  https://arxiv.org/pdf/1512.07679.pdf.

\bibitem{QNetsforBinaryActions}
N.~Yoshida, ``{Q}-networks for binary vector actions,'' 2015, [Online]
  https://arxiv.org/pdf/1512.01332.pdf.

\bibitem{Kyriakos_large_space_2022}
K.~Stylianopoulos and G.~C. Alexandropoulos, ``Online {RIS} configuration
  learning for arbitrary large numbers of $1$-bit phase resolution elements,''
  Apr. 2022, [Online] https://arxiv.org/pdf/2204.08367.pdf.

\bibitem{NIPS_bayes_opt}
J.~Snoek, H.~Larochelle, and R.~P. Adams, ``Practical {B}ayesian optimization
  of machine learning algorithms,'' in \emph{Proc. NIPS}, vol.~25, 2012.

\end{thebibliography}

\end{document}